\pdfoutput=1
\newcommand*{\ATLASLATEXPATH}{latex/}

\documentclass[cernpreprint,texlive=2016,txfonts,UKenglish]{\ATLASLATEXPATH atlasdoc}

\usepackage[backend=biber]{\ATLASLATEXPATH atlaspackage}
\usepackage{longtable}
\usepackage{multirow}
\usepackage{rotating}
\usepackage{graphicx}
\usepackage{hyperref}
\usepackage{tabularx}
\usepackage{setspace}
\newcolumntype{Y}{>{\centering\arraybackslash}X}
\usepackage{\ATLASLATEXPATH atlasbiblatex}
\usepackage{amssymb}
\usepackage{pifont}
\usepackage{\ATLASLATEXPATH atlascontribute}

\usepackage[BSM]{\ATLASLATEXPATH atlasphysics}

\graphicspath{{logos/}{figures/}}

\usepackage{stop2L_MORIOND17-defs}
\usepackage{comment}

\AtlasTitle{Search for direct top squark pair production in final states with two leptons in $\sqrt{s} = 13$ TeV $pp$ collisions with the ATLAS detector}

\author{The ATLAS Collaboration}

\PreprintIdNumber{CERN-EP-2017-150}

\arXivId{1708.03247}

\AtlasJournal{EPJC}
\AtlasJournalRef{\EPJC\ 77 (2017) 898}
\AtlasDOI{10.1140/epjc/s10052-017-5445-x}

\AtlasAbstract{%
The results of a search for direct pair production of top squarks in events with two opposite-charge leptons (electrons or muons) are reported, using 36.1 fb$^{-1}$ of integrated luminosity from proton--proton collisions at $\sqrt{s}=13$ TeV collected by the ATLAS detector at the Large Hadron
Collider. To cover a range of mass differences between the top squark $\tilde{t}$ and lighter supersymmetric particles, four possible decay modes of the top squark are targeted with dedicated selections: the decay $\tilde{t} \rightarrow b \tilde{\chi}_{1}^{\pm}$ into a $b$-quark and the
lightest chargino with $\tilde{\chi}_{1}^{\pm} \rightarrow W \tilde{\chi}_{1}^{0}$, the decay $\tilde{t} \rightarrow t \tilde{\chi}_{1}^{0}$ into an on-shell top quark and the lightest neutralino, the three-body decay $\tilde{t} \rightarrow b W \tilde{\chi}_{1}^{0}$ and the four-body decay
$\tilde{t} \rightarrow b \ell \nu \tilde{\chi}_{1}^{0}$. No significant excess of events is observed above the Standard Model background for any selection, and limits on top squarks are set as a function of the $\tilde{t}$ and $\tilde{\chi}_{1}^{0}$  masses. The results exclude at 95\%
confidence level $\tilde{t}$ masses up to about 720 GeV, extending the exclusion region of supersymmetric parameter space covered by previous searches.}

\hypersetup{pdftitle={ATLAS document},pdfauthor={The ATLAS Collaboration}}

\begin{document}

\newcolumntype{L}[1]{>{\raggedright\arraybackslash}p{#1}}
\newcolumntype{C}[1]{>{\centering\arraybackslash}p{#1}}
\newcolumntype{R}[1]{>{\raggedleft\arraybackslash}p{#1}}

\maketitle

\section{Introduction}
\label{sec:intro}

The Standard Model (SM) of particle physics is extremely successful in describing the phenomena of elementary particles and their interactions. Nevertheless, it is believed to be only a low-energy realisation of a more general theory.
In its current form, it fails to explain several observations, such as the nature of dark matter, the baryon asymmetry of the universe and the stabilisation of the Higgs boson mass against radiative corrections from the Planck scale. 
These shortcomings could be remedied by the existence of new particles at the \TeV\ scale, which motivates extensive  searches at the Large Hadron Collider (LHC).

One of the most compelling theories beyond the SM is Supersymmetry (SUSY)~\cite{Golfand:1971iw,Volkov:1973ix,Wess:1974tw,Wess:1974jb,Ferrara:1974pu,Salam:1974ig}. SUSY is a space-time symmetry that for each SM particle postulates the existence of a partner particle whose spin ($S$) differs by one-half unit. The introduction of gauge-invariant and renormalisable interactions into SUSY models can violate the conservation of baryon number ($B$) and lepton number ($L$), resulting in a proton lifetime shorter than current experimental limits~\cite{Regis:2012sn}. This is usually solved by assuming that the multiplicative quantum number $R$-parity \cite{Farrar:1978xj}, defined as $R~=~(-1)^{3(B-L)+2S}$, is conserved.

In the framework of a generic $R$-parity-conserving model, SUSY particles are produced in pairs, and the lightest supersymmetric particle (LSP) is stable and a candidate for dark matter~\cite{Goldberg:1983nd,Ellis:1983ew}. 
The scalar partners of right-handed and left-handed quarks (squarks), \squarkR and \squarkL, can mix to form two mass eigenstates, $\tilde q_1$ and $\tilde q_2$, with $\tilde q_1$ defined to be the lighter one. 
In the case of the supersymmetric partner of the top quark, \stop, large mixing effects can lead to one top squark mass eigenstate, \stopone, that is significantly lighter than the other squarks. 
The charginos and neutralinos are mixtures of the bino, winos and Higgsinos that are superpartners of the U(1) and SU(2) gauge bosons and the Higgs bosons, respectively.
Their mass eigenstates are referred to as $\tilde{\chi}_{i}^{\pm}$ $(i=1,2)$ and $\tilde{\chi}_{j}^{0}$ $(j=1,2,3,4)$ in order of increasing masses.
In a large variety of models, the  LSP is the lightest neutralino \ninoone.

\begin{figure}[!htb]
\centering
\subfloat[two-body $b\chinoonepm$ decay]{\label{subfig:stopbC1} \includegraphics[width=0.32\textwidth]{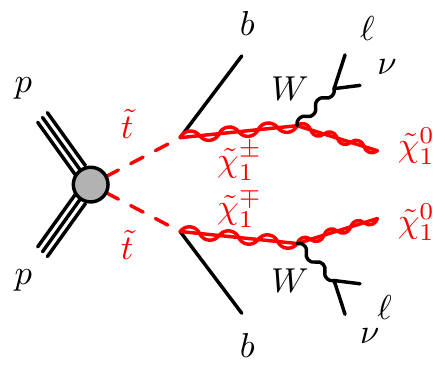}}
\subfloat[two-body $t\ninoone$ decay]{\label{subfig:stoptN1}\includegraphics[width=0.32\textwidth]{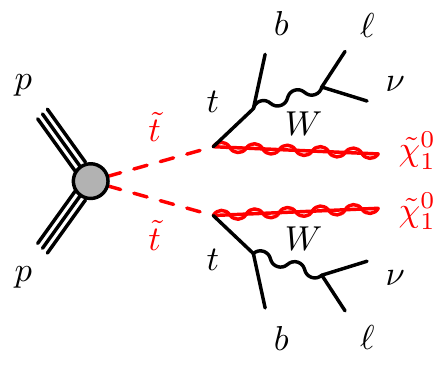}}\\
\subfloat[three-body $b\Wboson\ninoone$ decay]{\label{subfig:stoptB}\includegraphics[width=0.32\textwidth]{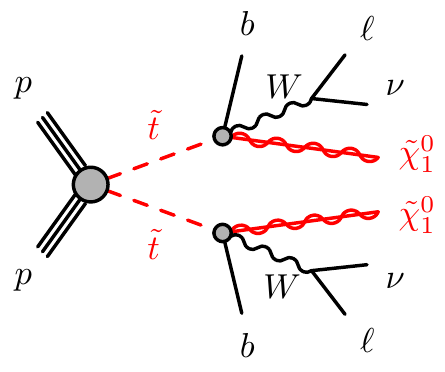}}
\subfloat[four-body $b\ell\nu\ninoone$ decay]{\label{subfig:stopfB}\includegraphics[width=0.32\textwidth]{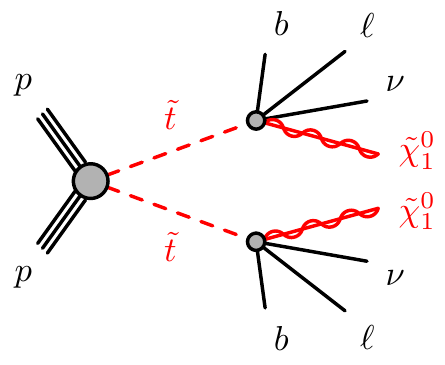}}
\caption{Diagrams representing the four main signals targeted by the analyses: (a) the decay of the top squark via the lightest chargino ($\stop \rightarrow b \chinoonepm$), (b) the two-body decay into an on-shell top quark and the lightest neutralino ($\stop \rightarrow t \ninoone$), (c) the three-body decay mode into an on-shell $W$ boson, a $b$-quark and the lightest neutralino ($\stop \rightarrow b W \ninoone$) and (d) the four-body decay mode ($\stop \rightarrow b f f' \ninoone$) where the two fermions $f$ and $f'$ are a lepton with its neutrino in this article. 
}
\label{fig:diagram}
\end{figure} 

In this paper a search for direct pair production of the top squark is reported, in final states with two isolated leptons (electrons or muons) and missing transverse momentum. The search utilises 36.1~\ifb\ of proton--proton collision data collected by the ATLAS experiment in 2015 and 2016 at a centre-of-mass energy $\sqrt{s}=13$~\TeV.

The top squark is assumed to decay into either the lightest chargino or the lightest neutralino. Depending on the mass difference between the top squark and the lighter SUSY particles, different decay modes are relevant. The decays $\stop \rightarrow t \ninoone$ and $\stop \rightarrow b \chinoonepm$ (where $t$ and $b$ represent either the quark or the anti-quark, depending on the charge conjugation) with $\chinoonepm \rightarrow \Wboson \ninoone$ dominate when they are kinematically accessible.
For intermediate mass differences, $m_{\ninoone}+m_{W}+m_{b} < m_{\stop} < m_{\ninoone}+m_{t}$, the three-body decay $\stop \rightarrow b W \ninoone$ is considered. For smaller mass differences, the four-body decay channel $\stop \rightarrow b f f' \ninoone$, where $f$ and $f'$ are two fermions from the $\Wboson^{*}$ decay, is assumed to occur. In this search, $f$ and $f'$ are a lepton and its associated neutrino. For each of these decay modes, shown by the diagrams in Figure~\ref{fig:diagram}, a dedicated event selection is performed to optimise the search significance, as detailed in Table~\ref{tab:index}.

The results of the searches are interpreted in simplified models~\cite{Alwall:2008ve,Alwall:2008ag,Alves:2011wf} as a function of the top squark and lightest neutralino masses. Additionally, results are also interpreted in one phenomenological Minimal Supersymmetric Standard Model (pMSSM)~\cite{Fayet:1976et,Fayet:1977yc,Djouadi:1998di,Berger:2008cq} model including the following decay modes: $\stop \rightarrow t \ninoone$, $\stop \rightarrow b \chinoonepm$ with $\chinoonepm \rightarrow \Wboson \ninoone$ and $\stop \rightarrow t \ninotwo$, with $\ninotwo \rightarrow h/\Zboson \ninoone$.
Previous ATLAS~\cite{Aad:2015pfx,Aaboud:2016lwz} and CMS~\cite{CMS-SUS-13-011,CMS-SUS-13-004,CMS-SUS-13-019,CMS-SUS-13-023,CMS-SUS-14-001,CMS-SUS-14-006,CMS-SUS-14-015,CMS-SUS-14-021,CMS-SUS-14-015,CMS-SUS-15-003,CMS-SUS-15-004,CMS-SUS-15-005,CMS-SUS-16-008,CMS-SUS-16-009} analyses have set exclusion limits at 95\% confidence level (CL) on the signal scenarios considered here. When considering simplified models including the $\stop \rightarrow t \ninoone$ decay, top squark masses up to about 700~\GeV\ have been excluded for a nearly massless lightest neutralino. For the same assumptions about the lightest neutralino mass, if the $\stop \rightarrow b \chinoonepm$ decay is dominant, top squark masses up to about 500~\GeV\ have been excluded.

\begin{table}
\caption{\label{tab:index} Summary of the sections dedicated to the two-body, three-body and four-body selections and signal types targeted by each
selection.}
\begin{center}
\begin{tabular}{l| c | c | c >{\centering}m{2.5cm} }
\toprule
            & Two-body  & Three-body & Four-body \\
\midrule
Variables   & \multicolumn{3}{c}{\,\,\hyperref[sec:variables]{Section~\ref*{sec:variables}}} \\
\midrule
Event selection   & \hyperref[sec:twobody_selection]{Section~\ref*{sec:twobody_selection}} & \hyperref[sec:threebody_selection]{Section~\ref*{sec:threebody_selection}} & \hyperref[sec:fourbody_selection]{Section~\ref*{sec:fourbody_selection}} \\
Background determination    & \hyperref[sec:twobody_backgrounds]{Section~\ref*{sec:twobody_backgrounds}} & \hyperref[sec:threebody_backgrounds]{Section~\ref*{sec:threebody_backgrounds}} & \hyperref[sec:fourbody_backgrounds]{Section~\ref*{sec:fourbody_backgrounds}} \\
Results    & \hyperref[sec:twobody_results]{Section~\ref*{sec:twobody_results}} & \hyperref[sec:threebody_results]{Section~\ref*{sec:threebody_results}} & \hyperref[sec:fourbody_results]{Section~\ref*{sec:fourbody_results}} \\
\midrule
Interpretation   & \multicolumn{3}{c}{\,\,\hyperref[sec:interpretations]{Section~\ref*{sec:interpretations}}} \\
\midrule
Targeted decay modes  & $b\chinoonepm$ and $t\ninoone$ &  $bW\ninoone$  & $b \ell \nu \ninoone$ \\
Signal diagram         & Figures~\ref{subfig:stopbC1} and~\ref{subfig:stoptN1} & Figure~\ref{subfig:stoptB} & Figure~\ref{subfig:stopfB} \\
Targeted $m_{\stop}$ range  & $> m_b+m_{\chinoonepm}$  & $\geq m_b+m_W+ m_{\ninoone}$ & $< m_b+m_W+ m_{\ninoone}$ \\
                       		& or $> m_t+m_{\ninoone}$ & and $< m_t+m_{\ninoone}$ & \\
  \bottomrule
\end{tabular}
\end{center}
\end{table}

\FloatBarrier
\section{ATLAS detector}
\label{sec:detector}

\newcommand{\AtlasCoordFootnote}{%
ATLAS uses a right-handed coordinate system with its origin at the nominal interaction point (IP) in the centre of the detector and the $z$-axis along the beam pipe.
The $x$-axis points from the IP to the centre of the LHC ring, and the $y$-axis points upwards.
Cylindrical coordinates $(r,\phi)$ are used in the transverse plane, $\phi$ being the azimuthal angle around the $z$-axis.
The pseudorapidity is defined in terms of the polar angle $\theta$ as $\eta = -\ln \tan(\theta/2)$.
Rapidity is defined as $y=0.5 \ln \left[ \left( E + p_{z} \right) / \left( E - p_{z} \right) \right]$ where $E$ denotes the energy and $p_{z}$ is the component of the momentum along the beam direction.
}

The ATLAS detector~\cite{PERF-2007-01} at the LHC is a multi-purpose particle detector with a cylindrical forward-backward symmetric geometry\footnote{\AtlasCoordFootnote} and an approximate $4\pi$ coverage in solid angle.
It consists of an inner tracking detector (ID) surrounded by a thin superconducting solenoid providing a \SI{2}{\tesla} axial magnetic field, electromagnetic and hadron calorimeters, and a muon spectrometer.
The inner tracking detector covers the pseudorapidity range $|\eta| < 2.5$. It consists of silicon pixel, silicon microstrip, and transition radiation tracking detectors. 
The newly installed innermost layer of pixel sensors~\cite{ATLAS-TDR-19} was operational for the first time during the 2015 data-taking. Lead/liquid-argon (LAr) sampling calorimeters provide electromagnetic (EM) energy measurements with high granularity. A hadron (steel/scintillator-tile) calorimeter covers the central pseudorapidity range ($|\eta| < 1.7$). The end-cap and forward regions are instrumented with LAr calorimeters for both the EM and hadronic energy measurements up to $|\eta| = 4.9$. The muon spectrometer surrounds the calorimeters and features three large air-core toroid superconducting magnets with eight coils each. It includes a system of precision tracking chambers and fast detectors for triggering. 
The field integral of the toroids ranges between 2.0 and \SI{6.0}{\tesla m} across most of the detector.

\FloatBarrier
\section{Data samples and event reconstruction}
\label{sec:objects}

The data were collected by the ATLAS detector in 2015 and 2016 during $pp$ collisions at a centre-of-mass energy of $\sqrt{s} = 13$ TeV, with a peak instantaneous luminosity of ${\cal L} =1.4 \times 10^{34}$~cm$^{-2}$s$^{-1}$, a bunch spacing of 25~ns, and an average number of $pp$ interactions per bunch crossing (pile-up) of $\langle \mu \rangle = 14$ in 2015 and $\langle \mu \rangle = 24$ in 2016. Only events taken in stable beam conditions, and for which all relevant detector systems were operational, are considered in this analysis. The integrated luminosity of the resulting data set is 36.1~\ifb, with an uncertainty of $\pm3.2$\%. This uncertainty is derived, following a methodology similar to that detailed in Ref.~\cite{Lumi:8TeV}, from a preliminary calibration of the luminosity scale using $x$--$y$ beam-separation scans performed in August 2015 and May 2016.

Candidate events are required to have a reconstructed vertex with at least two associated tracks with transverse momentum \pT~>~400~MeV. The vertex with the highest scalar sum of the squared transverse momenta of the associated tracks is considered the primary vertex of the event. 

Electron (\textit{baseline}) candidates are reconstructed from three-dimensional electromagnetic calorimeter energy depositions matched to ID tracks, and are required to have pseudorapidity $|\eta|<2.47$, $\pT>\SI{7}{\GeV}$, and to pass a loose likelihood-based identification requirement~\cite{Aaboud:2016vfy}. The likelihood input variables include measurements of calorimeter shower shapes and of track properties from the ID. 

Muon (\textit{baseline}) candidates are reconstructed in the pseudorapidity region $|\eta|<2.4$ from muon spectrometer tracks matching ID tracks.
They must have $\pT>\SI{7}{\GeV}$ and must pass the medium identification requirements defined in Ref.~\cite{PERF-2015-10}, which are based on requirements on the number of hits in the different ID and muon spectrometer subsystems, and on the significance of the charge-to-momentum ratio ($q/p$) measurement ~\cite{PERF-2015-10}.

Jets are reconstructed from three-dimensional energy clusters in the calorimeter~\cite{PERF-2014-07} with the anti-$k_t$ jet clustering algorithm~\cite{Cacciari:2008,Cacciari:2011ma} with a radius parameter $R=0.4$.
Only jet candidates with $\pT>\SI{20}{\GeV}$ and $|\eta|<2.8$ are considered. Jets are calibrated as described in Refs.~\cite{PERF-2015-05,PERF-2016-04}, and the expected average energy contribution from pile-up clusters is subtracted according to the jet area~\cite{ATL-PHYS-PUB-2015-015}.
Additional selections are applied to jets with $\pt<\SI{60}{GeV}$ and $|\eta|<2.4$ in order to reject jets produced in pile-up collisions~\cite{ATLAS-CONF-2014-018}.
The ``medium'' working point is used for the pile-up rejection, which has an efficiency of about 92\% for jets produced by the hard scatter. 
Jets resulting from the hadronisation of $b$-quarks are identified using a multivariate $b$-tagging algorithm ({\textsc MV2c10}), which is based on quantities such as impact parameters of associated tracks and reconstructed secondary vertices \cite{ATL-PHYS-PUB-2016-012,PERF-2012-04}. This algorithm is used at a working point that provides 77\% $b$-tagging efficiency in simulated \ttbar events, and a rejection factor of 134 for light-quark flavours and gluons and 6 for charm jets. The jets satisfying the $b$-tagging requirements are referred to as $b$-jets.

Events are discarded if they contain any jet with $\pT>\SI{20}{\GeV}$ failing to satisfy basic quality selection criteria that reject detector noise and non-collision backgrounds~\cite{ATLAS-CONF-2010-038}.

To resolve reconstruction ambiguities, an overlap removal algorithm is applied to candidate leptons and jets. Non-$b$-tagged jets which lie within $\Delta R =\sqrt{(\Delta y)^2+(\Delta\phi)^2}$ < 0.2 (here $y$ stands for the rapidity) from an electron candidate are removed, and the same is done for jets which lie close to a muon candidate and are consistent with the characteristics of jets produced by muon bremsstrahlung. Finally, any lepton candidate which lies within $\Delta R < 0.4$ from the direction of a surviving jet candidate is removed, in order to reject leptons from the decay of a $b$- or $c$-hadron. Electrons which share an ID track with a muon candidate are also removed. 

Additional selections are then applied to the remaining lepton and jet candidates. Tighter requirements on the lepton candidates are imposed, which are then referred to as ``signal'' electrons or muons.
Signal electrons must satisfy the \textit{medium} likelihood-based identification requirement as defined in Ref.~\cite{Aaboud:2016vfy}. 
Signal electrons must have a transverse impact parameter with respect to the reconstructed primary vertex, $d_0$, with a significance of $\vert d_0\vert/\sigma(d_0) < 5$.
For signal muons, the corresponding requirement is $\vert d_0\vert/\sigma(d_0) < 3$. The tracks associated with the signal leptons must have a longitudinal impact parameter with respect to the reconstructed primary vertex, $z_0$, satisfying $\vert z_0 \sin\theta\vert  < 0.5$~mm. 
Isolation criteria are applied to both electrons and muons by placing an upper limit on the sum of the transverse energy of the calorimeter energy clusters in a cone of $\Delta R_\eta = \sqrt{(\Delta \eta)^2+(\Delta\phi)^2} = 0.2$ around the electron (excluding the deposit from the electron itself), and the scalar sum of the \pt of tracks within a variable-size cone around the lepton (excluding its own track). The track isolation cone radius for electrons (muons) is given by the smaller of $\Delta R = \SI{10}{~GeV}/\pt$ and $\Delta R_\eta = 0.2\,(0.3)$. The isolation criteria are optimised such that the isolation selection efficiency is uniform across $\eta$, and it increases from 95\% for $\pt = \SI{25}{~GeV}$ to 99\% for $\pt = \SI{60}{~GeV}$ in $Z\to\ell\ell$ events.

Jets are required to have $| \eta | < 2.5$.

The missing transverse momentum ($\mathbf p^{\mathrm{miss}}_{\mathrm{T}}$), whose magnitude is denoted by \MET, is defined as the negative vector sum of the transverse momenta of all identified baseline objects (electrons, muons, jets) and an additional soft term.
The soft term is constructed from all tracks that are not associated with any reconstructed electron, muon or jet, but which are associated with the primary vertex. In this way, the \MET value is adjusted for the best calibration of the jets and the other identified objects above, while maintaining pile-up independence in the soft term~\cite{ATL-PHYS-PUB-2015-027, ATL-PHYS-PUB-2015-023}.

\FloatBarrier
\section{Event selection}
\label{sec:selection-common}

For the two-body and three-body selections, events are accepted if they pass an online selection (trigger) requiring a minimum of two electrons, two muons or an electron and a muon matched to the trigger objects. The offline selection requires that the leading lepton has a \pT larger than 25~GeV and the subleading lepton a \pT larger than 20~GeV, ensuring that trigger efficiencies are constant in the relevant phase space. The four-body selection accepts events passing an \MET-based trigger and having offline \MET > 200 GeV. This ensures that the trigger efficiency is constant in the relevant phase space. Using this trigger permits the use of a reduced lepton \pT threshold of 7~GeV, increasing acceptance for the low lepton \pT produced in the four-body $\stop \rightarrow b\ell\nu\ninoone$ decay. 

Events are required to have exactly two signal leptons which must be of opposite charge (electrons, muons, or one of each) with an invariant mass (regardless of the flavour of the leptons in the pair) \mll greater than $20~\GeV$ ($10~\GeV$ for the four-body selection) in order to remove leptons from low-mass resonances. Except for the four-body selection, events with same-flavour (SF) lepton pairs with \mll\ between 71.2 and 111.2 GeV are rejected, in order to reduce the backgrounds with leptons produced by $Z$ bosons. No additional selection is applied to the \mll\ value of different-flavour (DF) lepton pairs. In the following, the requirements described in the preceding part of this section are referred to as ``common selection''.

\subsection{Discriminators and kinematic variables}
\label{sec:variables}

For the different decay modes considered, dedicated sets of discriminating variables are used to separate the signal from the SM backgrounds.

The missing transverse momentum and the \pT of the leading leptons and jets are used to define three useful ratio variables :

\begin{equation*}
 \Rone = \MET / (\MET + \pT(\ell_1) + \pT(\ell_2) + \pT(j_1) + \pT(j_2)),
\end{equation*}
\begin{equation*}
 \Rll = \MET / (\pT(\ell_1) + \pT(\ell_2)),
\end{equation*}
and 
\begin{equation*}
  R_{2\ell 4j} = {\MET/(\MET + \pT(\ell_1) + \pT(\ell_2) + \sum_{i=1,...,N\leq 4}\pT(j_i))},
\end{equation*}
where $\pT(\ell_1)$ and $\pT(\ell_2)$ are the leading and subleading lepton transverse momenta and $\pT(j_{i =1,...,N\leq 4})$ are the transverse momenta in decreasing order of up to the four leading jets. The variables \Rone and \Rll are used to reject backgrounds, e.g. \zjets, which peak at lower values than the signal. Similarly, $R_{2\ell 4j}$ is a powerful discriminant against multi-jet events.

Other variables employed are :

\begin{itemize} 

\item[-] \pbll: defined as the vector
\begin{equation*}
\pbll = \mathbf p^{\mathrm{miss}}_{\mathrm{T}} +
  \mathbf \pT(\ell_1) + \mathbf \pT(\ell_2).
\end{equation*}  
The \pbll variable, with magnitude $p^{\ell\ell}_{\mathrm{T,boost}}$, can be interpreted as the opposite of the vector sum of all the transverse hadronic activity in the event.

\item[-] \dphib: the azimuthal angle between the $ \mathbf p^{\mathrm{miss}}_{\mathrm{T}} $ vector and the \pbll\ vector.

\item[-] \DX: defined as 
\begin{equation*}
\DX = \frac{2\cdot \left( p_{\textrm z} (\ell_1)+p_{\textrm z} (\ell_2)\right)}{E_{\textrm CM}}
\end{equation*} where $E_{\textrm CM}=13$ TeV is used and $p_{\textrm z} (\ell_1)$,$p_{\textrm z} (\ell_2)$ are respectively the leading and subleading lepton longitudinal momenta. This variable helps to discriminate between gluon- and quark-initiated processes. The former tend to peak towards zero, while the latter tend to peak at higher values.

\item[-] \cosThetaB: the cosine of the angle between the direction of motion of either of the two
    leptons and the beam axis in the centre-of-mass frame of the two leptons~\cite{Melia:2011cu}.
    This variable is sensitive to the spin of the pair-produced particle, providing additional rejection against diboson backgrounds.

  \item[-] \mttwoL: lepton-based ``stransverse'' mass. The stransverse mass defined in Refs.~\cite{Lester:1999tx,Barr:2003rg} is a kinematic variable used to bound the masses of a pair of identical particles which have each decayed into a visible and an invisible particle. This quantity is defined as
\begin{equation*} 
\mttwo
( \mathbf p_{\mathrm{T,1}}, \mathbf p_{\mathrm{T,2}}, \mathbf
q_{\mathrm{T}}) = \min_{\mathbf q_{\mathrm{T,1}} + \mathbf q_{\mathrm{T,2}} = 
\mathbf q_{\mathrm{T}} }
\left\{ \max [\; 
    m_{\mathrm{T}}( \mathbf p_{\mathrm{T,1}}, \mathbf q_{\mathrm{T,1}} ),  
    m_{\mathrm{T}}( \mathbf p_{\mathrm{T,2}}, \mathbf q_{\mathrm{T,2}} 
    ) 
\;] 
\right\} ,
\label{mt2eq}
\end{equation*}
\noindent where $m_{\mathrm T}$ indicates the transverse mass,\footnote{The transverse mass is defined by the equation $m_\textup{T}(\mathbf p_{\mathrm{T}}, \mathbf q_{\mathrm{T}})=\sqrt{2|\mathbf p_{\mathrm{T}}||\mathbf q_{\mathrm{T}}|(1-\mathrm{cos}(\Delta\phi))}$, where $\Delta\phi$ is the angle between the particles of negligible mass with transverse momenta $\mathbf p_{\mathrm{T}}$ and $\mathbf q_{\mathrm{T}}$.} $\mathbf p_{\mathrm{T,1}}$ and $\mathbf p_{\mathrm{T,2}}$ are the transverse momentum vectors of two particles, and $\mathbf
q_{\mathrm{T,1}}$ and $\mathbf q_{\mathrm{T,2}}$ are transverse momentum vectors with $ \mathbf q_{\mathrm{T}} = \mathbf q_{\mathrm{T,1}} + \mathbf q_{\mathrm{T,2}}$.
The minimisation is performed over all the possible decompositions 
of  $\mathbf q_{\mathrm{T}}$. For $t\bar{t}$ or $WW$ decays with $t\rightarrow b \ell \nu$ and $\Wboson \rightarrow \ell \nu$, when the transverse momenta of the two leptons in each event are taken as  $\mathbf p_{\mathrm{T,1}}$ and
$\mathbf p_{\mathrm{T,2}}$, and $\mathbf p^{\mathrm{miss}}_{\mathrm{T}}$ as $\mathbf q_\mathrm{T}$, $m_\textup{T2}(\mathbf \pT(\ell_1), \mathbf \pT(\ell_2), \mathbf p^{\mathrm{miss}}_{\mathrm{T}})$ is bounded sharply from above by the mass of the \Wboson\ boson~\cite{Cho:2007dh,Burns:2008va}. In the $\stop \rightarrow b \chinoonepm$ decay mode the upper bound is strongly correlated with the mass difference between the chargino and the lightest neutralino. 
In this paper, $m_\textup{T2}(\mathbf \pT(\ell_1), \mathbf \pT(\ell_2), \mathbf p^{\mathrm{miss}}_{\mathrm{T}})$ is referred to simply as \mttwoL.
    
\end{itemize}

The three-body selection uses a number of ``super-razor'' variables that are defined in Ref.~\cite{Buckley:2013kua}.
They are designed to identify events with two massive parent particles (i.e. top squarks) each decaying into a set of visible (only leptons are considered in this case, all other particles including jets are ignored) and invisible particles (i.e.~neutrinos and neutralinos).
These variables are:

\begin{itemize}
    \item[-] \RPT : defined as
    \begin{equation*}
         \RPT = \frac{|\vec{J}_{\textrm T}|}{|\vec{J}_{\textrm T}| + \sqrt{\hat{s}}_{\textrm R}/4},
    \end{equation*} 
	where $\vec{J}_{\textrm T}$ is the vector sum of the transverse momenta of the visible particles and the missing transverse momentum,
	and $\sqrt{\hat{s}_{\textrm R}}$ is a measure of the system's energy in the razor frame $R$ as defined in Ref.~\cite{Buckley:2013kua} as the frame in which the two visible leptons have equal and opposite $p_{\textrm z}$. In the case where all possible visible particles are considered, the razor frame $R$ becomes an approximation of the 
	pair production centre-of-mass frame with the centre-of-mass energy $\sqrt{\hat{s}}_{\textrm R}$.
	In this analysis, only leptons are considered in the visible system. Therefore, \RPT tends towards zero in
	events that do not contain additional activity (i.e. dibosons) due to vanishing $|\vec{J}_{\textrm T}|$, whereas in events
	that contain additional activity (i.e. \ttbar) this variable tends towards unity, thus providing separation power between the two cases. 
    \item[-] \GAMr: The Lorentz factor associated with the boosts from the razor frame $R$ to the approximations of the two decay frames of the parent particles. It is a measure of how the two visible systems are distributed, tending towards unity when the visible particles are back-to-back or have different momenta, while preferring lower values when they are equal in momenta and collinear. 
    \item[-] \MDR: defined as
    \begin{equation*}
        \MDR\ = \frac{\sqrt{\hat{s}_{\textrm R}}}{\gamma_{{\textrm R}+1}}.
    \end{equation*}
    This variable has a kinematic end-point that is proportional to the mass-splitting between the parent particle
    and the invisible particle. Therefore, it provides rejection against both the top quark and diboson production 
    processes when it is required to be greater than the mass of the \Wboson\ boson, and in this case it also helps to reject the residual \zjets\ background.
    \item[-] \DPB: The quantity  \DPB is the azimuthal angle between the razor boost from the laboratory to the $R$ frame 
    and the sum of the visible momenta as evaluated in the $R$ frame.
    For systems where the invisible particle has a mass that is comparable to the pair-produced massive particle, 
    this variable has a pronounced peak near $\pi$, making it, in general, a good discriminator in searches for models with small mass differences.
\end{itemize}

\FloatBarrier
\subsection{Two-body event selection}
\label{sec:twobody_selection}

This selection targets the top squark two-body decays (Figures~\ref{subfig:stopbC1},\ref{subfig:stoptN1}) into either a bottom quark and a chargino, with the chargino decaying into the lightest neutralino and a \Wboson\ boson, or a near-mass-shell top quark and a neutralino.

In these decays, the kinematic properties of signal events are similar to those of \ttbar events.
In particular, when the top squarks are produced at rest the momenta carried by the neutralinos in the final state are small and the discrimination difficult.
Better separation between signal events and the \ttbar background can be obtained for top squark pairs which recoil from initial-state radiation (ISR).

Three signal regions (SRs), summarised in Table~\ref{tab:SRchargino_def} and denoted by \ensuremath{\mathrm{SR(A,B,C)}^{\mathrm{2-body}}_{x}}, where $x$ stands for the lower bound of the \mttwoL interval, were optimised to target different scenarios:
\begin{itemize}

\item \SRA\ targets the decays into $b \chinoonepm$ in scenarios where $m_{\stopone}-m_{\chinoonepm}$ is below 10~\GeV\ and the $b$-jets from the decay of the \stopone are too low in energy to be reconstructed. For this reason, $b$-jets with $\pT>25$~\GeV\ are vetoed to reduce the contamination from SM processes including top quarks. No further requirement is imposed on the hadronic activity of the event. Events with SF leptons are required to have $\mll>111.2$~\GeV\ and $\Rone>0.3$ to reduce the contamination from \zjets events. The contribution from diboson production is expected to be the dominant background in the SR and it is reduced by requiring the events to have $\DX<0.07$. Furthermore, events are required to have $\mttwoL>180$~\GeV. 

\item \SRB\ targets the decays into $b \chinoonepm$ in scenarios with a mass-splitting between the top squark and the chargino larger than 10~\GeV, such that the jets from the hadronisation of $b$-quarks are expected to be detectable. At least two jets with $\pT>25$~\GeV\ are required, with at least one of them being identified as a $b$-jet. Events from \ttbar and \zjets production are suppressed by requiring $\dphib<1.5$. The main expected SM processes satisfying this selection are \ttbar and \ttbar+\Zboson with the \Zboson boson decaying into neutrinos. A final selection of $\mttwoL>140$~\GeV\ is applied. Because of the similar final state, this selection is the most sensitive to signal scenarios in which the \stopone decays into $t+\ninoone$, with large $m_{\stopone} - m_{\ninoone}$.

\item \SRtN\ targets the decays into $t+\ninoone$, in scenarios where $m_{\stopone} \sim m_{\ninoone}+m_{t}$. Candidate events are required to have $\MET > 200$~GeV and at least three jets with $\pT>25$~\GeV, where one of the jets is interpreted as ISR. The other two jets are expected to arise from the decay of the top quarks in the final state.
One of the jets in the event is required to be $b$-tagged, effectively separating the signal events from SM diboson production. The \zjets background is suppressed by requiring \Rll\ to be larger than 1.2. Events are finally required to have $\mttwoL>110$~\GeV. 

\end{itemize}

For the model-dependent exclusion limits, a shape fit of the \mttwoL distribution is performed for the \SRA\ and \SRB\ selections: the distribution is divided into bins of width 20 GeV, starting from $\mttwoL=120$~GeV; the last bin's low boundary corresponds to the requirement on the same variable in the definitions of \SRA\ and \SRB; each bin is referred to as \ensuremath{\mathrm{SR(A,B)}^{\mathrm{2-body}}_{x,y}}, where $x$ and $y$ denote the low and high edges of the bin.

\begin{table}[!htbp]
\centering
\caption{\label{tab:SRchargino_def} \textit{Two-body selection} signal region definitions.}
\begin{tabularx}{0.95\textwidth}{l | *{2}{Y} | *{2}{Y} | *{2}{Y} }
\toprule
                                & \multicolumn{2}{c}{\SRA}   & \multicolumn{2}{c}{\SRB}  &  \multicolumn{2}{c}{\SRtN} \\
\midrule
Lepton flavour                             & SF         & DF            & SF                & DF                    & SF    & DF    \\
$\pT(\ell_1),\pT(\ell_2)$ [GeV]            & \multicolumn{2}{c|}{$> 25$, $> 20$}&  \multicolumn{2}{c|}{$> 25$, $> 20$} 			    &   \multicolumn{2}{c}{$> 25$, $> 20$}      \\[7pt]
                               			   &     &          & [20, 71.2]   &              & [20, 71.2] 		&          \\
$\mll$ [GeV]                               & $> 111.2$    & $> 20$          & or   & $> 20$              &  or 		&   $> 20$         \\
                               			   &   &           &  $> 111.2$  &               & $> 111.2$ 		&        \\[7pt]
$\Rone$                                    & $> 0.3$       & --            & \multicolumn{2}{c|}{--}                    & \multicolumn{2}{c}{--}  \\
$\Rll$                                     & \multicolumn{2}{c|}{--}     & \multicolumn{2}{c|}{--}                    & \multicolumn{2}{c}{$>1.2$}  \\
$\DX$                                      & \multicolumn{2}{c|}{$< 0.07$} & \multicolumn{2}{c|}{--}                    & \multicolumn{2}{c}{--}  \\
\dphib                                     & \multicolumn{2}{c|}{--}     & \multicolumn{2}{c|}{$< 1.5$}                & \multicolumn{2}{c}{--}  \\
\njet                         			   & \multicolumn{2}{c|}{--}     & \multicolumn{2}{c|}{$\geq 2$}              & \multicolumn{2}{c}{$\geq 3$}  \\
\nbjet                     				   & \multicolumn{2}{c|}{$= 0$}      & \multicolumn{2}{c|}{$\geq 1$}              & \multicolumn{2}{c}{$\geq 1$}  \\
\MET [GeV]                                 & \multicolumn{2}{c|}{--}     & \multicolumn{2}{c|}{--}                    & \multicolumn{2}{c}{$> 200$}  \\
$\mttwoL$ [GeV]                            & \multicolumn{2}{c|}{$> 180$}   & \multicolumn{2}{c|}{$> 140$}                  & \multicolumn{2}{c}{$> 110$}  \\
\hline
\bottomrule
\end{tabularx}

\end{table}

\FloatBarrier
\subsection{Three-body event selection}
\label{sec:threebody_selection}

This selection targets the top squark three-body decay mode (Figure~\ref{subfig:stoptB}), which is expected to be the dominant decay mode when the two-body decay mode into the lightest chargino or neutralino is kinematically forbidden, i.e. for $m_{\ninoone}+m_{W}+m_{b} < m_{\stopone} < m_{\ninoone} + m_{t}$ and $m_{\stopone} < m_{\chinoonepm} + m_{b}$. 

Two orthogonal signal regions, \SRTBW\ and \SRTBT, are summarised in Table~\ref{tab:threebodysrdefinitions}. The \SRTBW\ targets the region where $\Delta m(\stop,\ninoone) \sim m_{W}$ in which the produced $b$-jets have low transverse momentum, and hence are often not reconstructed. The second signal region \SRTBT\ targets the region in which $\Delta m(\stop,\ninoone) \sim m_{t}$. 

The two regions make use of a common set of requirements on \RPT, \GAMr, and in the two-dimensional (\cosThetaB, \DPB) plane. 
In addition, \SRTBW\ requires that no $b$-jet is identified in the event and that \MDR~$>95$~\GeV. The large \MDR\ requirement suppresses the top quark and diboson backgrounds.  In the case of \SRTBT, the requirements are: at least one $b$-jet and \MDR$>110$~\GeV. The $b$-jet requirement makes the selection orthogonal to \SRTBW, so that the two SRs can be statistically combined.
Furthermore, a slightly tighter \MDR\ requirement is necessary to eliminate the background that originates from top quark production processes.

\begin{table}[!htb]
    \centering
    \caption{\textit{Three-body selection} signal region definitions. }
    \label{tab:threebodysrdefinitions}
    \begin{tabularx}{0.75\textwidth}{l | *{2}{Y} | *{2}{Y} }
    \toprule
                                                & \multicolumn{2}{c}{\SRTBW}		& \multicolumn{2}{c}{\SRTBT} 		\\
    \midrule
    Lepton flavour                              & SF 			 & DF 				& SF 			 & DF 				\\
    $\pT(\ell_1),\pT(\ell_2)$ [GeV]             & \multicolumn{2}{c|}{$>25$, $>20$} 	& \multicolumn{2}{c}{$>25$, $>20$}  	\\[7pt]
                               			    	& [20, 71.2]   	 &              	& [20, 71.2] 		&          		\\
	$\mll$ [GeV]                              	& or   			 & $> 20$           &  or 				&   $> 20$      \\
                               			   		& $> 111.2$  	 &               	& $> 111.2$ 		&        		\\[7pt]
    \nbjet                                      & \multicolumn{2}{c|}{$=0$}     	& \multicolumn{2}{c}{$\geq 1$} 			\\
    \MDR\ [GeV]                                  & \multicolumn{2}{c|}{$>95$}       	& \multicolumn{2}{c}{$>110$} 		\\
    \RPT                                        & \multicolumn{2}{c|}{$>0.7$} 		& \multicolumn{2}{c}{$>0.7$} \\
    \GAM                                        & \multicolumn{2}{c|}{$>0.7$} 		& \multicolumn{2}{c}{$>0.7$} \\
    \DPB                                        & \multicolumn{2}{c|}{$> 0.9 |\cosThetaB| + 1.6$} & \multicolumn{2}{c}{$> 0.9 |\cosThetaB| + 1.6$} \\
    \bottomrule
    \end{tabularx}
\end{table}

\FloatBarrier
\subsection{Four-body event selection}
\label{sec:fourbody_selection}

The selection described here targets the four-body decay mode of the top squark (Figure~\ref{subfig:stopfB}) for scenarios where $m_{\stopone} < m_{\ninoone} + m_{b} + m_{W}$ and $m_{\stopone} < m_{\chinoonepm} + m_{b}$. In this region the top squark decay into $c\ninoone$ might be dominant, depending on various SUSY model parameters. The branching ratio into this final state is here assumed to be negligible. For these small mass splittings, the leptons in the final state, originating from the virtual \Wboson\ boson decays, are expected to have low \pT.

Signal events can be distinguished from SM processes if a high-\pT jet from ISR leads to a large transverse boost of the sparticle pair system and enhances the \MET value. At least two jets with \pT>25~\GeV\ are required in the event. The leading jet is considered to be the ISR jet and required to have $\pT>150$~GeV. Since the jets resulting from \stop decays tend to have low \pT in this scenario, at most one more energetic jet with $\pT>25~\GeV$ is permitted in the event and the transverse momentum of the third jet (if present) must satisfy $\pT(j_3)/\MET$< 0.14.

In order to remove events originating from low-mass resonances, the invariant mass of the two leptons, \mll, is required to be greater than 10~\GeV.
Furthermore, upper limits on \ptleplead\ and \ptlepsubl, respectively of 80 GeV and 35 GeV, are applied.

The signal region  \SRfB\ is defined as summarised in Table~\ref{tab:fourbodysrdefinitions}.
The two variables $R_{2\ell 4j} $ and \Rll\ must be larger than $0.35$ and $12$ to reject multi-jet and \ttbar backgrounds, respectively.
Finally, the two most energetic jets in the event must not be tagged as $b$-jets.

\begin{table}[!h]
\centering
\caption{\label{tab:fourbodysrdefinitions}\textit{Four-body selection} signal region definition.}
    \begin{tabular}{l | c }
        \toprule
       & \SRfB \\
        \midrule
        Lepton flavour          & SF and DF \\
        \MET [GeV] & > 200\\
        $\pT(\ell_1)$ [GeV] & [7, 80] \\
        $\pT(\ell_2)$ [GeV] & [7, 35] \\
        \mll [GeV] & > 10 \\
        \njet  & $\geq2$ \\
        $\pT(j_1)$ [GeV] &  > 150\\
        $\pT(j_2)$ [GeV] &  > 25\\     
        $\pT(j_3)/\MET $ & < 0.14\\
        $R_{2\ell 4j}$ & > 0.35\\
        $\Rll$ &  > 12\\
        \nbjet & veto on $j_1$ and  $j_2$  \\
        \bottomrule
    \end{tabular}
\end{table}

\FloatBarrier
\section{Samples of simulated events}
\label{sec:mcsamples}

Monte Carlo (MC) simulated event samples are used to aid in the estimation of the background from SM processes and to model the SUSY signal. 
The event generator, parton shower and hadronisation generator, cross-section normalisation, parton distribution function (PDF) set and underlying-event parameter set (tune) of these samples are given in Table~\ref{tab:MC}, and more details of the event generator configurations can be found in Refs.~\cite{ATL-PHYS-PUB-2016-004,ATL-PHYS-PUB-2016-003,ATL-PHYS-PUB-2016-002,ATL-PHYS-PUB-2016-005}.
Cross-sections calculated at next-to-next-to-leading order (NNLO) in QCD including resummation of next-to-next-to-leading logarithmic (NNLL) soft gluon terms were used for top quark production processes. 
For production of top quark pairs in association with vector or Higgs bosons, cross-sections calculated at next-to-leading order (NLO) were used, and the event generator cross-sections calculated by \SHERPA (at NLO for most of the processes) are used when normalising the multi-boson backgrounds.
In all MC samples, except those produced by \SHERPA, the \textsc{EvtGen}~v1.2.0 program~\cite{EvtGen} was used to model the properties of the bottom and charm hadron decays.
Additional MC samples are used when estimating systematic uncertainties, as detailed in Section~\ref{sec:systematics}.

\begin{table*}[t]
\caption{Simulated signal and background event samples: the corresponding event generator, parton shower generator, cross-section normalisation, PDF set and underlying-event tune are shown.\vspace{-0.2cm}}
\label{tab:MC}
\begin{center}
\resizebox{\textwidth}{!}
{\normalsize
\begin{tabular}{l l c c c c}
\hline
Physics process    & Event generator & Parton shower generator & Cross-section & PDF set & Tune \\
                   &          &           & normalisation &     &      \\
\hline
SUSY Signals           & \AMCATNLO 2.2.3~\cite{Alwall:2014hca} & \PYTHIA 8.186~\cite{Sjostrand:2007gs} & NLO+NLL~\cite{Beenakker:1996ch,Kulesza:2008jb,Kulesza:2009kq,Beenakker:2009ha,Beenakker:2011fu,Borschensky:2014cia} & NNPDF23LO~\cite{Lai:2010vv} & A14~\cite{pub-2014-021} \\
\hline
  \zjets & \SHERPA 2.2.1~\cite{Gleisberg:2008ta}   & \SHERPA 2.2.1       & NNLO~\cite{Catani:2009sm}  & NLO CT10~\cite{Lai:2010vv}         & \SHERPA default\\ \hline
  
  $t\bar{t}$        & \textsc{powheg-box} v2~\cite{Alioli:2010xd} & \PYTHIA6.428~\cite{Sjostrand:2006za}  & NNLO+NNLL~\cite{Czakon:2013goa,Czakon:2012pz,Czakon:2012zr,Baernreuther:2012ws,Cacciari:2011hy,Czakon:2011xx}  & NLO CT10  &\textsc{Perugia2012}~\cite{Skands:2010ak}\\
  $Wt$              & \textsc{powheg-box} v2 & \PYTHIA6.428  & NNLO+NNLL~\cite{Kidonakis:2010ux}  & NLO CT10   & \textsc{Perugia2012}\\ \hline

$t\bar{t} W/Z/\gamma^{*}$ & \AMCATNLO 2.2.2   & \PYTHIA 8.186   & NLO~\cite{Alwall:2014hca}   & NNPDF23LO & A14    \\
Diboson        & \SHERPA 2.2.1       & \SHERPA 2.2.1    & Generator NLO  & NLO CT10      & \SHERPA default \\
\hline
$t\bar{t}h$    & \AMCATNLO 2.2.2        & \HERWIG 2.7.1~\cite{Corcella:2000bw}          & NLO~\cite{Dittmaier:2012vm}   & CTEQ6L1~\cite{Pumplin:2002vw} & A14  \\
$Wh$, $Zh$     & \AMCATNLO 2.2.2       & \PYTHIA 8.186   & NLO~\cite{Dittmaier:2012vm}   & NNPDF23LO     & A14  \\
$t\bar{t} WW$, $\ttbar\ttbar$  & \AMCATNLO 2.2.2       & \PYTHIA 8.186   & NLO~\cite{Alwall:2014hca}     & NNPDF23LO     & A14  \\
$tZ$, $tWZ$, $t\bar{t}t$    & \AMCATNLO 2.2.2       & \PYTHIA 8.186   & LO                          & NNPDF23LO     & A14  \\
Triboson       & \SHERPA 2.2.1        & \SHERPA 2.2.1    & Generator LO, NLO            & CT10       & \SHERPA default \\
\hline
\end{tabular}
}
\end{center}
\end{table*}

SUSY signal samples were generated from leading-order (LO) matrix elements with up to two extra partons, using the \AMCATNLO~\cite{Alwall:2014hca} event generator. The two-body signals used \PYTHIA 8.186~\cite{Sjostrand:2007gs} for the modelling of the SUSY decay chain, parton showering, hadronisation and the description of the underlying event. The three-body and four-body signals were decayed with \textsc{Pythia}8+\textsc{MadSpin} \cite{Artoisenet:2012st} instead. Parton luminosities were provided by the \textsc{NNPDF23LO} PDF set. 
Jet--parton matching was realised following the CKKW-L prescription~\cite{Lonnblad:2011xx}, with a matching scale set to one quarter of the pair-produced superpartner mass. In all cases, the mass of the top quark was fixed at 172.5~\GeV.  
Signal cross-sections were calculated to next-to-leading order in the strong coupling constant, adding the resummation of soft gluon emission at next-to-leading-logarithmic accuracy (NLO+NLL)~\cite{Beenakker:1997ut,Beenakker:2010nq,Beenakker:2011fu}. The nominal cross-sections and their uncertainties were taken from an envelope of cross-section predictions using different PDF sets and factorisation and renormalisation scales, as described in Ref.~\cite{Borschensky:2014cia}. All two-, three- and four-body samples were generated assuming a 100\% branching ratio into the respective final states.
 
For the pMSSM inspired models, the mass spectrum of sparticles was calculated using \textsc{Softsusy} 3.7.3~\cite{Allanach:2001kg} and cross-checked with \textsc{SPheno} 3.3.8~\cite{Porod:2003um,Porod:2011nf} and \textsc{Suspect} 2.5~\cite{Djouadi:2002ze}. \textsc{Hdecay} and \textsc{Sdecay}, included in \textsc{Susy-Hit}~\cite{Djouadi:2006bz} were used to generate decay tables of the SUSY particles. 

To simulate the effects of additional $pp$ collisions in the same and nearby bunch crossings, additional interactions were generated using the soft QCD processes of \PYTHIA 8.186 
with the A2 tune~\cite{ATLAS-PHYS-PUB-2012-003} and the MSTW2008LO PDF set~\cite{Martin:2009iq}, and they were overlaid onto each simulated hard-scatter event. 
The MC samples were reweighted to the pile-up distribution observed in the data. The MC samples were processed through an ATLAS detector simulation~\cite{SOFT-2010-01} based on \textsc{Geant4}~\cite{Agostinelli:2002hh} or, in the case of $\ttbar t$ and the SUSY signal samples, a fast simulation using a parameterisation of the calorimeter response and \textsc{Geant4} for the other parts of the detector~\cite{ATL-PHYS-PUB-2010-013}.
All MC samples are reconstructed in the same manner as the data. Corrections derived from data control samples are applied to simulated events to account for differences between data and simulation in reconstruction efficiencies, momentum scale and resolution of leptons and in the efficiency and false positive rate for identifying jets resulting from the hadronisation of $b$-quarks. 

\FloatBarrier
\section{Background estimation}
\label{sec:backgrounds}

The dominant SM background processes satisfying the SR requirements are estimated by simulation, which is normalised to data and verified in separate regions of the phase space. Dedicated control regions (CRs), described in Sections~\ref{sec:twobody_backgrounds}--\ref{sec:fourbody_backgrounds}, enhanced in a particular background component are used for the normalisation. Subdominant background yields are taken directly from MC simulation or from additional independent studies in data. For each signal region, a simultaneous ``background fit'' is performed to the number of events found in the CRs, using a statistical minimisation based on a likelihood implemented in the HistFitter package \cite{Baak:2014wma}. In each fit, the normalisations of the background contributions having dedicated CRs are allowed to float, while the MC simulation is used to describe the shape of distributions of kinematical variables. The level of agreement between the background prediction and data is compared in dedicated validation regions (VRs), which are not used to constrain the background normalisation or nuisance parameters in the fit.

In order to keep the background control region kinematically as close as possible to the SR, the two-body, three-body and four-body selections use different sets of CRs. The definitions of the regions used in each analysis and the results of the fits are described in the following subsections. 

The background due to jets misidentified as leptons (hereafter referred to as ``fake'' leptons) and non-prompt leptons is collectively referred to as ``FNP'': it consists of semileptonic $\ttbar$, $s$-channel and $t$-channel single-top-quark, $W$+jets and light- and heavy-flavour multi-jet events.
It is estimated from data with a method similar to that described in Refs.~\cite{TOPQ-2010-01,TOPQ-2011-01}.
Two types of lepton identification criteria are defined for this evaluation: ``tight'' and ``loose'', corresponding to signal and baseline leptons described in Section~\ref{sec:objects}. The method makes use of the number of observed events containing loose--loose, loose--tight, tight--loose and tight--tight lepton pairs in a given SR. The probability for prompt leptons satisfying the loose selection criteria to also pass the tight selection is measured using a $Z\to\ell \ell$ ($\ell = e, \mu$) sample. 
The equivalent probability for fake or non-prompt leptons is measured in data from multi-jet- and \ttbar-enriched control samples. The number of events containing a contribution from one or two fake or non-prompt leptons is calculated from these probabilities. 

Systematic uncertainties in the samples of simulated events affect the expected yields in the different regions and are taken into account to determine the uncertainties in the background predictions. The systematic uncertainties are described by nuisance parameters, which are not constrained by the fit, since the number of floating background normalisation parameters is equal to the number of CRs. Each uncertainty source is described by a single nuisance parameter, and all correlations between background processes and selections are taken into account. 
A list of systematic uncertainties considered in the fits is provided in Section~\ref{sec:systematics}.

\FloatBarrier
\subsection{Two-body selection background determination}
\label{sec:twobody_backgrounds}

The main background sources for the two-body selection are respectively diboson production in \SRA\\ and \ttbar and \ttbar+Z in \SRB\ and \SRtN. These processes are normalised to data in dedicated CRs, summarised in Table~\ref{tab:twobody_CRdef} together with the corresponding VRs: \CRTbC\ (included in the background fits of \SRA\ and \SRB), \CRTtN\ (included in the background fit of \SRtN), \CRVVbC\ (included in the background fits of \SRA\ and \SRB), \CRTZ\ (included in the background fits of \SRA, \SRB\ and \SRtN) and \CRVZbC\ (included in the background fits of \SRA\ and \SRB). The control and validation regions are labelled using the targeted background process as subscript, which can also include additional selection details, and the associated selection as superscript. For example, the ``$3j$'' subscript of \CRTtN refers to the minimum jet multiplicity which is required in this control region.
In \CRTZ\ and \CRVZbC, events with three charged leptons including one same-flavour opposite-charge pair with $|\mll-m_{\Zboson}| < 20$~GeV are selected. In order to mimic the kinematics of the \ttbar+\Zboson events with invisible \Zboson decays, a corrected \MET variable, \METcorr, is defined by vectorially adding the momentum of the same-flavour opposite-charge lepton pair to the $\mathbf p^{\mathrm{miss}}_{\mathrm{T}}$ vector. 

In order to test the reliability of the background prediction, the results of the simultaneous fit are cross-checked in VRs which are disjoint from both the corresponding control and signal regions. Overlapping regions, e.g. \CRTbC\ and \CRTtN, are only included in independent background fits, so that no correlation is introduced. The expected signal contamination in the CRs is generally below 5\%. The highest signal contamination in the VRs, of about $18\%$, is expected in \VRTtN\ for a top squark mass of 400~\GeV\ and a lightest neutralino mass of 175~\GeV.

\begin{table}[!htb]
  \caption{\textit{Two-body selection} control and validation regions definition. The common selection defined in Section~\ref{sec:selection-common} also applies to all regions except \CRTZ\ and \CRVZbC, which require three leptons including one same-flavour opposite-charge pair with $|\mll-m_{\Zboson}| < 20$~GeV.}
    \label{tab:twobody_CRdef}
    \begin{center}
    \footnotesize
    \begin{tabular}{l | c c c c c | c c c }
    \toprule
                    			& \CRTbC    & \CRTtN    & \CRVVbC  & \CRTZ     & \CRVZbC    &   \VRTbC      & \VRTtN\    & \VRVVbC \\
    \midrule                                               
    Leptons                     & 2, DF     &   2       & 2, SF    & 3         & 3          & 2, DF         & 2         & 2, DF      \\
    \mttwoL    [GeV]            & [100, 120]  & [60, 100] & [100, 120] & --        & --         & > 120         & > 100     & [100, 120]  \\
\hskip-5pt $\begin{array}{l}\nbjet \\\njet \end{array}$ & $\begin{array}{c} \geq 1 \\ \mathrm{-} \end{array}$ & $\begin{array}{c} \geq 1 \\ \geq 3 \end{array}$ & $\begin{array}{c} 0 \\ \mathrm{-} \end{array}$ & 
$\begin{array}{c} \geq 2 \\ \geq 3 \end{array}$ or $\begin{array}{c} = 1 \\ \geq 4 \end{array}$ & $\begin{array}{c} 0 \\ \mathrm{-} \end{array}$& $\begin{array}{c} \geq 1 \\ \geq 2 \end{array}$ & $\begin{array}{c} \geq 1 \\ \geq 3 \end{array}$ & $\begin{array}{c} 0 \\ \mathrm{-} \end{array}$ \\
    $p^{\ell\ell}_{\mathrm{T,boost}}$ [GeV]                 & --        & --        & < 25      & --       & --         & --            &  --         & < 25 \\
    \dphib                      & --        & --        & --        & --       & --         & > 1.5         &  --         & -- \\
    \Rone                       & --        & --        & > 0.3     & --       & --         & --            & --          & -- \\
    \METcorr [GeV]              & --        & --        & --        & $> 120$  & $> 120$    & --            & --          & -- \\
    \MET [GeV]                  &           & $> 200$   & --        & --       & --         & --            & $>200$      & -- \\
    \Rll                        & --        & $<1.2$    & --        & --       & --         & --            & $<1.2$      & -- \\
    \bottomrule
    \end{tabular}
    \end{center}
\end{table}

Figure~\ref{fig:var_CRbC} shows the distributions of some of the kinematic variables used to define the four control regions after the \SRA\ background fit, so that the plots illustrate the modelling of the shape of each variable. In general, good agreement is found between the data and the background model within uncertainties. The other selection variables are equally well described by the background prediction.

\begin{figure}[htb!]
\begin{center}
\subfloat[]{\includegraphics[width=0.5\textwidth]{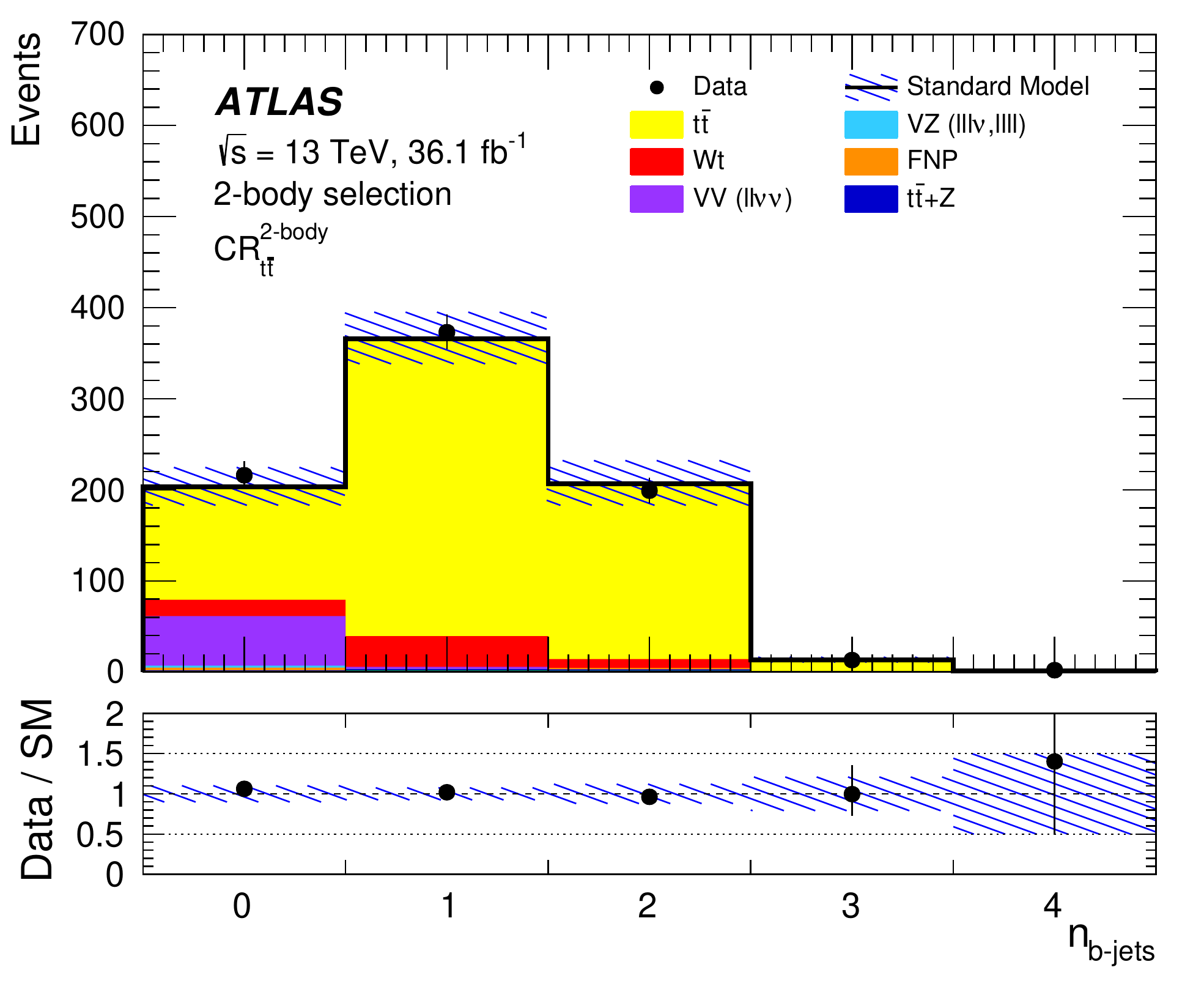}}
\subfloat[]{\includegraphics[width=0.5\textwidth]{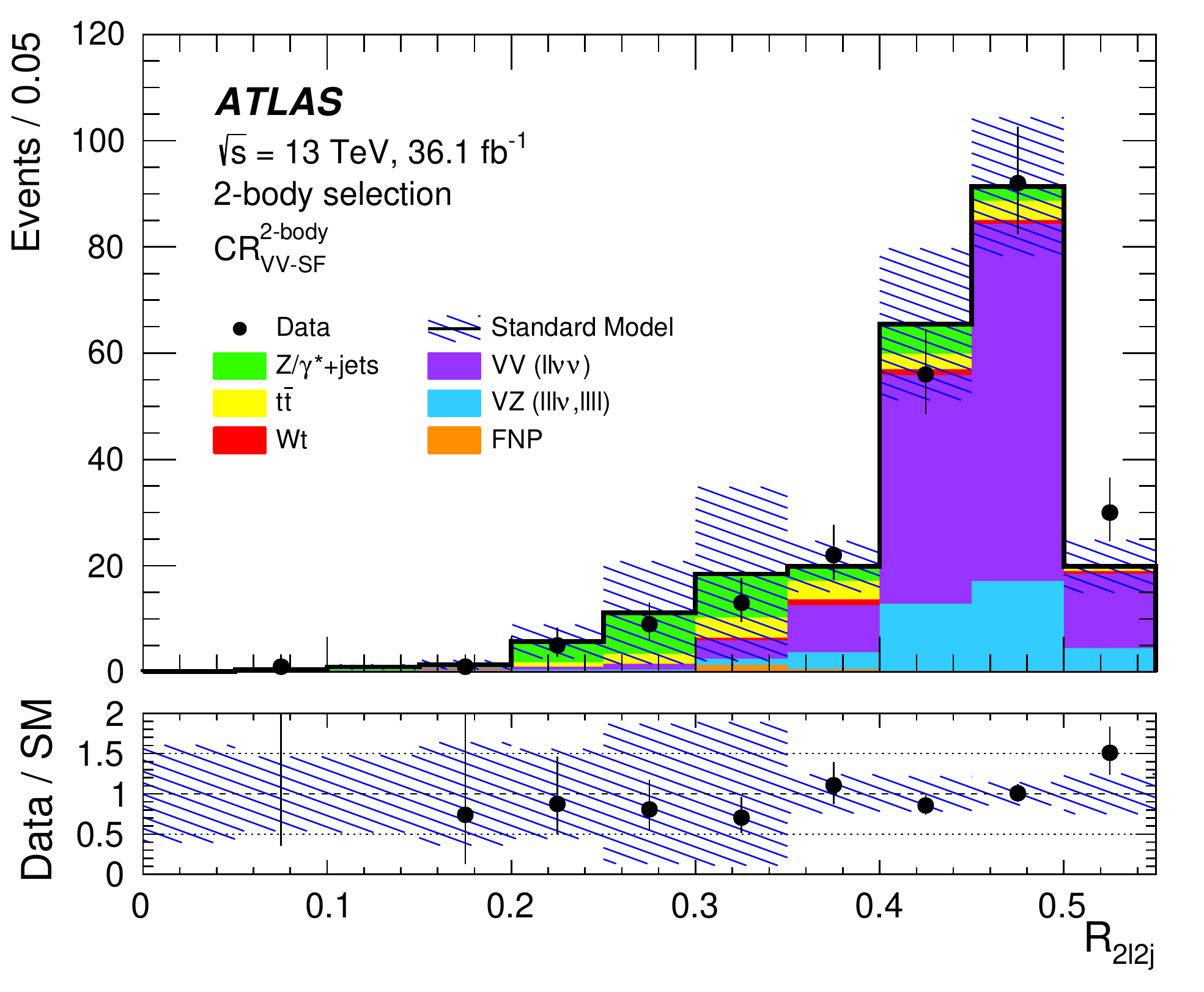}}\\
\subfloat[]{\includegraphics[width=0.5\textwidth]{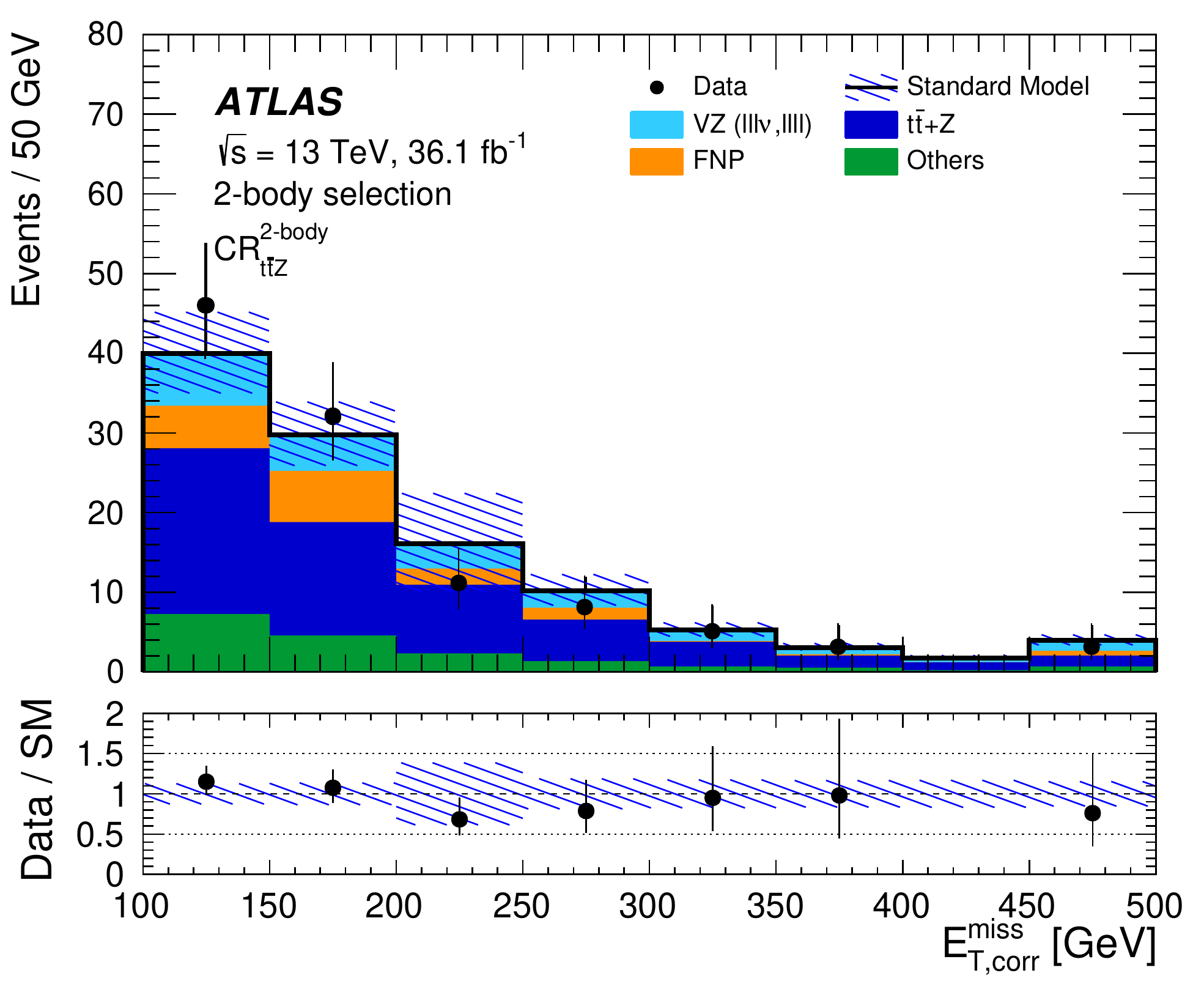}}
\subfloat[]{\includegraphics[width=0.5\textwidth]{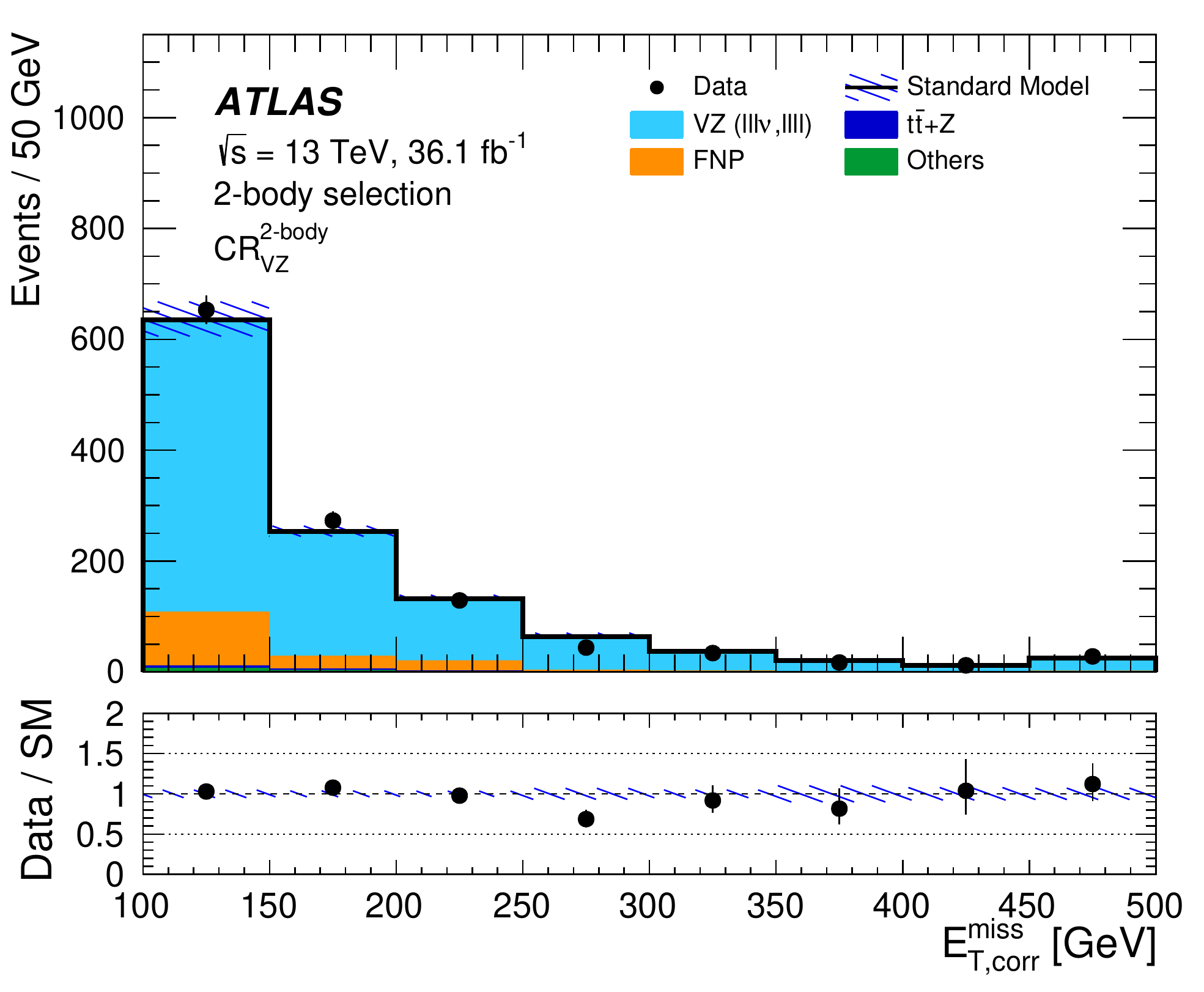}}
\end{center}
\caption{\textit{Two-body selection} distributions of (a) \nbjet\ in \CRTbC, (b) \Rone in \CRVVbC\ and (c,d) \METcorr\ in \CRTZ\ and \CRVZbC\ after the \SRA\ background fit. The contributions from all SM backgrounds are shown as a histogram stack; the hatched bands represent the total uncertainty in the background predictions after the fit to the data has been performed. The counting uncertainty on data is also shown by the black error bars. The rightmost bin of each plot includes overflow events.} 
\label{fig:var_CRbC}
\end{figure}

The results of the background fits, as well as the MC expected background composition before the fit, are reported in Table~\ref{tab:CRfit_bC1} for the CRs used in the \SRA\ and \SRB\ background fits, and in Table~\ref{tab:tN1fit} for the CRs used in the \SRtN\ background fit. The normalisations for fitted backgrounds are found to be consistent with the theoretical predictions, when uncertainties are considered. By construction, in the CRs the yields observed and predicted by the fits are the same. Good agreement, within one standard deviation from the SM background prediction, is observed in the VRs and summarised in Figure~\ref{fig:VRSRpulls}.

\begin{table}
\caption{\textit{Two-body selection} background fit results for the CRs of the \SRA\ and \SRB\ background fits. The nominal predictions from MC simulation are given for comparison for those backgrounds (\ttbar, $VV$-SF, $\ttbar Z$ and $VZ$) that are normalised to data in dedicated CRs. The ``Others'' category contains the contributions from $\ttbar W$, $\ttbar h$, $\ttbar WW$, $\ttbar t$, $\ttbar \ttbar$, $Wh$, $ggh$ and $Zh$ production. Combined statistical and systematic uncertainties are given. Entries marked ``--'' indicate a negligible background contribution.
\label{tab:CRfit_bC1}
}
\begin{center}
\setlength{\tabcolsep}{0.0pc}
{\small
\begin{tabular*}{\textwidth}{@{\extracolsep{\fill}}lrrrr}
\noalign{\smallskip}\hline\noalign{\smallskip}
                                   & \CRTbC                 & \CRVVbC              & \CRTZ          & \CRVZbC               \\[-0.05cm]
\noalign{\smallskip}\hline\noalign{\smallskip}
Observed events                    & $587$                  & $213$                & $91$           & $836$                    \\
\noalign{\smallskip}\hline\noalign{\smallskip}
Estimated SM Events     		   & $587 \pm 24$           & $213 \pm 15$         & $91 \pm 10$    & $836 \pm 29$              \\
\noalign{\smallskip}\hline\noalign{\smallskip}
\ttbar                 & $532 \pm 25$           & $14 \pm 4$           & --          	& -- 			  \\
$Wt$                               & $44 \pm 6$             & $4.0 \pm 1.5$        & --             & --              \\
\zjets                  		   & $0.02_{-0.02}^{+0.05}$ & $19 \pm 10$          & --             & --              \\
$VV$-SF                & --                     & $135 \pm 18$         & --             & --              \\
$VV$-DF                            & $2.2 \pm 0.8$          & --                   & --             & --              \\
$VZ$                   & $0.18 \pm 0.12$        & $38 \pm 7$           & $17.5 \pm 2.5$ & $730 \pm 50$              \\
\ttbar+\Zboson         & $2.2 \pm 0.8$          & $0.07 \pm 0.07$      & $47 \pm 12$    & $8.9 \pm 2.5$              \\
Others                             & $3.8 \pm 0.4$          & $0.41 \pm 0.18$      & $14.5 \pm 1.4$ & $10.3 \pm 0.9$ \\
Fake and non-prompt                & $1.6 \pm 0.9$          & $0_{-0}^{+5}$ 	   & $12 \pm 7$  	& $86 \pm 34$              \\
\noalign{\smallskip}\hline\noalign{\smallskip}
Nominal MC,  \ttbar         & $504$                  & $14$                 & --             & --              \\
Nominal MC,  $VV$-SF        & --                     & $122$                & --             & --            \\
Nominal MC,  $VZ$           & $0.18$                 & $39$                 & $18$           & $735$              \\
Nominal MC,  \ttbar+\Zboson & $3.57$                 & $0.08$               & $56$           & $11$              \\
\noalign{\smallskip}\hline\noalign{\smallskip}
\end{tabular*}
}
\end{center}
\end{table}

\begin{table}
\caption{\textit{Two-body selection} background fit results for the CRs of the \SRtN\ background fit. The nominal predictions from MC simulation, are given for comparison for those backgrounds (\ttbar and $\ttbar Z$) that are normalised to data in dedicated CRs. 
The ``Others'' category contains the contributions from $\ttbar W$, $\ttbar h$, $\ttbar WW$, $\ttbar t$, $\ttbar \ttbar$, $Wh$, $ggh$ and $Zh$ production.
Combined statistical and systematic uncertainties are given. Entries marked ``--'' indicate a negligible background contribution.
\label{tab:tN1fit}
}
\begin{center} 
\setlength{\tabcolsep}{0.0pc}
{\small
\begin{tabular*}{0.7\textwidth}{@{\extracolsep{\fill}}lrr}
\noalign{\smallskip}\hline\noalign{\smallskip}
                              & \CRTtN            & \CRTZ           \\[-0.05cm]
\noalign{\smallskip}\hline\noalign{\smallskip}
\noalign{\smallskip}\hline\noalign{\smallskip}
Observed events               & $212$             & 91             \\
\noalign{\smallskip}\hline\noalign{\smallskip}
Estimated SM Events              & $212 \pm 15$      & $91 \pm 10$     \\
\noalign{\smallskip}\hline\noalign{\smallskip}
\ttbar            & $184 \pm 16$      & --             \\
\ttbar+Z          & $1.03 \pm 0.32$     & $47 \pm 12$    \\
$Wt$                            & $23 \pm 7$        & --             \\
$VV$                      	  & $1.69 \pm 0.30$     & $17.7 \pm 2.2$ \\
\zjets          			  & $0.05 \pm 0.02$   & --             \\
Others                        & $1.91 \pm 0.12$   & $14.6 \pm 1.0$  \\
Fake and non-prompt           & --   & $12 \pm 7$ \\
\noalign{\smallskip}\hline\noalign{\smallskip}
Nominal MC,  \ttbar    & $201$    & -- \\
Nominal MC,  \ttbar+Z  & $1.23$   & $55.7$ \\
\noalign{\smallskip}\hline\noalign{\smallskip}
\end{tabular*}
}
\end{center}
\end{table}

\FloatBarrier
\subsection{Three-body selection background determination}
\label{sec:threebody_backgrounds}

In the three-body signal regions defined in Section~\ref{sec:threebody_selection}, the SM background is dominated by diboson and \ttbar production.
A single control region is used for \ttbar production, while two CRs are defined to target diboson events with either same-flavour or different-flavour lepton pairs.
The background predictions are tested in VRs that are defined to be kinematically adjacent to, yet disjoint from, the signal regions. 
The definitions of the control and validation regions are shown in Table~\ref{tab:threebodycrvrdef}. The overlap between \VRTtB\ and \VRVVDFtB\ does not affect the final results as these regions are not used to constrain the background normalisations. The signal contamination in the CRs and VRs is generally small, with the maximum found to be about 12\% in \VRVVDFtB\ for a top squark mass of 220~\GeV\ and a lightest neutralino mass of 110~\GeV.

\begin{table}[!htb]
    \caption{\textit{Three-body selection} control and validation regions definitions. The common selection defined in Section~\ref{sec:selection-common} also applies to all regions.}
    \label{tab:threebodycrvrdef}
    \begin{center}
    {\footnotesize
    \begin{tabular}{l | c c c | c c c}
    \toprule
                             & \CRTtB    & \CRVVDFtB  & \CRVVSFtB\ & \VRTtB\ & \VRVVDFtB\ & \VRVVSFtB  \\
    \midrule                               
Lepton flavour               & DF        & DF         & SF        & DF     & DF        & SF \\
$|m_{\ell\ell}-m_{Z}|$ [GeV] & --        & --         & > 20      & --     & --        & >20 \\
\nbjet         & $>0$        & $=0$       & $=0$      & $=0$   & $=0$      & $=0$  \\
\MDR\ [GeV]                   & $>80$     & $>50$      & $>70$     & $>80$  & $[50,95]$ & $[60,95]$ \\
\RPT                         & $>0.7$    & $<0.5$     & $<0.5$    & $<0.7$ & $<0.7$    & $<0.4$ \\
\GAM                         & --        & $>0.7$     & $>0.7$    & --     & $>0.7$    & $>0.7$ \\
(\cosThetaB, \DPB)           & \multicolumn{3}{c|}{$\DPB<(0.9\times|\cosThetaB|+1.6)$} & \multicolumn{3}{c}{$\,\,\DPB>(0.9\times|\cosThetaB|+1.6)$} \\
    \bottomrule
    \end{tabular}
    }
    \end{center}
\end{table}

Table~\ref{tab:threebodybkgonly_CRVR} shows the expected and observed numbers of events in each of the control regions after the background fit.
The total number of fitted background events in the validation regions is in agreement with the observed number of data events.
Figure~\ref{fig:threebodykinCRPlots} shows three distributions in the control regions after the background fit, so that the plots illustrate the MC modelling of the shape of each variable. In general, good agreement between the data and the background model is found within uncertainties. The other selection variables are equally well described by the background prediction. Good agreement, within one standard deviation from the SM background prediction, is observed in the VRs and summarised in Figure~\ref{fig:VRSRpulls}.

\begin{table}[!htb]
\caption{\textit{Three-body selection} background fit results for the CRs of the \SRTBW\ and \SRTBT\ background fit. The nominal predictions from MC simulation, are given for comparison for those backgrounds (\ttbar, $VV$-DF and $VV$-SF) that are normalised to data in dedicated CRs. 
Combined statistical and systematic uncertainties are given. Entries marked ``--'' indicate a negligible background contribution.}
\label{tab:threebodybkgonly_CRVR}
\begin{center}
\setlength{\tabcolsep}{0.0pc}
{\small
\begin{tabular*}{0.7\textwidth}{@{\extracolsep{\fill}}lrrr}
\noalign{\smallskip}\hline\noalign{\smallskip}
             & \CRTtB  & \CRVVDFtB       & \CRVVSFtB\ \\[-0.05cm]
\noalign{\smallskip}\hline\noalign{\smallskip}
Observed events             & $951$         & $2046$              & $1275$         \\
\noalign{\smallskip}\hline\noalign{\smallskip}
Estimated SM Events & $951\pm31$    & $2046\pm50$         & $1275\pm40$    \\
\noalign{\smallskip}\hline\noalign{\smallskip}
$t\bar{t}$      & $833\pm33$    & $620\pm110$         & $330\pm60$     \\
$VV$-DF          & $11.5\pm2.4$  & $1090\pm130$        & --             \\
$VV$-SF          & --            & --                  & $380\pm90$     \\
$Wt$                        & $101\pm10$    & $186\pm28$          & $103\pm17$     \\
$t\bar{t}$+$V$              & $4.3\pm0.4$   & $0.39\pm0.06$       & $0.36\pm0.07$  \\
$Z/\gamma*$+jets            & $0.70\pm0.22$  & $1.8_{-1.8}^{+2.5}$ & $430\pm50$     \\
Higgs bosons                & $0.31\pm0.08$ & $79\pm9$            & $6.2\pm0.8$    \\
Fake and non-prompt         & $0.00_{-0.00}^{+0.30}$   & $65.4\pm2.2$        & $24.0\pm1.3$   \\
\noalign{\smallskip}\hline\noalign{\smallskip}
Nominal MC, $t\bar{t}$       & $787$    & $590$         & $320$     \\
Nominal MC, $VV$-DF          & $11.3$  & $1069$         & --             \\
Nominal MC, $VV$-SF          & --            & --                  & $370$     \\
\noalign{\smallskip}\hline\noalign{\smallskip}
\end{tabular*}
}
\end{center}
\end{table}

\begin{figure}[!htb]
\begin{center}
\subfloat[]{\includegraphics[width=0.48\textwidth]{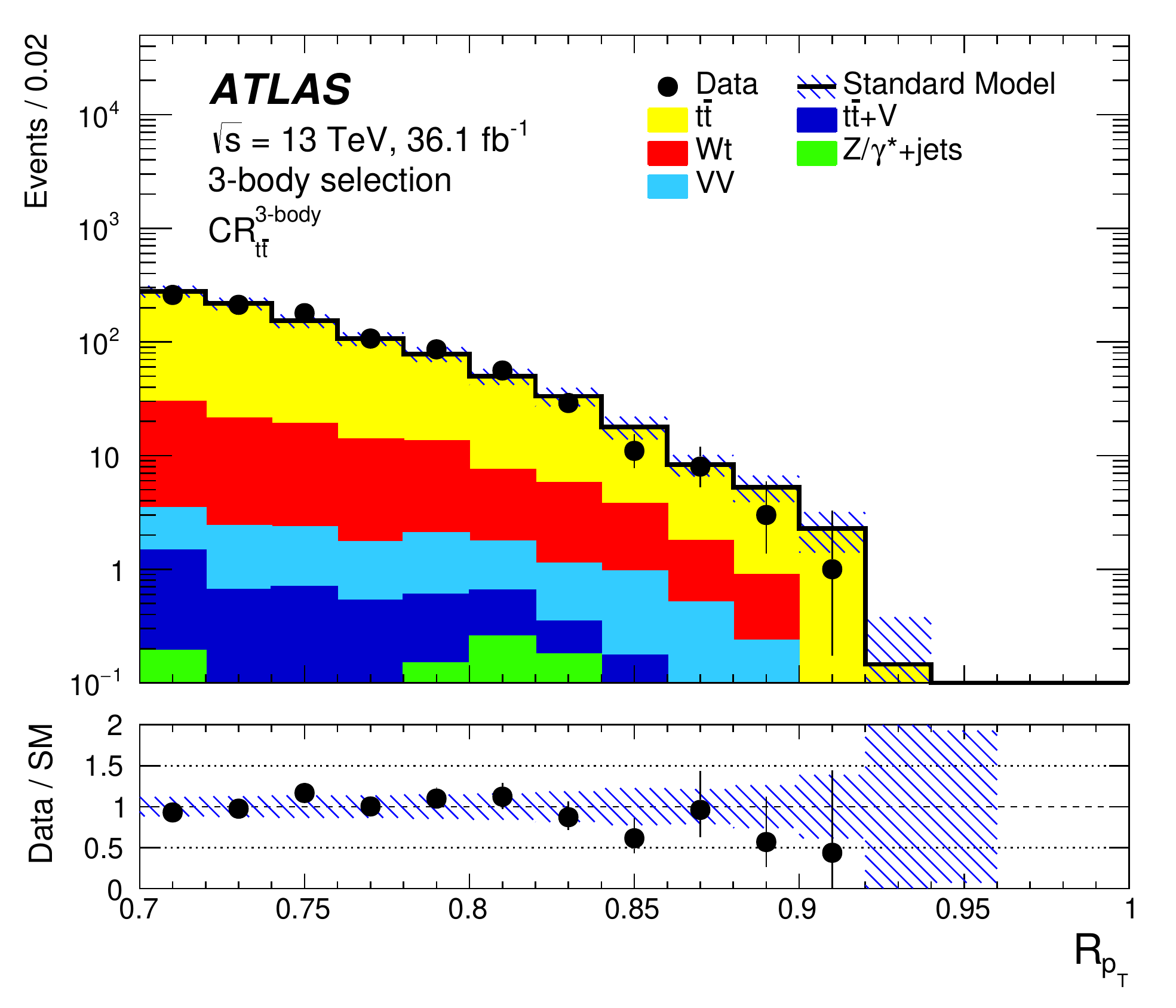}}
\subfloat[]{\includegraphics[width=0.48\textwidth]{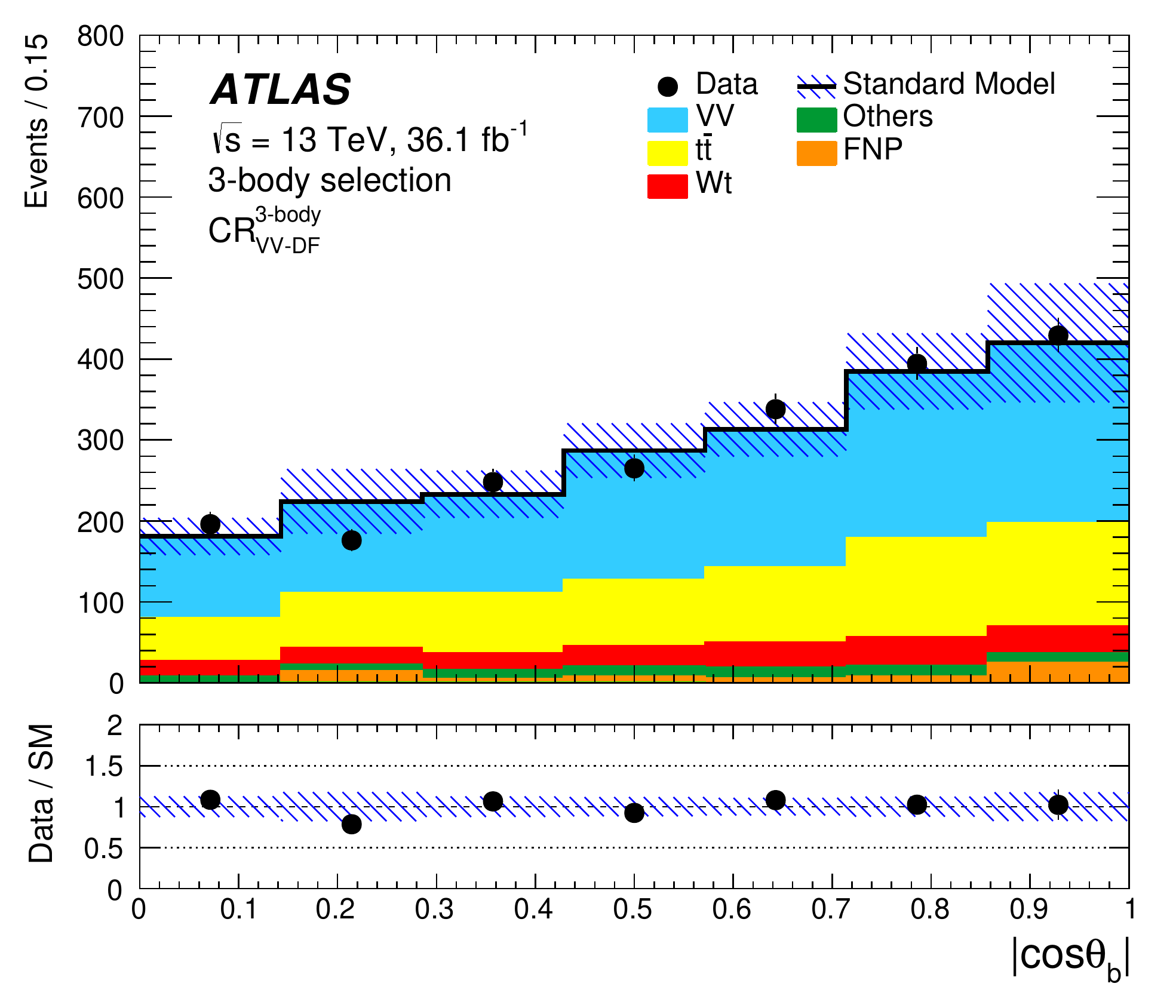}}\\
\subfloat[]{\includegraphics[width=0.48\textwidth]{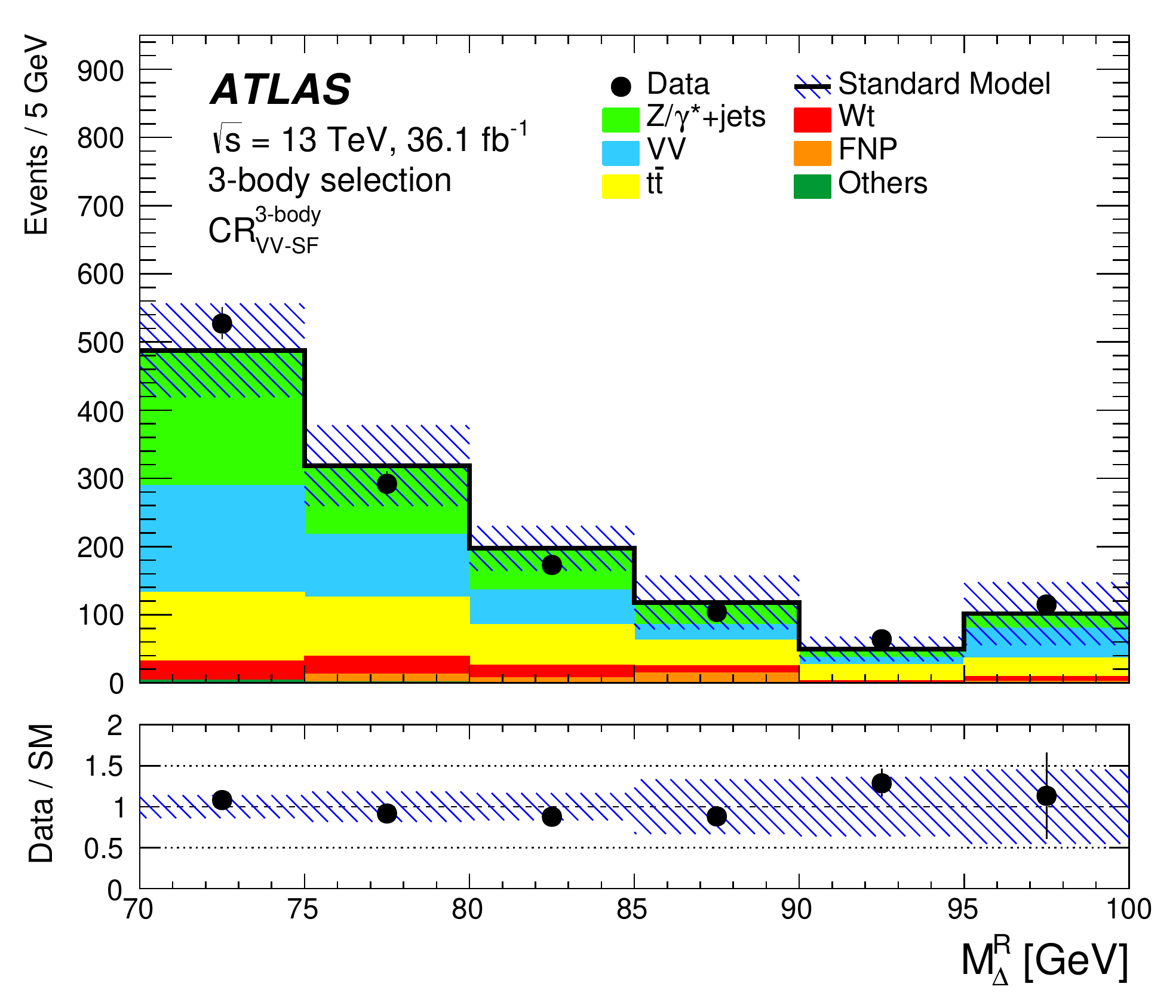}}
\end{center}
\caption{\textit{Three-body selection} distributions of (a) \RPT in \CRTtB, (b) \cosThetaB in \CRVVDFtB, and (c) \MDR\ in \CRVVSFtB\ after the background fit. The contributions from all SM backgrounds are shown as a histogram stack; the hatched bands represent the total uncertainty in the background predictions after the fit to the data has been performed. The counting uncertainty on data is also shown by the black error bars. The rightmost bin of each plot includes overflow events.} 
\label{fig:threebodykinCRPlots}
\end{figure}

\FloatBarrier
\subsection{Four-body selection background determination}
\label{sec:fourbody_backgrounds}

In the four-body SR, the largest SM background contributions stem from \ttbar and diboson production, as well as \zjets\ production with the \Zboson\ boson decaying into $\tau\tau$ with both $\tau$ leptons decaying leptonically. Three dedicated control regions are defined: \CRTfB, \CRVVfB\ and \CRZfB.
The background predictions are tested in three validation regions that are defined to be kinematically similar to, but disjoint from, both the control and signal regions. The definitions of the control and validation regions are shown in Table~\ref{tab:fourbodycrvrdef}.
In the \ttbar control region the signal contamination is less than $\sim 6\%$, while in \CRVVfB\ and \CRZfB\ the highest signal contamination, for a top squark mass of 260~\GeV\ and a lightest neutralino mass of 180~\GeV, is respectively $\sim 30\%$ and $\sim 9\%$.

\begin{table}[!htb]
    \caption{\textit{Four-body selection} control and validation regions definition. The common selection reported in Table~\ref{tab:fourbodysrdefinitions} also applies to all regions.}
    \label{tab:fourbodycrvrdef}
    \begin{center}
    {\footnotesize
    \begin{tabular}{l | c c c | c c c}
    \toprule
                             & \CRTfB    & \CRVVfB & \CRZfB  & \VRTfB   & \VRVVfB & \VRZfB  \\
    \midrule                                                                              
Leading lepton \pT [GeV]     & [7, 80]    & [7, 80]  & > 20    & [7, 80]   & [7, 80]  & > 50 \\
Subleading lepton \pT [GeV] & [7, 35]    & [7, 35]  & > 20    & [7, 35]   & [7, 35]  & [7, 20]\\
 \njet            & $\geq 2$  & $=1$    & $=1$    & $\geq 2$ & $=1$    & $=1$\\
Leading jet \pT [GeV]        & [100, 150] & > 150   & > 150   & > 150    & > 150   & > 150 \\
\mll [GeV]                   & > 10      & > 45    & [10, 45] & > 10     & > 45    & [10, 45]\\
$R_{2\ell 4j}$                 & -         & -       & -       & < 0.35   & -       & - \\
\Rll                         & -         & < 5     & -       & < 12     & > 5     & - \\
\nbjet         & -         & $=0$    & $=0$    & -        & $=0$    & $=0$ \\
    \bottomrule
    \end{tabular}
    }
    \end{center}
\end{table}

Table~\ref{tab:fourbodybkgonly_CRVR} shows the expected and observed numbers of events in each of the control regions after the background fit.
Good agreement between data and the SM predictions is observed in the validation regions and shown in Figure~\ref{fig:VRSRpulls}. Figure~\ref{fig:fourbodykinCRPlots} shows three distributions in the control regions for this analysis after applying the normalisation factors provided by the background fit. Good agreement between data and the SM predictions is observed. The other selection variables are equally well described by the background prediction. The largest observed deviation ($1.4 \sigma$) from the SM background prediction is found in \VRZfB. The yields in the other VRs are found to be compatible with the SM predictions within one standard deviation.

\begin{table}[!htb]
\caption{\textit{Four-body selection} background fit results for the CRs of the \SRfB\ background fit. The nominal predictions from MC simulation, are given for comparison for those backgrounds (\ttbar, $VV$ and $Z_{\tau\tau}$) that are normalised to data in dedicated CRs. 
Combined statistical and systematic uncertainties are given. }
\label{tab:fourbodybkgonly_CRVR}
\begin{center}
\setlength{\tabcolsep}{0.0pc}
{\small
\begin{tabular*}{\textwidth}{@{\extracolsep{\fill}}lrrr}
\noalign{\smallskip}\hline\noalign{\smallskip}
                                   & \CRTfB        & \CRVVfB                 & \CRZfB  \\[-0.05cm]
\noalign{\smallskip}\hline\noalign{\smallskip}
Observed events                    & $1251$        & $110$                   & $106$                  \\
\noalign{\smallskip}\hline\noalign{\smallskip}
Estimated SM Events   				& $1251 \pm 35$ & $110 \pm 10$            & $106 \pm 10$           \\
\noalign{\smallskip}\hline\noalign{\smallskip}  
\ttbar                 & $960\pm 50$   & $47 \pm 20$             & $10 \pm 6$             \\
$VV$                   & $37  \pm 22$  & $40 \pm 22$             & $18 \pm 11$             \\ 
$Z_{\tau\tau}$         & $22  \pm 8$   & $0.00_{-0.00}^{+0.17}$  & $54 \pm 16$            \\
\ttbar+Z                           & $5.6 \pm 0.8$ & $0.08 \pm 0.01$         & $0.05 \pm 0.02$        \\
$Wt$                                 & $62  \pm 19$  & $9.0 \pm 2.7$           & $2.7\pm 2.4$           \\
$Z_{ee}$, $Z_{\mu\mu}$             & $0.7 \pm 0.5$ & $0.2 _{-0.2}^{+0.4}$    & $1.6\pm 0.6$           \\
Others                             & $11.2\pm 1.6$ & $0.51 \pm 0.12$         & $3.2 \pm 0.6$          \\
Fake and non-prompt                & $154\pm 14$   & $13.1 \pm 2.0$          & $16 \pm 7$             \\
\noalign{\smallskip}\hline\noalign{\smallskip}
Nominal MC,  \ttbar         & $931$ & $46$             & $10$             \\
Nominal MC,  $VV$             & $47$  & $51$              & $23$         \\ 
Nominal MC,  $Z_{\tau\tau}$ & $20$  & $0$ & $51$ \\
\noalign{\smallskip}\hline\noalign{\smallskip}
\end{tabular*}
}
\end{center}
\end{table}

\begin{figure}[!htb]
\begin{center}
\subfloat[]{\includegraphics[width=0.48\textwidth]{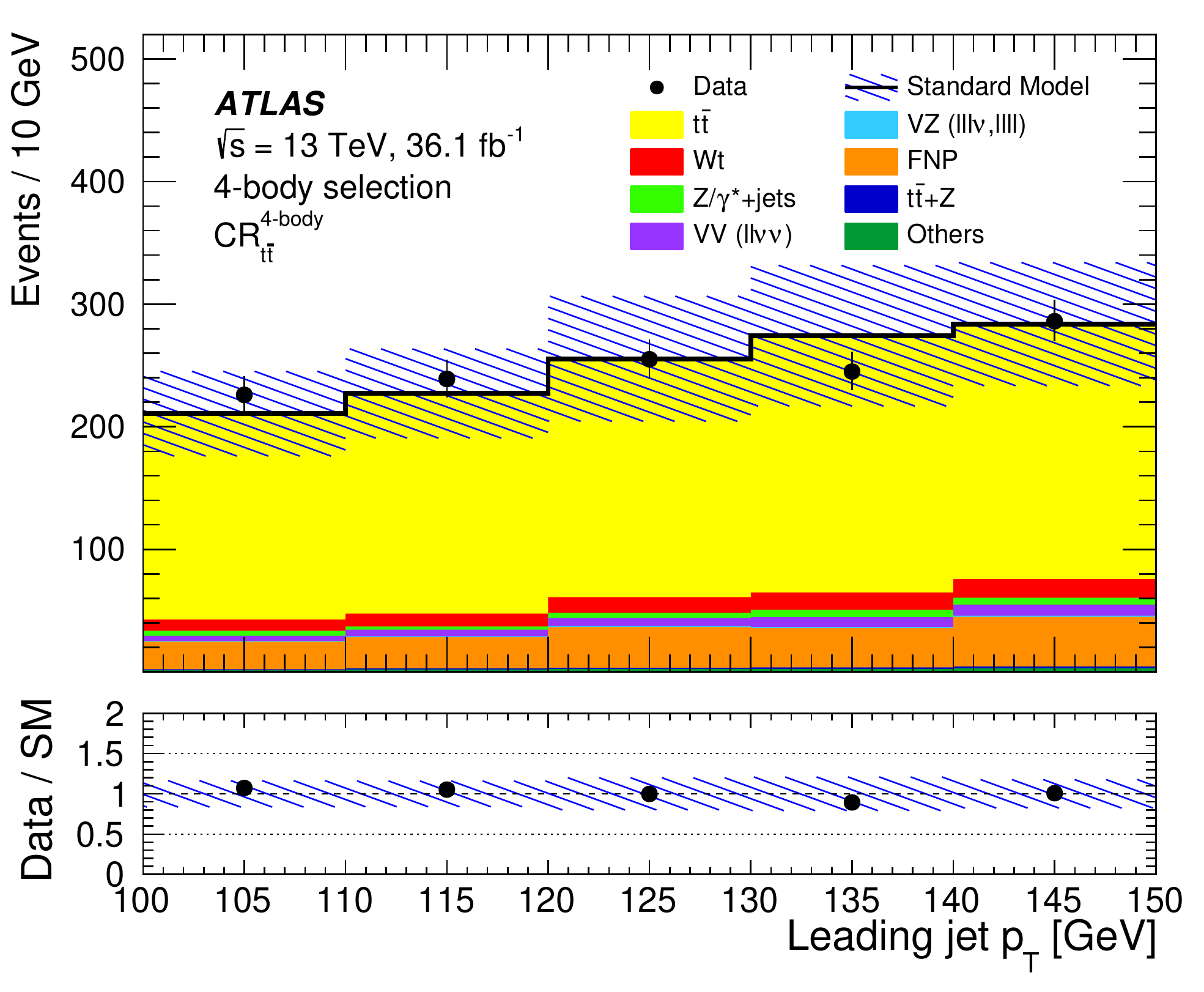}}
\subfloat[]{\includegraphics[width=0.48\textwidth]{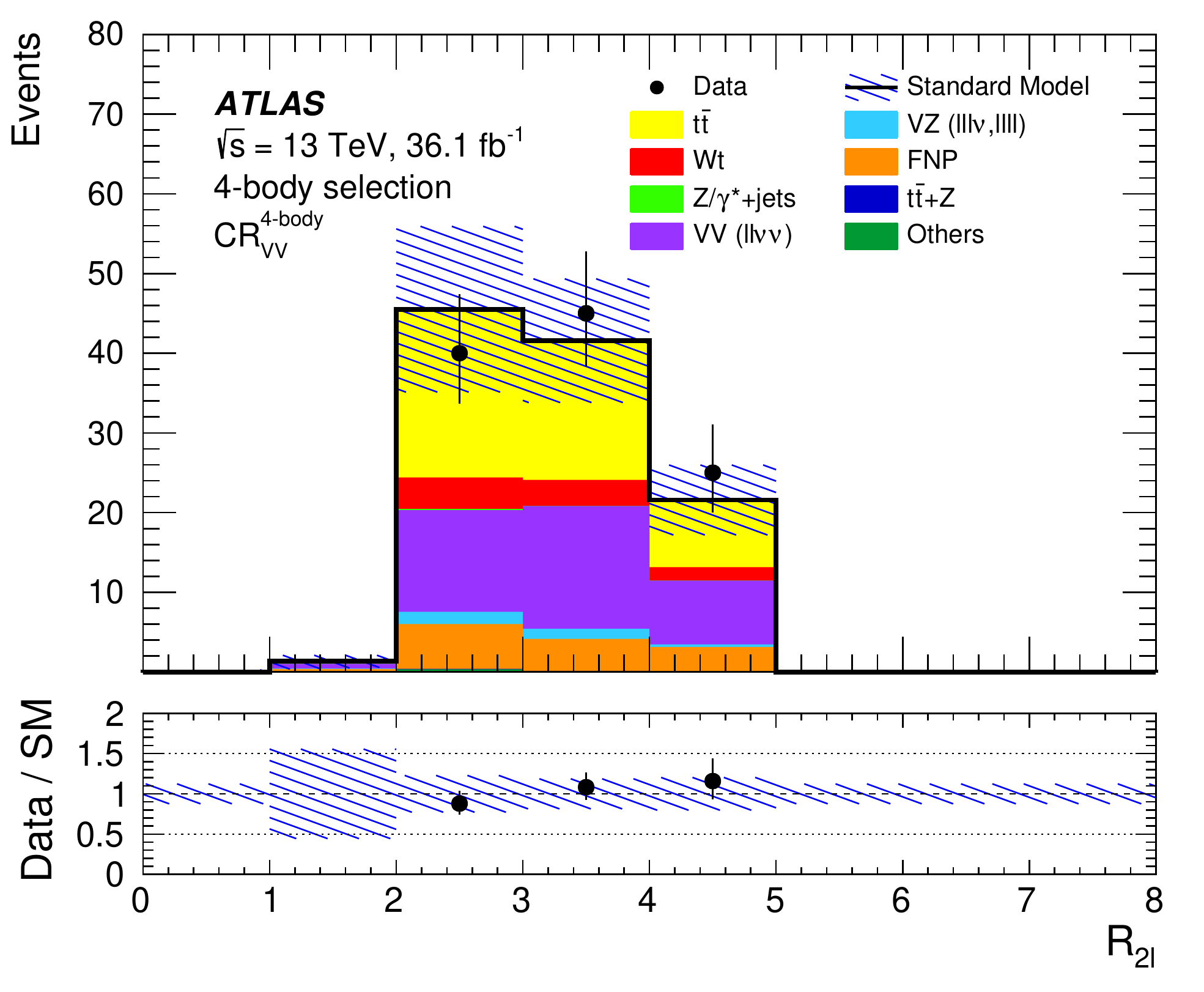}}\\
\subfloat[]{\includegraphics[width=0.48\textwidth]{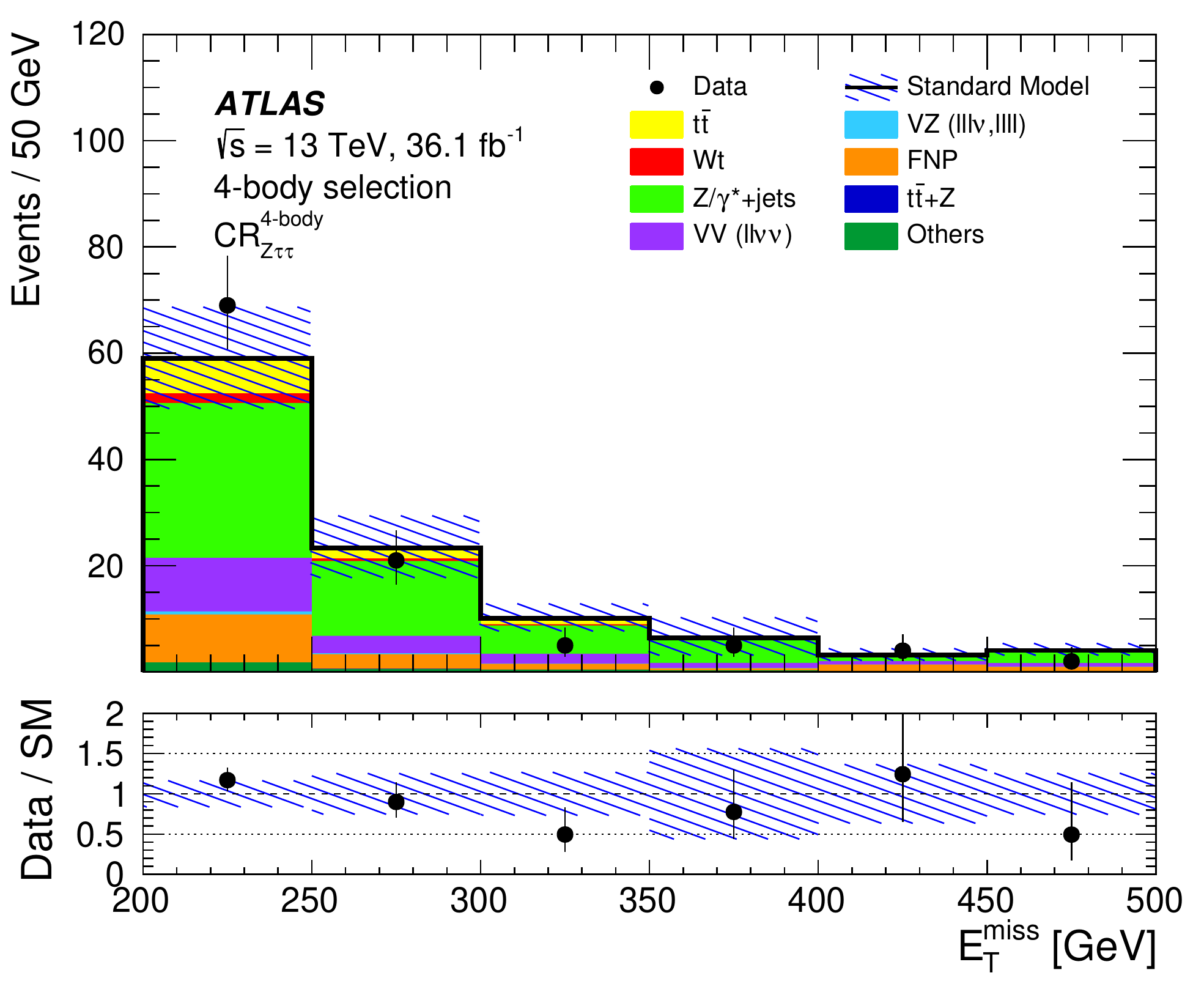}}
\end{center}
\caption{\textit{Four-body selection} distributions of the (a) $\pT(j_1)$ in \CRTfB, (b) \Rll in \CRVVfB\ and (c) \MET in \CRZfB after the background fit. The contributions from all SM backgrounds are shown as a histogram stack; the hatched bands represent the total uncertainty in the background predictions after the fit to the data has been performed. The counting uncertainty on data is also shown by the black error bars. The rightmost bin of each plot includes overflow events.}
\label{fig:fourbodykinCRPlots}
\end{figure}

\FloatBarrier 
\section{Systematic uncertainties}
\label{sec:systematics}

The primary sources of systematic uncertainty are related to: the jet energy scale (JES), jet energy resolution (JER), and the theoretical and MC modelling uncertainties in the backgrounds. The statistical uncertainties of the simulated event samples are also taken into account. The effect of the systematic uncertainties is evaluated for all signal samples and background processes. Since the normalisation of the dominant background processes is extracted in dedicated control regions, the systematic uncertainties only affect the extrapolation to the signal regions in these cases. Statistical uncertainties due to the limited number of data events in the CRs are also included in the fit for each region.

The JES and JER uncertainties are derived as a function of the \pT and $\eta$ of the jet, as well as of the pile-up conditions and the jet flavour composition of the selected jet sample~\cite{ATL-PHYS-PUB-2015-015}. Uncertainties associated to the modelling of the $b$-tagging efficiencies for $b$-jets, $c$-jets and light-flavour jets~\cite{ATLAS-CONF-2014-004,ATLAS-CONF-2014-046} are also considered. 

The systematic uncertainties related to the modelling of \MET in the simulation are estimated by propagating the uncertainties in the energy and momentum scale of electrons, muons and jets, as well as the uncertainties in the resolution and scale of the soft term~\cite{ATL-PHYS-PUB-2015-023}.

Other detector-related systematic uncertainties, such as those in lepton reconstruction efficiency, energy scale, energy resolution and in the modelling of the trigger efficiency~\cite{Aaboud:2016vfy,PERF-2015-10}, are found to have a small impact on the results and are generally negligible compared to the other detector-related uncertainties.

The uncertainties in the modelling of the $\ttbar$ and single-top backgrounds in simulation are estimated by varying the renormalisation and factorisation scales by a factor of two, as well as the amount of initial- and final-state radiation used to generate the samples~\cite{ATL-PHYS-PUB-2016-004}. Uncertainties in the parton shower modelling are assessed as the difference between the predictions from \POWHEG showered with \PYTHIA and \HERWIG, and those due to the event generator choice by comparing \POWHEG and \AMCATNLO~\cite{ATL-PHYS-PUB-2016-004}.
An uncertainty in the acceptance due to the interference between \ttbar and single top quark $Wt$ production is assigned by comparing the predictions of dedicated LO \MADGRAPH~2.5 samples. These samples are used to compare the predictions for \ttbar and $Wtb$ with the inclusive $WWbb$ process, where the same production diagrams are included, but top quarks are not required to be on-shell. 

The diboson background MC modelling uncertainties are estimated by varying up and down by a factor of two the renormalisation, factorisation and resummation scales used to generate the sample~\cite{ATL-PHYS-PUB-2016-002}.
For $\ttbar Z$ production, the predictions from the \AMCATNLO and \SHERPA event generators are compared and the full difference between the respective predictions is assigned as an uncertainty. Uncertainties related to the choice of renormalisation and factorisation scales are assessed by varying the corresponding event generator parameters up and down by a factor of two around their nominal values~\cite{ATL-PHYS-PUB-2016-005}. 

The uncertainties related to the choice of QCD renormalisation and factorisation scales in \zjets\ events are assessed by varying the corresponding event generator parameters up and down by a factor of two around their nominal values. Uncertainties due to our choice of the resummation scale and the matching scale between the matrix element and the parton shower are estimated by varying up and down by a factor of two the corresponding parameters in \SHERPA.

The cross-sections used to normalise the MC samples are varied according to the uncertainty in the cross-section calculation, i.e., 5.3\% uncertainty for single top quark $Wt$-channel~\cite{Kant:2014oha}, 6\% for diboson, 13\% for $\ttbar W$ and 12\% for $\ttbar Z$ production~\cite{Alwall:2014hca}. For $\ttbar WW$, $tZ$, $tWZ$, $\ttbar h$, $\ttbar t$, $\ttbar\ttbar$, and triboson production processes, which constitute a small background, a 50\% uncertainty in the event yields is assumed.

Systematic uncertainties are assigned to the FNP background estimate to account for potentially different compositions (heavy flavour, light flavour or photon conversions) between the signal and control regions, as well for the contamination from prompt leptons in the regions used to measure the probabilities for loose fake or non-prompt leptons to satisfy the tight signal criteria. Parameterisations of these probabilities are independently derived from \ttbar- and multi-jet-enriched same-charge dilepton samples. The \ttbar-enriched sample is used to derive the parameterisation from which the central prediction for the FNP background is obtained. The full difference between the predictions derived from the \ttbar and the multi-jet parameterisation is assigned as the systematic uncertainty in the central FNP prediction and symmetrised.

A 3.2\% uncertainty in the luminosity measurement is also taken into consideration for all signal and background estimates that are directly derived from MC simulations.

Table~\ref{tab:SR_syst} summarises the contributions of the different sources of systematic uncertainty in the total SM background predictions in the signal regions. The total systematic uncertainty ranges between 15\% and 46\%, with the dominant sources being the size of the MC event samples, the JES and \MET modelling, the numbers of events in the CRs and the \ttbar theoretical uncertainties.

Theory uncertainties in the signal acceptance are taken into account. These are computed by varying the strong coupling constant $\alpha_{s}$, the renormalization and factorization scales, the CKKW scale used to match the parton shower and matrix element descriptions and the parton shower tunes. These uncertainties are mostly relevant for the four-body selection and range between 10\% and 30\% depending on the mass difference $m_{\stopone}-m_{\ninoone}$. 

\begin{sidewaystable}[ht]
  \caption{Sources of systematic uncertainty in the SM background estimates, estimated after the background fits. The values are given as relative uncertainties in the total expected background event yields in the SRs. Entries marked ``--'' indicate either a negligible contribution or an uncertainty that does not apply (for example the normalisation uncertainty for a background whose normalisation is not fitted for that specific signal region). MC statistics refer to the statistical uncertainty from the simulated event samples. The individual components can be correlated and therefore do not necessarily add up in quadrature to the total systematic uncertainty.
\label{tab:SR_syst}}
{\scriptsize
\begin{center}
\setlength{\tabcolsep}{0.0pc}
\begin{tabular*}{0.98\textwidth}{@{\extracolsep{\fill}}lccccccccccc}
\noalign{\smallskip}\hline\noalign{\smallskip}
Signal Region                                & \SRA\ SF     &\SRA\ DF  &\SRB\ SF  &\SRB\ DF 	 &\SRtN\ SF  &\SRtN\ DF  &\SRTBW\ SF 	   &\SRTBW\ DF 	   &\SRTBT\ SF      &\SRTBT\ DF      &\SRfB     \\
\noalign{\smallskip}\hline\noalign{\smallskip}
Total SM background uncertainty              &21\%         &32\%     &15\%     &21\%         &35\%      &38\%      &36\%           &39\%           &46\%           &42\%           &20\%           \\
\noalign{\smallskip}\hline\noalign{\smallskip}
\noalign{\smallskip}\hline\noalign{\smallskip}
Diboson theoretical uncertainties            &4.0\%		   &5.9\%	 &--       &-- 			 &--		&--		   &9.1\%		   &10\%		   &1.3\%		   &--		   	   &2.7\%		   \\
\ttbar theoretical uncertainties             &-- 		   &--		 &4.2\%	   &6.6\%		 &12\%		&13\%	   &13\%		   &18\%		   &25\%		   &24\%		   &8.1\%		   \\	 
\(Wt\) theoretical uncertainties 			 &-- 		   &--		 &--	   &1.9\%		 &--		&5.4\%	   &--		   	   &--		   	   &--			   &--			   &--		   \\	  
\ttbar-\(Wt\) interference                   &--           &--       &1.8\%    &7.9\%        &--        &--        &--             &--	           &-- 	           &--	           &--           \\
\noalign{\smallskip}\hline\noalign{\smallskip}
MC statistical uncertainties                 &13\%		   &28\%	 &12\%	   &13\%		 &15\%	    &15\%	   &16\%		   &14\%		   &20\%		   &22\%		   &10\%		   \\ 
\noalign{\smallskip}\hline\noalign{\smallskip}
$VV$ normalisation                           &14\%		   &--		 &--	   &--			 &--	    &--		   &12\%		   &4.3\%		   &1.3\%		   &--			   &9.2\%		   \\
\ttbar normalisation                         &--		   &--		 &--	   &--			 &16\%	    &15\%	   &1.8\%		   &2.5\%		   &3.5\%		   &3.5\%		   &8.6\%		   \\
\ttbar+\Zboson normalisation                 &--		   &--		 &7.6\%	   &9.9\%		 &8.5\%	    &10\%	   &--		       &--			   &--			   &--			   &--		   \\
\(Z_{\tau\tau}\) normalisation               &--		   &--		 &--	   &--			 &--	    &--		   &--		       &--			   &--			   &--			   &1.5\%		   \\
\noalign{\smallskip}\hline\noalign{\smallskip}
Jet energy scale                             &6.9\%        &3.1\%    &4.1\%    &6.4\%        &13\%      &22\%      &19\%           &18\%           &27\%           &11\%           &4.4\%           \\
Jet energy resolution                        &--           &--       &--       &--           &12\%      &16\%      &7.2\%          &18\%           &2.9\%          &22\%           &1.0\%           \\
\MET modelling                               &5.0\%        &13\%     &2.2\%    &3.2\%        &26\%      &23\%      &18\%           &11\%           &14\%           &6.5\%          &1.3\%           \\
\(b\)-tagging                                &--           &--       &3.0\%    &1.5\%        &--        &--        &2.7\%          &3.0\%          &1.0\%          &3.0\%          &2.2\%           \\
Pile-up reweighting                          &2.0\%        &3.2\%    &1.1\%    &4.3\%        &2.9\%     &4.6\%     &2.9\%          &5.0\%          &5.1\%          &4.9\%          &1.4\%           \\
Lepton modelling                             &1.3\%        &2.1\%    &--       &1.1\%        &--        &--        &1.1\%          &3.1\%          &4.6\%          &3.0\%          &2.5\%           \\
Fake and non-prompt leptons                  &--           &--       &7.4\%    &--           &4.0\%     &--        &2.8\%          &--	           &--	           &--             &14\%           \\
\noalign{\smallskip}\hline\noalign{\smallskip}
\end{tabular*}
\end{center}
}
\end{sidewaystable}

\FloatBarrier

\section{Results}
\label{sec:result}

The data are compared to background predictions in the signal regions of the different selections.
The number of observed events and the predicted number of SM background events from the background-only
fits in all SRs and VRs are shown in Figure~\ref{fig:VRSRpulls}. In all SRs, good agreement is observed between data and the SM background predictions. A detailed discussion of the results is given in the following sections.

\begin{figure}[htb!]
\begin{center}
\includegraphics[width=\textwidth]{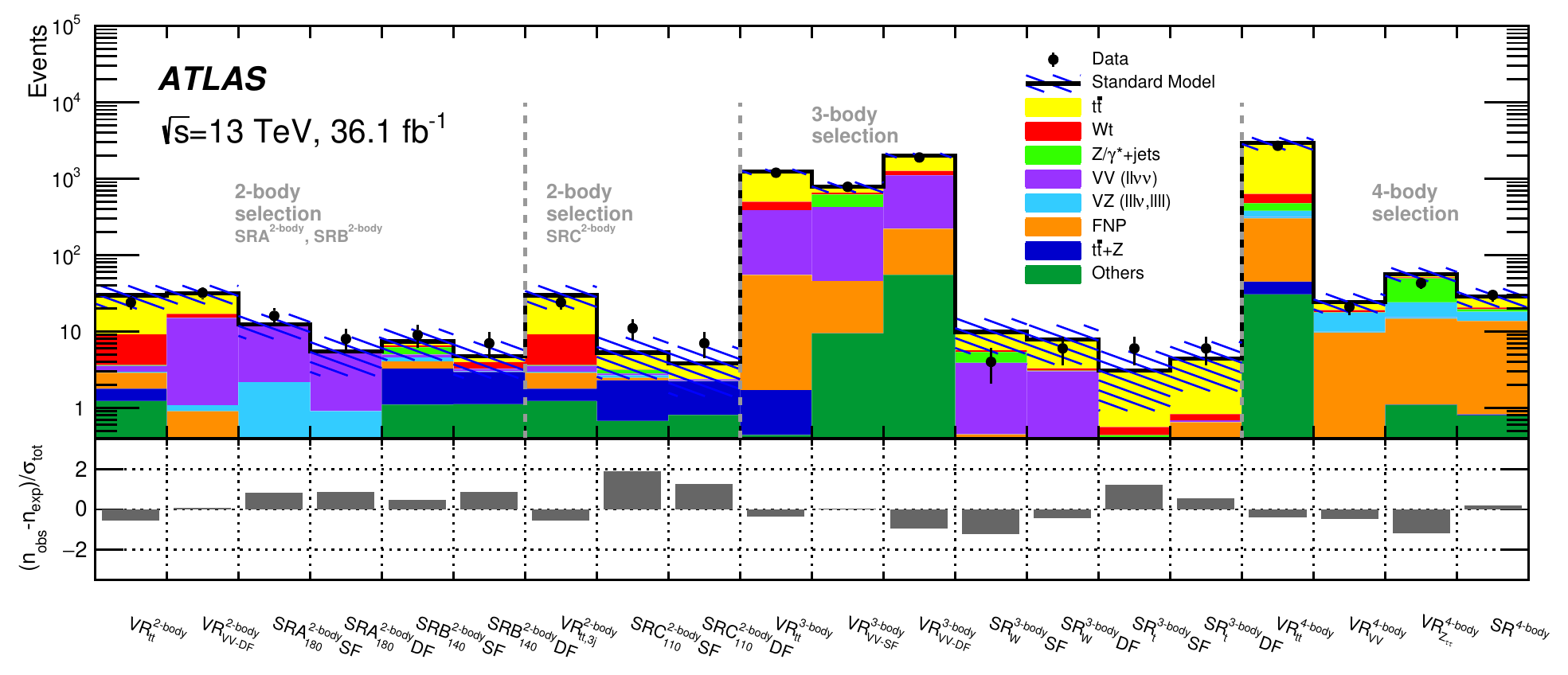}
\end{center}
\vspace{-0.8cm}
\caption{Comparison of the observed data ($n_{\textrm{obs}}$) with the predicted SM background ($n_{\textrm{exp}}$) in the SRs and associated VRs. The background predictions are obtained using the background-only fit configuration, and the hatched bands represent the total uncertainty in the background predictions after the fit to the data has been performed. The counting uncertainty on data is also shown by the black error bars. The bottom panel shows the difference between data and the predicted SM background divided by the total uncertainty ($\sigma_{\textrm{tot}}$).}
\label{fig:VRSRpulls}
\end{figure}

\FloatBarrier 
\subsection{Two-body results}
\label{sec:twobody_results}

Figure \ref{fig:twobody_SR} shows the \mttwoL distribution in each of the two-body signal regions, split between the same- and different-flavour lepton channels, omitting the selection on \mttwoL itself. The estimated SM yields in \SRA\ and \SRB\ are determined with a background fit simultaneously determining the normalisations of the background contributions from \ttbar, diboson with a SF lepton pair, \ttbar+\Zboson and diboson with more than two charged leptons by including \CRTbC, \CRVVbC, \CRTZ\ and \CRVZbC\ in the likelihood minimisation. The estimated SM yields in \SRtN\ are determined with a background fit simultaneously determining the normalisations of the background contributions from \ttbar\ and \ttbar+\Zboson by including \CRTtN\ and \CRTZ\ in the likelihood minimisation. No significant excess over the SM prediction is observed, as can be seen from the background-only fit results which are shown in Table~\ref{tab:bC1bkgonly_SR} for \SRA\ and \SRB, and Table~\ref{tab:tN1bkgonly_SR} for the \SRtN. Table~\ref{tab:bC1bkgonly_SRshape} reports the observed and expected yields for the SRs used for the computation of the exclusion limits.

\begin{figure}[t]
    \centering
    \includegraphics[width=0.48\textwidth]{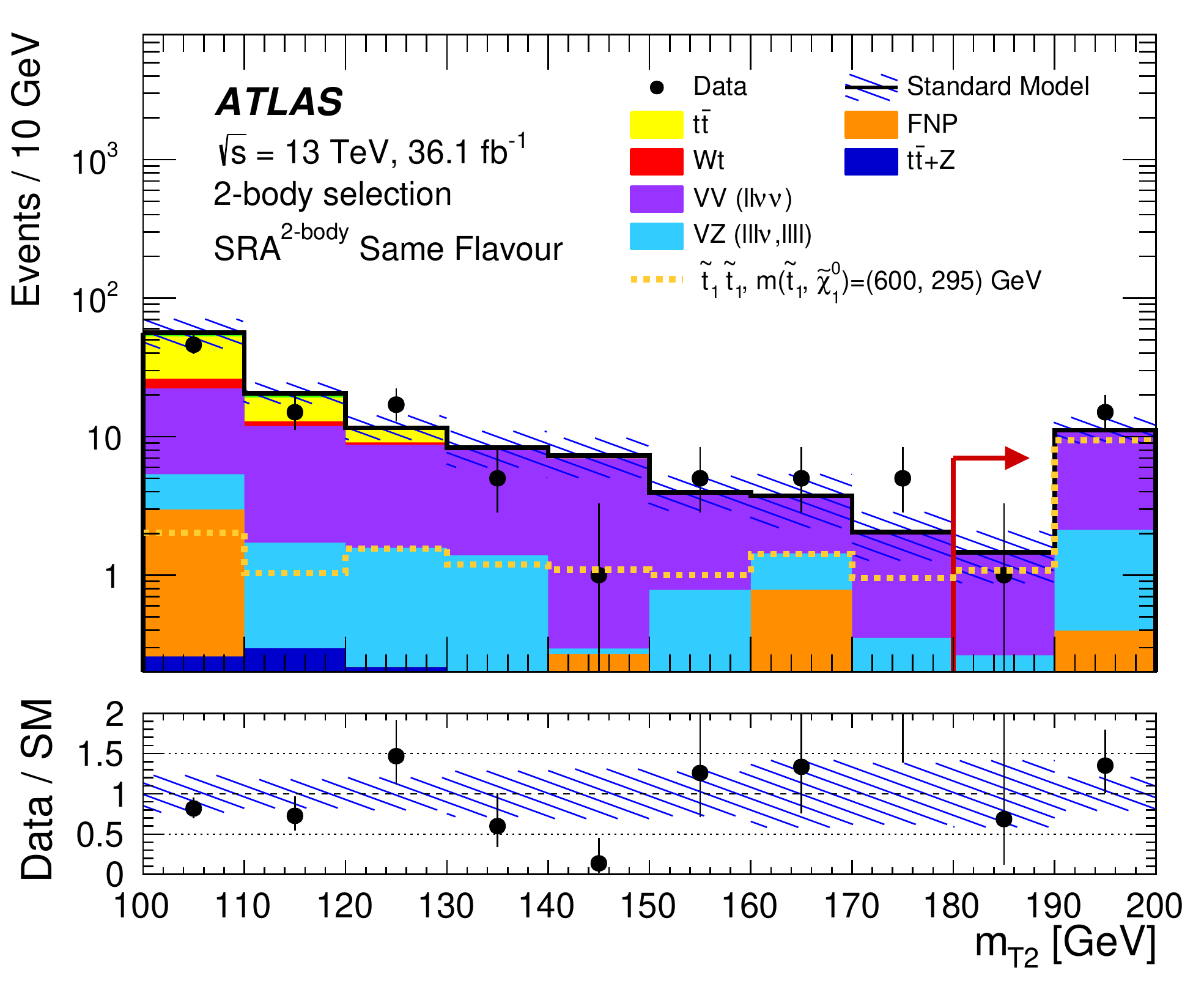}
    \includegraphics[width=0.48\textwidth]{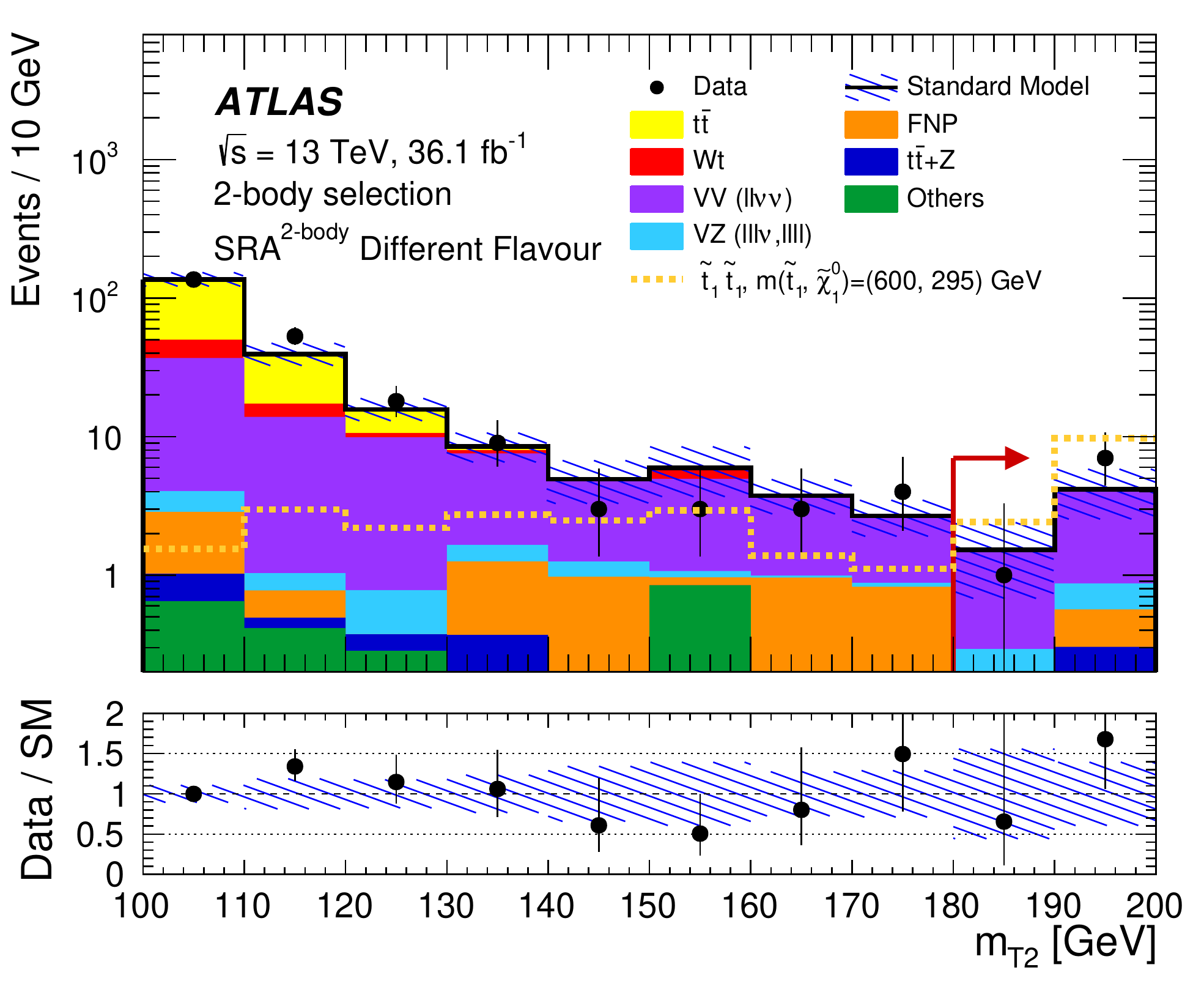}
    \includegraphics[width=0.48\textwidth]{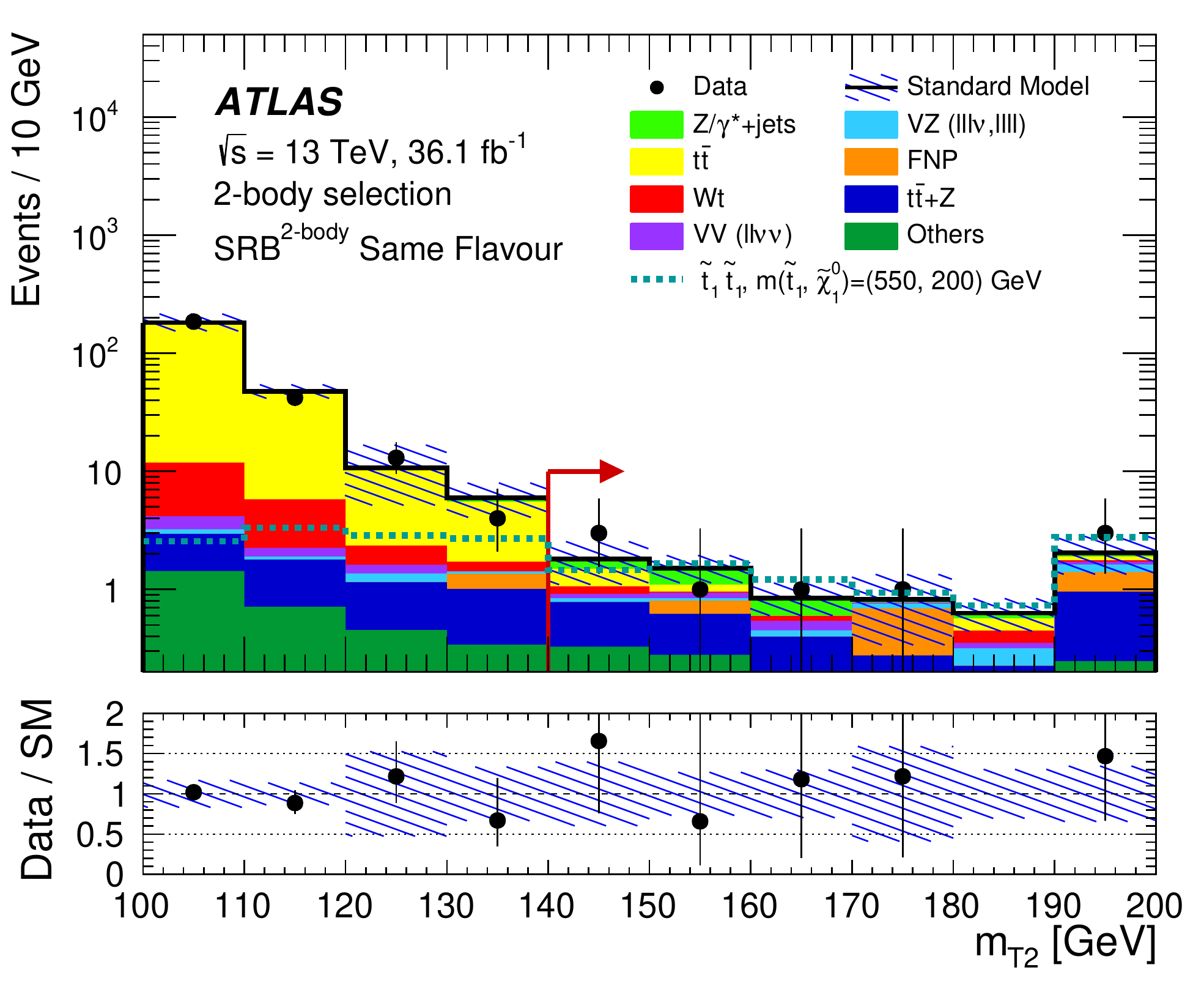}
    \includegraphics[width=0.48\textwidth]{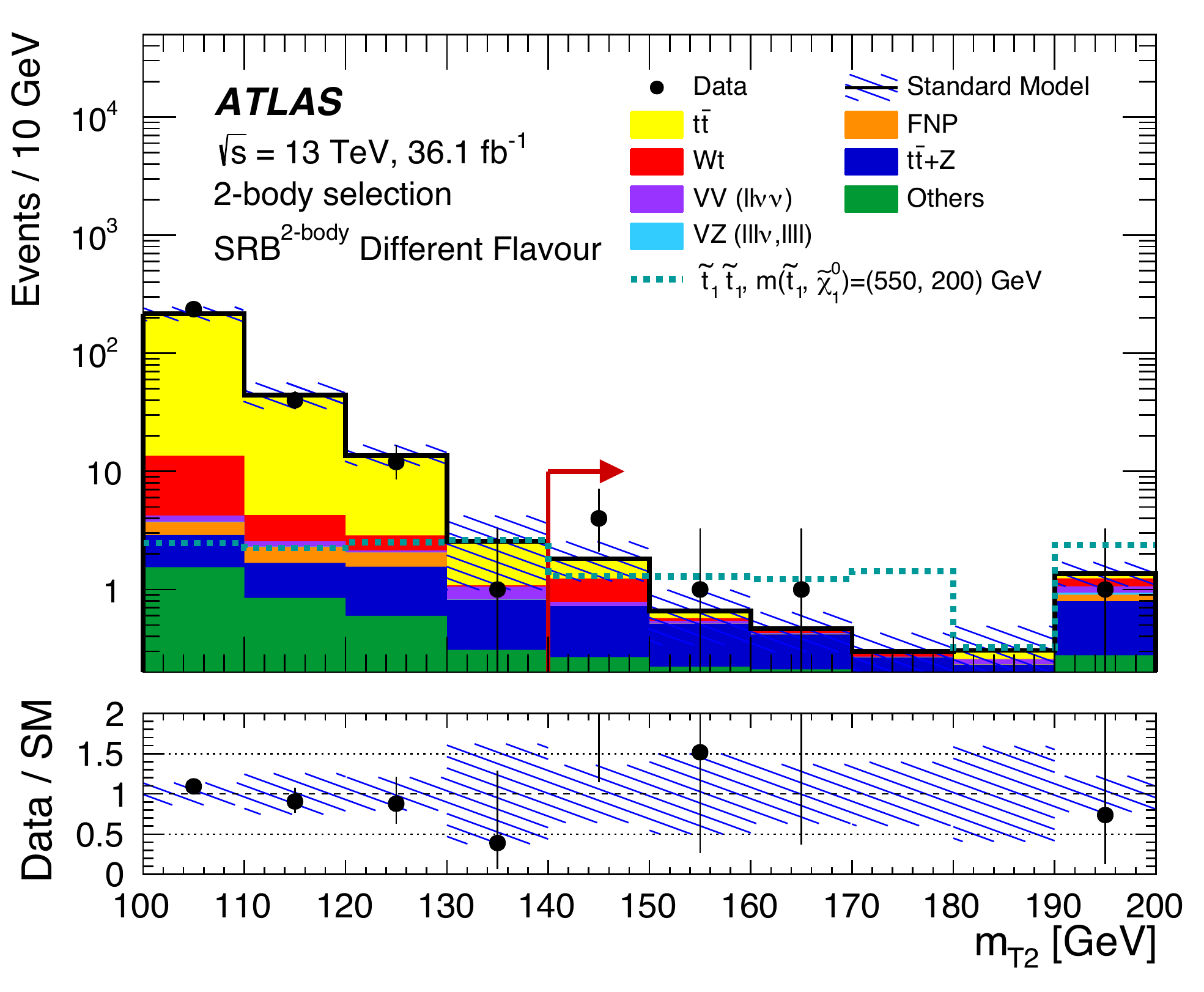}
    \includegraphics[width=0.48\textwidth]{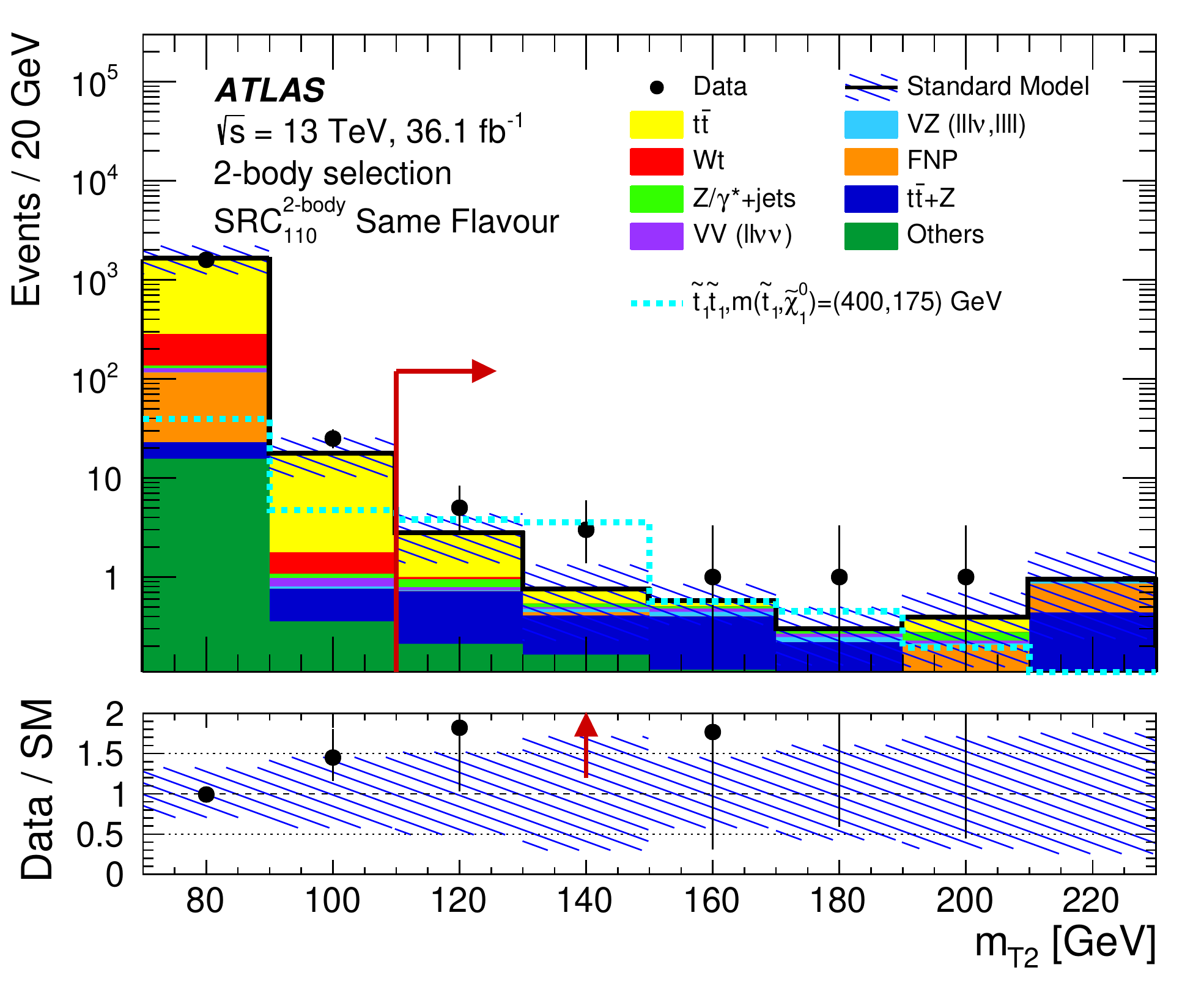}
    \includegraphics[width=0.48\textwidth]{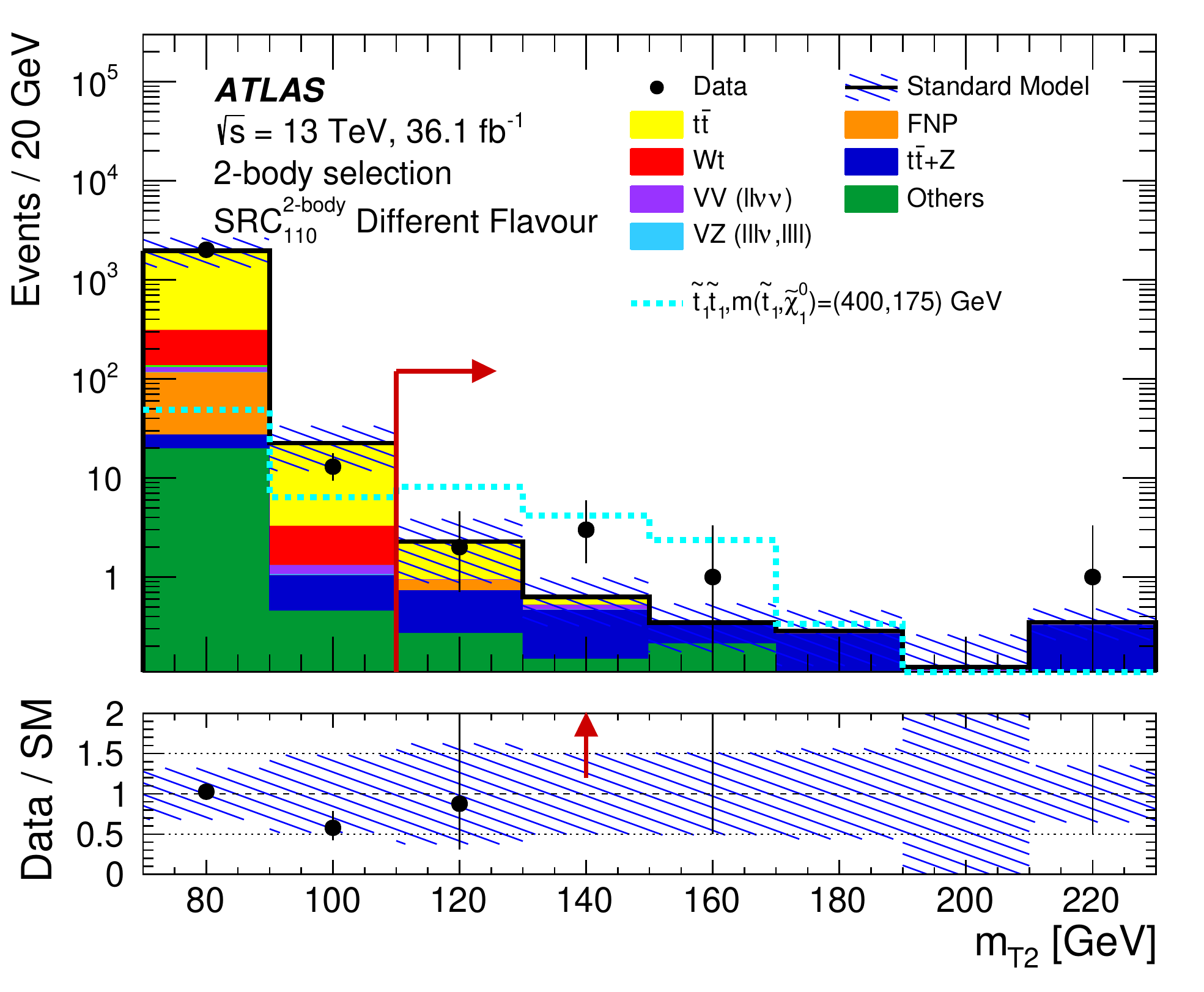}    
    \caption{\textit{Two-body selection} distributions of \mttwoL for events satisfying the selection criteria of the six SRs, except for the one on \mttwoL, after the background fit. The contributions from all SM backgrounds are shown as a histogram stack; the hatched bands represent the total uncertainty in the background predictions after the fit to the data has been performed. The counting uncertainty on data is also shown by the black error bars. The rightmost bin of each plot includes overflow events. Reference top squark pair production signal models are overlayed for comparison. Red arrows indicate the signal region selection criteria.
    }
    \label{fig:twobody_SR}
\end{figure}

\begin{table}[t]
\caption{\textit{Two-body selection} background fit results for \SRA\ and \SRB. The nominal predictions from MC simulation, are given for comparison for those backgrounds (\ttbar, $VV$-SF, $\ttbar Z$ and $VZ$) that are normalised to data in dedicated CRs. The ``Others'' category contains the contributions from $\ttbar W$, $\ttbar h$, $\ttbar WW$, $\ttbar t$, $\ttbar \ttbar$, $Wh$, $ggh$ and $Zh$ production. Combined statistical and systematic uncertainties are given. Entries marked ``--'' indicate a negligible background contribution.  The ``Others'' contribution to \SRB\ is dominated by $\ttbar W$.
}
\label{tab:bC1bkgonly_SR}
\begin{center}
\setlength{\tabcolsep}{0.1pc}
{\small
\resizebox{\textwidth}{!}{
\begin{tabular*}{\textwidth}{@{\extracolsep{\fill}}lrrrr}
\noalign{\smallskip}\hline\noalign{\smallskip}
                             & \SRA\ SF         & \SRA\ DF         & \SRB\ SF             & \SRB\ DF              \\[-0.05cm]
\noalign{\smallskip}\hline\noalign{\smallskip}
Observed events              & $16$            & $8$             & $9$                 & $7$                    \\
\noalign{\smallskip}\hline\noalign{\smallskip}
Estimated SM Events            &$12.3 \pm 2.6$   & $5.4 \pm 1.7$  & $7.4 \pm 1.1$       & $4.8 \pm 1.0$              \\
\noalign{\smallskip}\hline\noalign{\smallskip}
\ttbar           & --              & --              & $0.8 \pm 0.4$       & $0.8 \pm 0.5$              \\
$Wt$ events                  & --              & --              & $0.38 \pm 0.29$     & $0.7 \pm 0.5$              \\
\zjets            & $0.35 \pm 0.21$ & --              & $1.24 \pm 0.32$ 	   & $0.03 \pm 0.01$              \\
Fake and non-prompt          & $0.00_{-0.00}^{+0.30}$            & $0.00_{-0.00}^{+0.30}$              & $0.8 \pm 0.5$       & $0.00_{-0.00}^{+0.30}$					\\
$VV$-DF                      & --              & $4.5 \pm 1.5$   & --                  & $0.23 \pm 0.06$              \\
$VV$-SF          & $9.8 \pm 2.5$   & --              & $0.39 \pm 0.11$     & --              \\
$VZ$             & $1.91 \pm 0.31$ & $0.52 \pm 0.17$ & $0.53 \pm 0.14$     & $0.04 \pm 0.01$              \\
\ttbar + \Zboson & $0.08 \pm 0.03$ & $0.15 \pm 0.06$ & $2.3 \pm 0.6$     & $1.8 \pm 0.5$              \\
Others                       & $0.18 \pm 0.02$ & $0.24 \pm 0.07$ & $1.10 \pm 0.16$     & $1.11 \pm 0.16$              \\
 \noalign{\smallskip}\hline\noalign{\smallskip}
Nominal MC, \ttbar                       & --              & --              & $0.78$              & $0.8$              \\
Nominal MC, $VV$-SF           & $8.8$           & --              & $0.35$              & --              \\
Nominal MC, $VZ$              & $1.9$           & $0.52$          & $0.54$              & $0.04$              \\
Nominal MC, \ttbar + \Zboson  & $0.09$          & $0.17$          & $2.6$               & $2.2$              \\
\noalign{\smallskip}\hline\noalign{\smallskip}
\end{tabular*}
}
}
\end{center}
\end{table}

\begin{table}
\caption{\textit{Two-body selection} background fit results for \SRtN. The nominal predictions from MC simulation, are given for comparison for those backgrounds (\ttbar and $\ttbar Z$) that are normalised to data in dedicated CRs. 
The ``Others'' category contains the contributions from $\ttbar W$, $\ttbar h$, $\ttbar WW$, $\ttbar t$, $\ttbar \ttbar$, $Wh$, $ggh$ and $Zh$ production. 
Combined statistical and systematic uncertainties are given. Entries marked ``--'' indicate a negligible background contribution. }
\label{tab:tN1bkgonly_SR}
\begin{center}
\setlength{\tabcolsep}{2.0pc}
{\small
\begin{tabular*}{\textwidth}{@{\extracolsep{\fill}}lrr}
\noalign{\smallskip}\hline\noalign{\smallskip}
                             & \SRtN\ SF               & \SRtN\ DF \\ [-0.05cm]
\noalign{\smallskip}\hline\noalign{\smallskip}
Observed events              & 11                      & 7         \\
\noalign{\smallskip}\hline\noalign{\smallskip}
Estimated SM Events             & $5.3 \pm 1.8$           &   $3.8 \pm 1.5$  \\
\noalign{\smallskip}\hline\noalign{\smallskip}
\ttbar           & $2.1\pm 1.3$     		  &         $1.4\pm 1.2$    \\
\ttbar+Z         & $1.6 \pm 0.5$           &   $1.4 \pm 0.5$ \\
$Wt$                           &  $0.05_{-0.05}^{+0.09}$ &     $0.00_{-0.00}^{+0.23}$       \\
$VV$+$VZ$                    & $0.33 \pm 0.06$       & $0.12 \pm 0.04$  \\
\zjets                       & $0.3_{-0.3}^{+0.5}$                     &    --          \\
Others                       & $0.67 \pm 0.13$         &  $0.81 \pm 0.15$\\
Fake and non-prompt          & $0.18_{-0.18}^{+0.41}$& $0.00_{-0.00}^{+0.02}$  \\
\noalign{\smallskip}\hline\noalign{\smallskip}
Nominal MC, \ttbar   & $2.3$                  & $1.6$  \\
Nominal MC, \ttbar+Z & $1.9$                  & $1.70$ \\
\noalign{\smallskip}\hline\noalign{\smallskip}
\end{tabular*}
}
\end{center}
\end{table}

\begin{table}[b]
\caption{\textit{Two-body selection} background fit results for $\mathrm{SR(A,B)}^{\mathrm{2-body}}_{x,y}$ regions, where $x$ and $y$ denote the low and high edges of the bin. Combined statistical and systematic uncertainties are given. Uncertainties in the predicted background event yields are quoted as being symmetric.
}
\label{tab:bC1bkgonly_SRshape}
\begin{center}
\setlength{\tabcolsep}{0.1pc}
{\small
\resizebox{\textwidth}{!}{
\begin{tabular*}{\textwidth}{@{\extracolsep{\fill}}lrrrrrr}
\noalign{\smallskip}\hline\noalign{\smallskip}
                    &  Lepton flavour  & $\mathrm{SRA}^{\mathrm{2-body}}_{120,140}$ & $\mathrm{SRB}^{\mathrm{2-body}}_{120,140}$ & $\mathrm{SRA}^{\mathrm{2-body}}_{140,160}$ & $\mathrm{SRA}^{\mathrm{2-body}}_{160,180}$ \\[-0.05cm]
\noalign{\smallskip}\hline\noalign{\smallskip}
Observed events        & \multirow{2}{*}{SF} & $22$                                   & $17$                                   & $6$                                    & $10$                                   \\
Estimated SM Events &                     & $20.0 \pm 4.6$                         & $16.3 \pm 6.2$                         & $11.0 \pm 2.5$                         & $5.6 \pm 1.8$                          \\
\noalign{\smallskip}\hline\noalign{\smallskip}
Observed events        & \multirow{2}{*}{DF} & $27$                                   & $13$                                   & $6$                                    & $7$                                    \\
Estimated SM Events &                     & $23.8 \pm 4.2$                         & $16.1 \pm 5.3$                         & $10.8 \pm 2.1$                         & $6.4 \pm 1.3$                          \\
\noalign{\smallskip}\hline\noalign{\smallskip}
\end{tabular*}
}
}
\end{center}
\end{table}

\FloatBarrier
\subsection{Three-body results}
\label{sec:threebody_results}

Figure \ref{fig:threebodyNM1} shows the distributions of \RPT\ and \MDR\ in each of the signal regions, split between the same- and different-flavour channels, omitting the requirement on \RPT\ and on \MDR. The estimated SM yields in \SRTBW\ and \SRTBT\ are determined with a background fit simultaneously determining the normalisations of \ttbar, SF diboson production and DF diboson production by including \CRTtB, \CRVVSFtB\ and \CRVVDFtB\ in the likelihood minimisation. No excess over the SM prediction is observed. Table~\ref{tab:threebodybkgonly_SRwt} shows the background fit results. 

\begin{figure}[!htb]
\begin{center}
\subfloat[]{\includegraphics[width=0.48\textwidth]{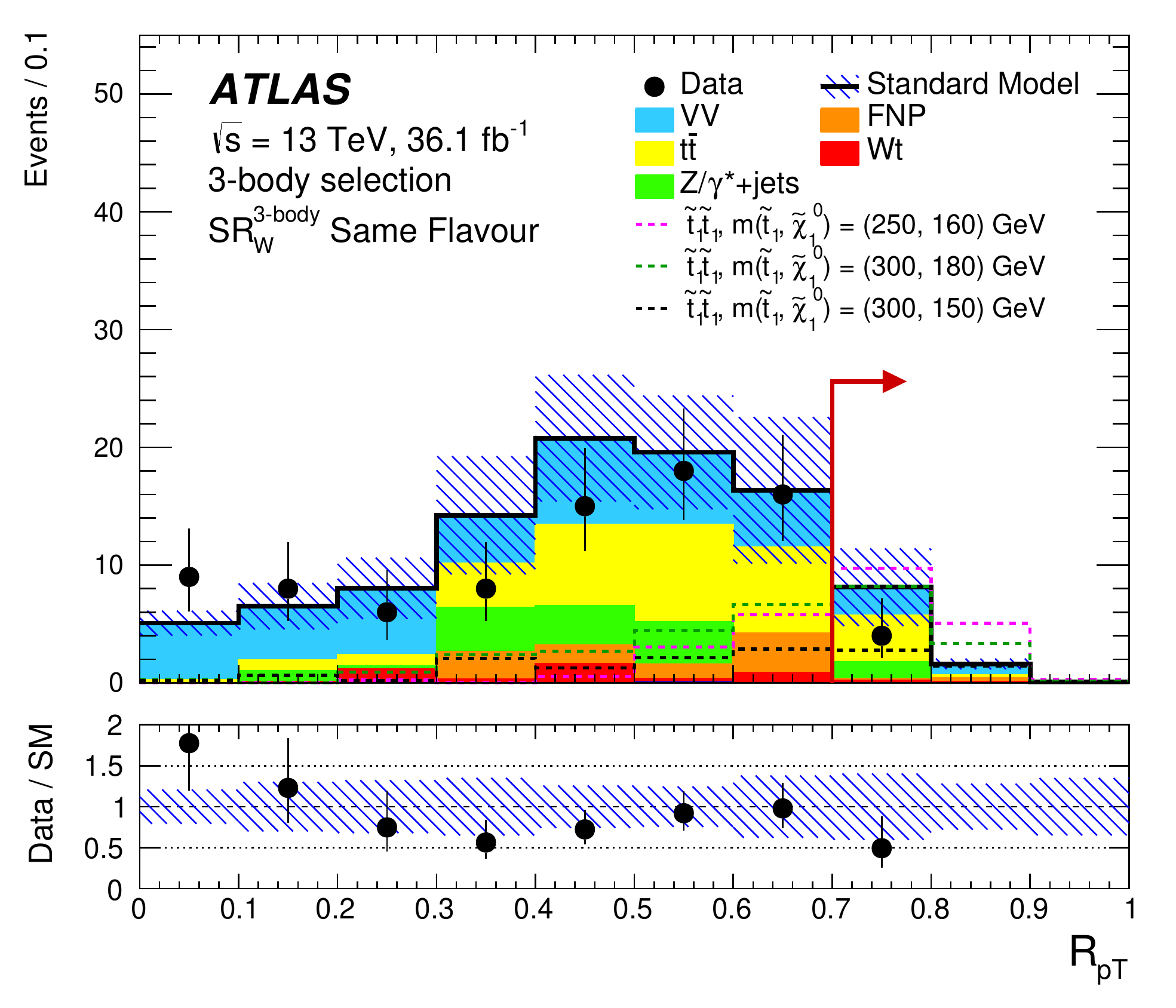}}
\subfloat[]{\includegraphics[width=0.48\textwidth]{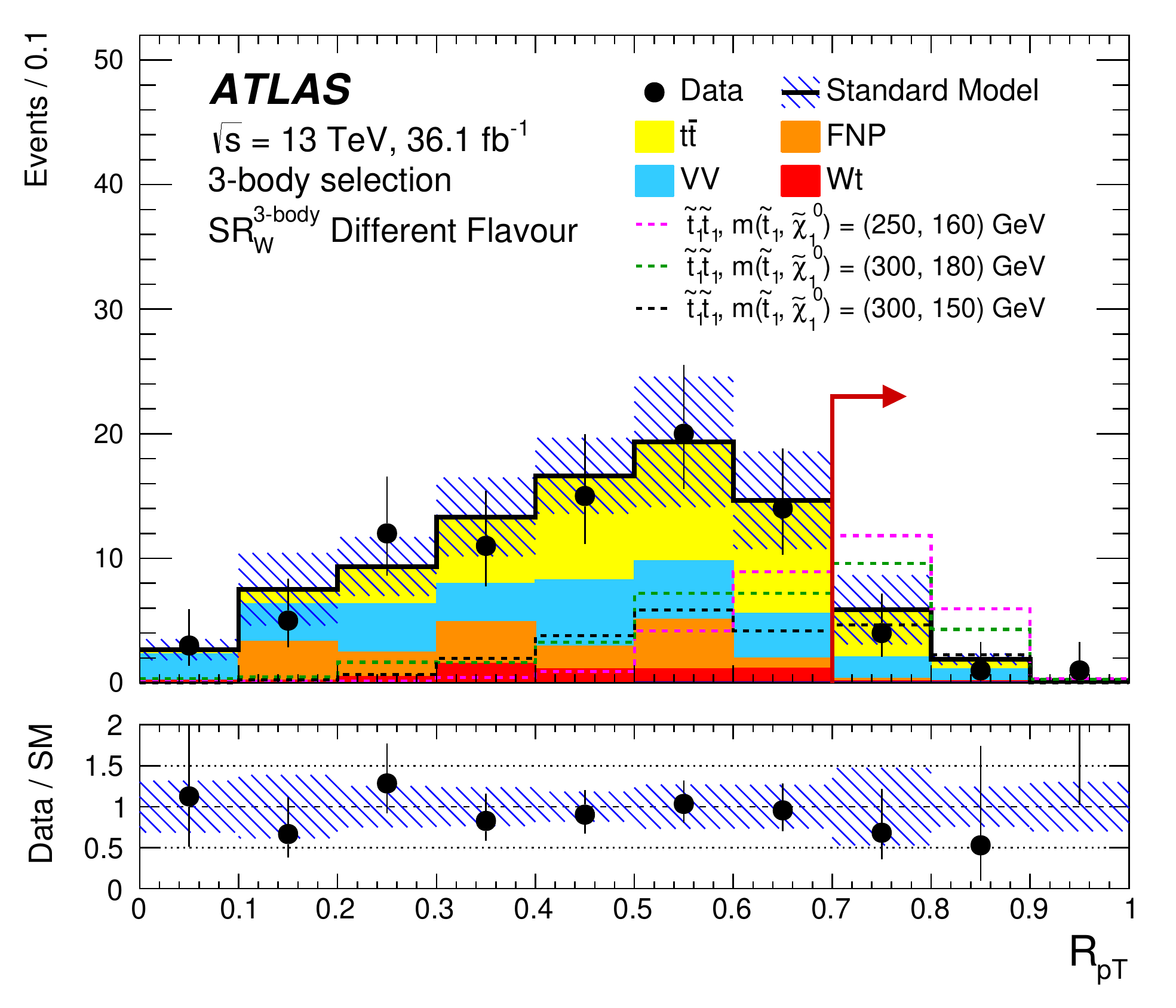}}\\
\subfloat[]{\includegraphics[width=0.48\textwidth]{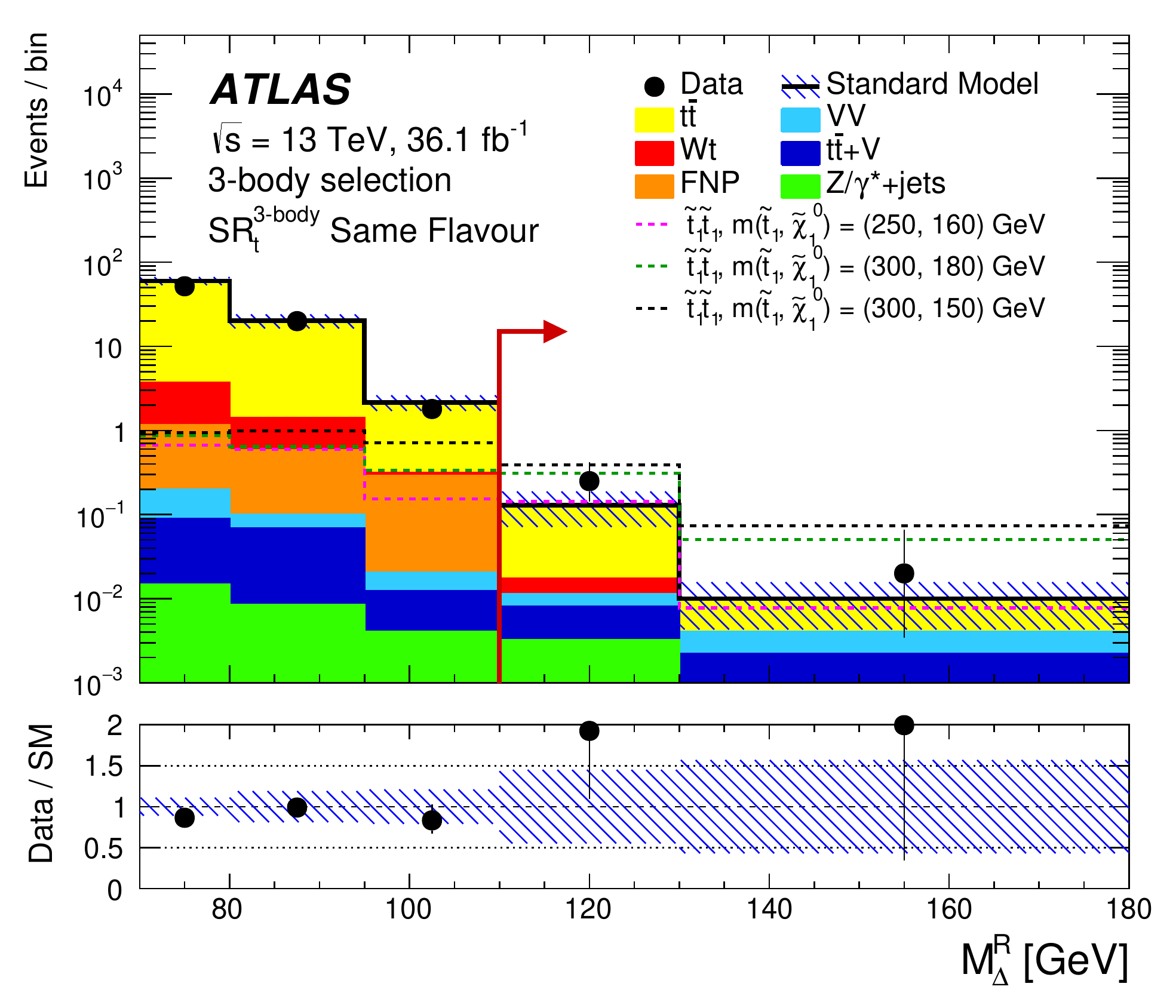}}
\subfloat[]{\includegraphics[width=0.48\textwidth]{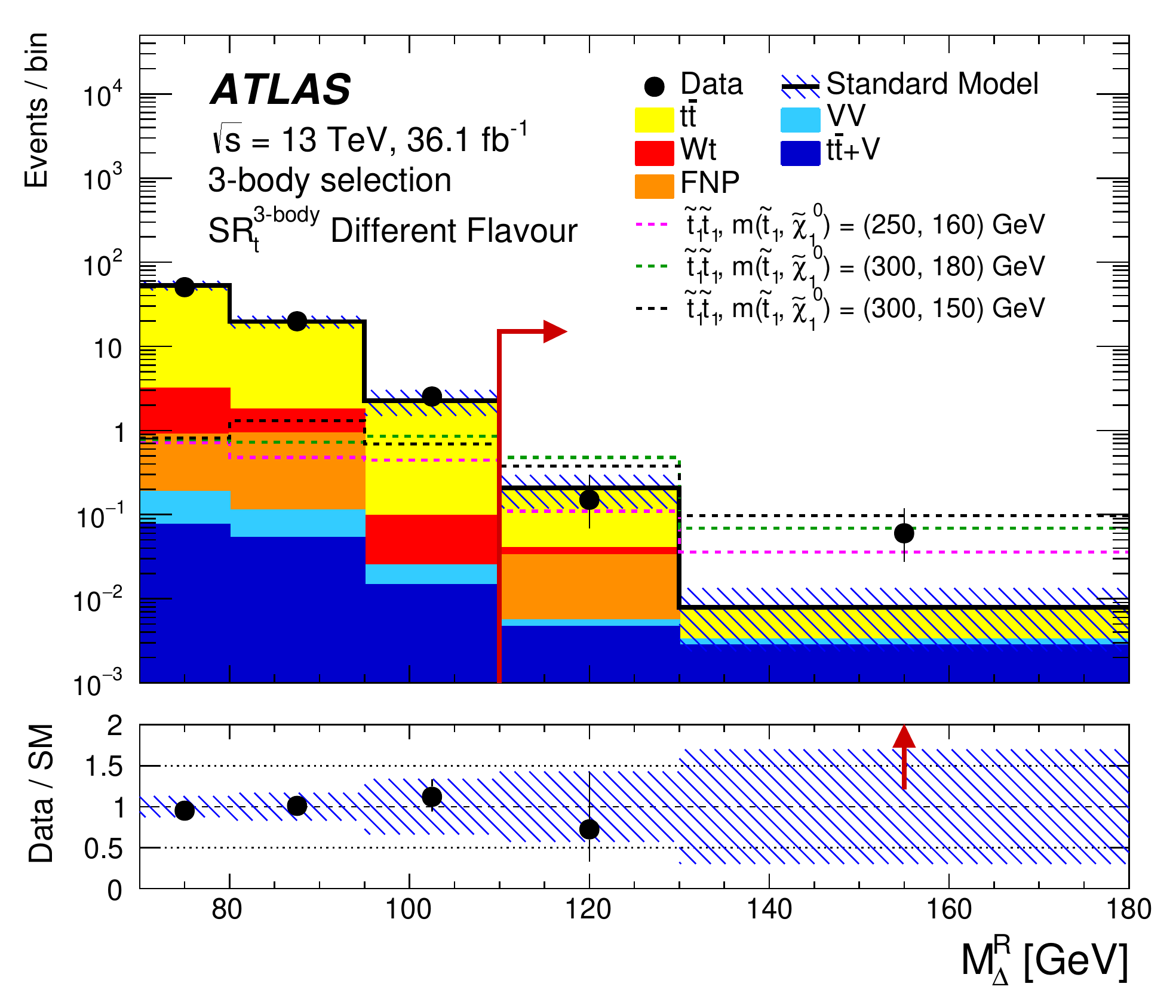}}
\end{center}
\caption{\textit{Three-body selection} distributions of \RPT\ in (a) same-flavour and (b) different-flavour events that satisfy all the \SRTBW\ selection criteria except for the one on \RPT, and of \MDR\ in the (c) same-flavour and (d)  different-flavour events that satisfy all the \SRTBT\ selection criteria except for the one on \MDR\ after the background fit. The contributions from all SM backgrounds are shown as a histogram stack; the hatched bands represent the total uncertainty in the background predictions after the fit to the data has been performed. The counting uncertainty on data is also shown by the black error bars. The rightmost bin of each plot includes overflow events. Reference top squark pair production signal models are overlayed for comparison. Red arrows indicate the signal region selection criteria.
}
\label{fig:threebodyNM1}
\end{figure}

\begin{table}[!htb]
\caption{\textit{Three-body selection} background fit results for \SRTBW\ and \SRTBT. The nominal predictions from MC simulation, are given for comparison for those backgrounds (\ttbar, $VV$-DF and $VV$-SF) that are normalised to data in dedicated CRs. Combined statistical and systematic uncertainties are given. Entries marked ``--'' indicate a negligible background contribution. }
\label{tab:threebodybkgonly_SRwt}
\begin{center}
\setlength{\tabcolsep}{0.0pc}
{\small
\begin{tabular*}{\textwidth}{@{\extracolsep{\fill}}lrrrr}
\noalign{\smallskip}\hline\noalign{\smallskip}
                       & \SRTBW\ SF     & \SRTBW\ DF     & \SRTBT\ SF     & \SRTBT\ DF              \\[-0.05cm]
\noalign{\smallskip}\hline\noalign{\smallskip}
Observed events        & $4$           & $6$           & $6$           & $6$            \\
\noalign{\smallskip}\hline\noalign{\smallskip}
Estimated SM Events      & $9.8\pm3.4$   & $7.8\pm3.0$   & $3.1\pm1.4$   & $4.4\pm1.8$    \\
\noalign{\smallskip}\hline\noalign{\smallskip}
$t\bar{t}$ & $4.2\pm1.6$   & $4.6\pm2.1$   & $2.5\pm1.3$   & $3.6\pm1.8$    \\
$VV$-DF    & --            & $2.9\pm1.4$   & --            & $0.04\pm0.03$    \\
$VV$-SF    & $3.4\pm2.1$   & --            & $0.16\pm0.08$ & --    \\
$Wt$                   & $0.31\pm0.22$ & $0.23\pm0.12$ & $0.12\pm0.05$ & $0.14\pm0.08$  \\
$t\bar{t}$+$V$         & $0.03\pm0.01$ & $0.06\pm0.02$ & $0.18\pm0.04$ & $0.24\pm0.07$  \\
$Z/\gamma^*$+jets       & $1.5\pm0.7$   & $0.05\pm0.01$ & $0.1\pm0.03$  & $0.0\pm0.0$    \\
Fake and non-prompt    & $0.42\pm0.28$ & $0.06\pm0.06$ & $0.00_{-0.00}^{+0.30}$   & $0.41\pm0.09$  \\
\noalign{\smallskip}\hline\noalign{\smallskip}
Nominal MC, $t\bar{t}$  & $4.0$         & $4.3$         & $2.4$         & $3.4$    \\
Nominal MC, $VV$-DF     & --            & $2.8$         & --            & $0.04$    \\
Nominal MC, $VV$-SF     & $3.4$         & --            & $0.16$        & --    \\
\noalign{\smallskip}\hline\noalign{\smallskip}
\end{tabular*}
}
\end{center}
\end{table}

\FloatBarrier 
\subsection{Four-body results}
\label{sec:fourbody_results}

Figure \ref{fig:fourbodySRplots} shows the distributions of $R_{2\ell 4j}$ and \Rll\ for events satisfying all the \SRfB\ selections. No significant excess over the SM prediction is visible. The estimated SM yields in \SRfB\ are determined with a background fit simultaneously determining the normalisations of \ttbar, diboson production, and \zjets where $\Zboson \rightarrow \tau\tau$, by including \CRTfB, \CRVVfB\ and \CRZfB\ in the likelihood minimisation. The background fit results are shown in Table~\ref{tab:fourbodybkgonly_SR}. The observed yield is less than one standard deviation from the background prediction in the SR. 

\begin{figure}[!htb]
\begin{center}
\subfloat[]{\includegraphics[width=0.48\textwidth]{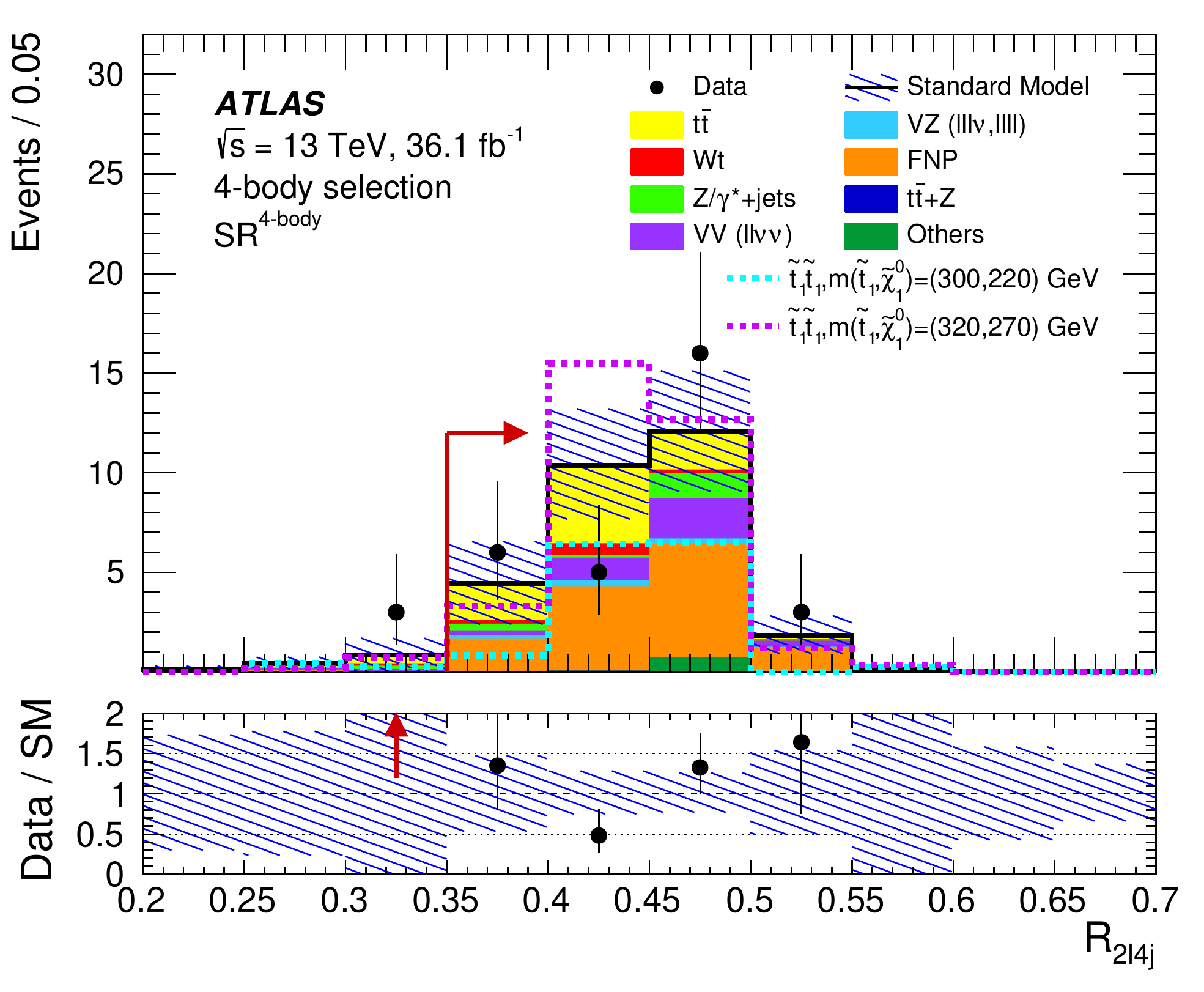}}
\subfloat[]{\includegraphics[width=0.48\textwidth]{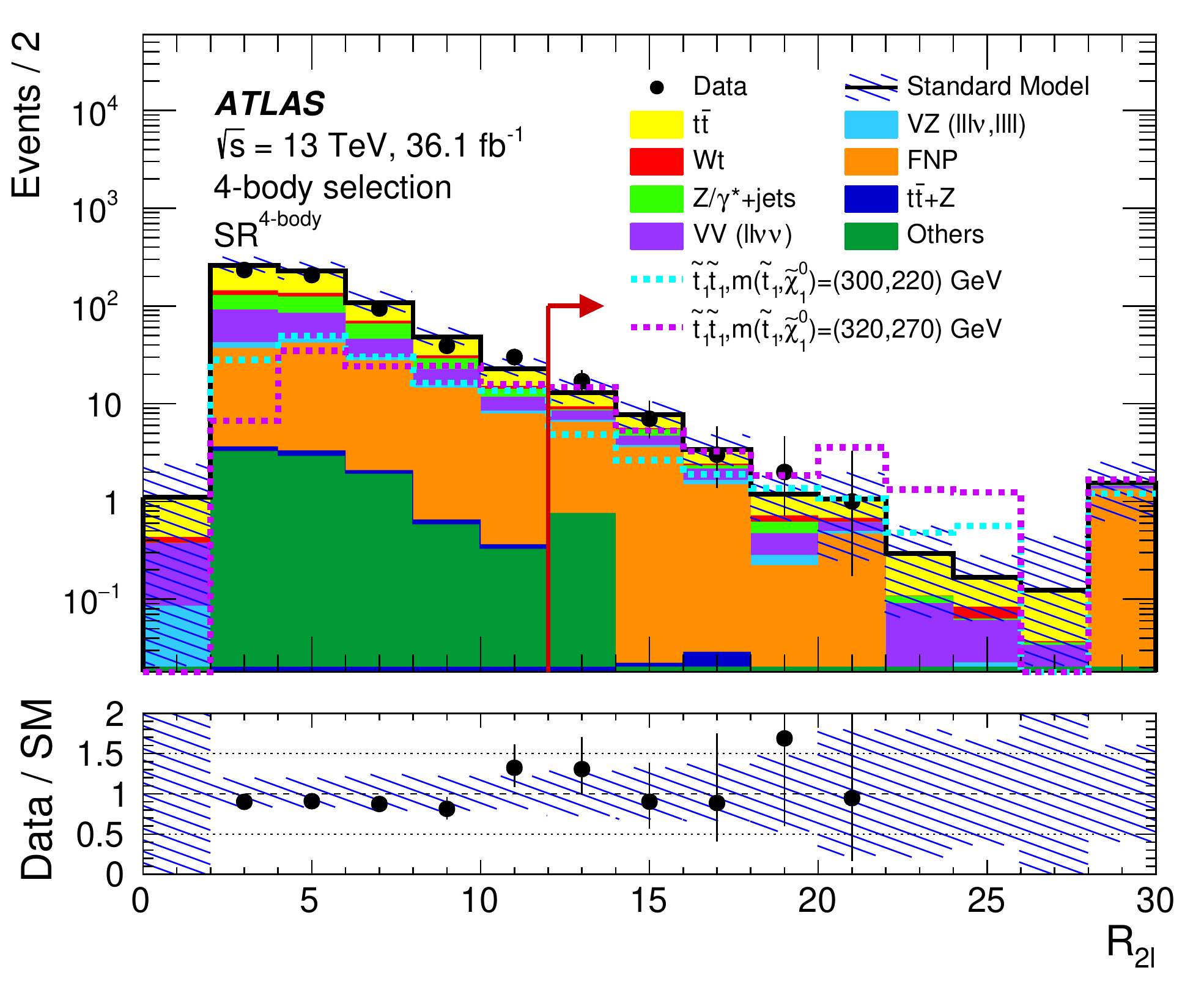}}
\end{center}
\caption{\textit{Four-body selection} distributions of (a) $R_{2\ell 4j}$ and (b) \Rll\ for events satisfying all the \SRfB\ selections except for the one on the variable shown in the figure, after the background fit. The contributions from all SM backgrounds are shown as a histogram stack; the hatched bands represent the total uncertainty in the background predictions after the fit to the data has been performed. The counting uncertainty on data is also shown by the black error bars. The rightmost bin of each plot includes overflow events. Reference top squark pair production signal models are overlayed for comparison. Red arrows indicate the signal region selection criteria.}
\label{fig:fourbodySRplots}
\end{figure}

\begin{table}[!htb]
\caption{\textit{Four-body selection} background fit results for \SRfB. The nominal predictions from MC simulation, are given for comparison for those backgrounds (\ttbar, VV and $Z_{\tau\tau}$) that are normalised to data in dedicated CRs. The ``Others'' category contains the contributions from $\ttbar W$, $\ttbar h$, $\ttbar WW$, $\ttbar t$, $\ttbar \ttbar$, $Wh$, $ggh$ and $Zh$ production. Combined statistical and systematic uncertainties are given.
}
\label{tab:fourbodybkgonly_SR}
\begin{center}
\setlength{\tabcolsep}{2.0pc}
{\small
\begin{tabular*}{0.55\textwidth}{@{\extracolsep{\fill}}lr}
\noalign{\smallskip}\hline\noalign{\smallskip}
                                            & \SRfB            \\[-0.05cm]
\noalign{\smallskip}\hline\noalign{\smallskip}
Observed events                                                 & 30                                  \\
\noalign{\smallskip}\hline\noalign{\smallskip}
Estimated SM Events                         						& $28 \pm 6$                      \\
\noalign{\smallskip}\hline\noalign{\smallskip}  
\ttbar                                                  & $7.9 \pm 2.0$                       \\
$VV$                                                      & $4.5 \pm 2.3$                       \\
$Z_{\tau\tau}$                                          & $1.2 \pm 0.6$                           \\
\ttbar+Z                                                            & $0.03\pm 0.01$                      \\
$Wt$                                                                  & $1.08 \pm 0.27$                     \\    
$\Zboson_{ee}$, $Z_{\mu\mu}$                                    & $0.21 \pm 0.09$                    \\
Others                                                              & $0.80 \pm 0.30$                  \\   
Fake and non-prompt                                                  & $12.8 \pm 4.3$                      \\
\noalign{\smallskip}\hline\noalign{\smallskip}
Nominal MC, \ttbar                                                   & $7.7$                   \\        
Nominal MC, $VV$                                                       & $5.7$                           \\
Nominal MC, $Z_{\tau\tau}$                                           & $1.1$                       \\  
\noalign{\smallskip}\hline\noalign{\smallskip}
\end{tabular*}
}
\end{center}
\end{table}

\FloatBarrier 
\subsection{Interpretation}
\label{sec:interpretations}

Two different sets of exclusion limits are derived for models of new physics beyond the SM. A model-independent upper limit on the visible cross-section $\sigma_{\mathrm{vis}}$ of new physics, defined as the ratio between the upper limit at 95\% CL on the number of signal events $S^{95}$ and the integrated luminosity, is derived in each SR by performing a fit which includes the observed yield in the SR as a constraint, and a free signal yield in the SR as an additional process. The CL$_\mathrm{s}$ method~\cite{Read:2002hq} is used to derive all the exclusion confidence levels. These limits assume negligible signal contamination in the CRs. This assumption leads to conservative results when comparing with model-dependent limits for models that predict a sizeable contamination in the CRs. Model-independent upper limits are presented in Table~\ref{tab:modindep}.

\begin{table}[!hbp]
\caption{\label{tab:modindep}
Model-independent 95\% CL upper limits on the visible cross-section ($\sigma_{\mathrm{vis}}$) of new physics, the visible number of signal events ($S^{95}_{\textrm obs}$), the visible number of signal events ($S^{95}_{\textrm exp}$) given the expected number of background events (and $\pm1\sigma$ excursions of the expected number), and the discovery $p$-value ($p(s = 0)$), all calculated with pseudo-experiments, are shown for each SR.}
\begin{center}
{\small
\begin{tabular*}{\textwidth}{@{\extracolsep{\fill}}llcccc}
\noalign{\smallskip}\hline\noalign{\smallskip}
                            & \textbf{Signal Region} & $\sigma_{\mathrm{vis}}$ [fb] & $S_{\textrm obs}^{95}$ & $S_{\textrm exp}^{95}$     & $p(s=0)$  \\
\noalign{\smallskip}\hline\noalign{\smallskip}
\multirow{8}{*}{Two-body}  & \SRA\ SF             & $0.37$   & $13.2$   & ${10}^{+4}_{-3}$ & $0.20$\\
\\[-1em]
                            & \SRA\ DF             & $0.26$   & $9.5$    & ${7.0}^{+3.0}_{-1.8}$  & $0.19$\\
\\[-1em]
                            & \SRB\ SF             & $0.24$   & $8.6$    & ${7.2}^{+2.7}_{-1.8}$ & $0.28$\\
\\[-1em]
                            & \SRB\ DF             & $0.23$   & $8.4$    & ${6.0}^{+2.7}_{-1.3}$ & $0.19$\\
\\[-1em]
                            & \SRtN\ SF            & $0.36$   & $13.0$   & $7.4^{+3.1}_{-2.0}$    & $0.05$\\
\\[-1em]
                            & \SRtN\ DF            & $0.26$   & $9.5$    & $6.3^{+2.5}_{-1.6}$    & $0.12$\\
\\[-1em]
\noalign{\smallskip}\hline\noalign{\smallskip}
\multirow{5}{*}{Three-body} & \SRTBW-SF           & $0.17$   & $6.1$    & ${9}^{+4}_{-2}$  & 0.72 \\
\\[-1em]
                            & \SRTBW-DF           & $0.21$   & $7.5$    & ${8.5}^{+3.5}_{-2.0}$  & 0.85\\
\\[-1em]
                            & \SRTBT-SF           & $0.24$   & $8.8$    & ${6.0}^{+2.4}_{-1.4}$  & 0.12\\
\\[-1em]
                            & \SRTBT-DF           & $0.23$   & $8.2$    & ${6.6}^{+2.8}_{-1.6}$  & 0.28\\
\\[-1em]
\noalign{\smallskip}\hline\noalign{\smallskip}
\multirow{1}{*}{Four-body}  & \SRfB      		  & $0.48$   &  $17.4$  & $16^{+7}_{-5}$         & $0.37$ \\
\\[-1em]
\noalign{\smallskip}\hline\noalign{\smallskip}
\end{tabular*}
}
\end{center}
\end{table}

Model-dependent limits are computed for various \stopone pair production scenarios. Profile likelihood fits are performed including the expected signal yield and its associated uncertainties in the CRs and SRs. All limits are quoted at 95\% CL. When setting limits, the regions included in the \mttwoL shape fits ($\mathrm{SRA}^{\mathrm{2-body}}_{x,y}$ and $\mathrm{SRB}^{\mathrm{2-body}}_{x,y}$) are statistically combined. Similarly, the \SRTBW\ and \SRTBT\ signal regions are statistically combined as well. For each signal model, the SR with the best expected limit is used for setting the final limit.

Limits for simplified models in which pair-produced \stopone decay with 100\% branching ratio into a top quark and \ninoone are shown in the \stopone--\ninoone mass plane in Figure~\ref{fig:tN1_exclusion}. The various SRs cover the different \stopone mass ranges, as described in Table~\ref{tab:index}. Top squark masses up to 720~\GeV\ are excluded for a massless lightest neutralino. Neutralino masses up to 300~\GeV\ are excluded for $m_{\stopone}=645$~\GeV. In the three-body decay hypothesis, top squark masses are excluded up to 430~\GeV\ for $m_{\stopone}-m_{\ninoone}$ close to the \Wboson boson mass. In the four-body decay hypothesis, top squark masses are excluded up to 400~\GeV\ for $m_{\stopone}-m_{\ninoone}=40$~\GeV.

\begin{figure}[!htb]
    \centering
   \includegraphics[width=0.7\textwidth]{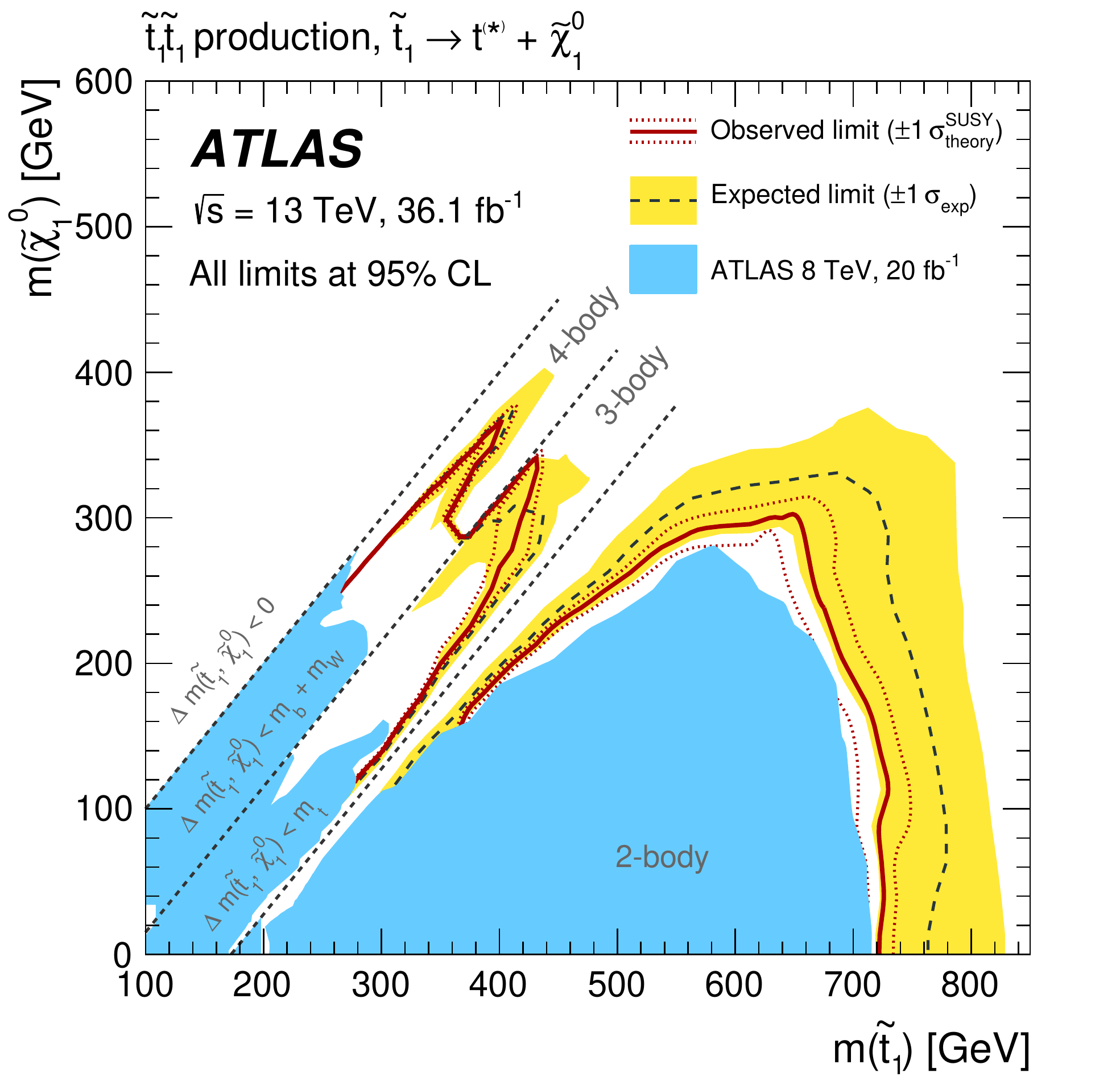}
\caption{Exclusion contour for a simplified model assuming \stopone pair production, decaying via $\stopone \rightarrow t^{(*)}\ninoone$ with 100\% branching ratio. 
  The dashed grey line and the shaded yellow band are the expected limit and its $\pm1\sigma$ uncertainty. 
  The thick solid red line is the observed limit for the central value of the signal cross-section. 
  The expected and observed limits do not include the effect of the theoretical uncertainties in the signal cross-section. 
  The dotted lines show the effect on the observed limit when varying the signal cross-section by $\pm1\sigma$ of the theoretical uncertainty. The shaded blue areas show the observed exclusion from the ATLAS $\sqrt{s}=8$~TeV analyses~\cite{Aad:2015pfx}.
  }
\label{fig:tN1_exclusion}
\end{figure}

Limits are shown for a class of simplified models in which only pair-produced \stopone decaying with 100\% branching ratio into the lightest chargino and a $b$-quark are considered. 
Figure~\ref{fig:bC1_exclusion} shows the interpretation in the \stopone--\ninoone mass plane assuming that $m_{\stopone} - m_{\chinoonepm} = 10$~\GeV. Top squark masses up to 700~\GeV\ are excluded for an LSP mass up to 200~\GeV.

\begin{figure}[!htb]
    \centering
   \includegraphics[width=0.7\textwidth]{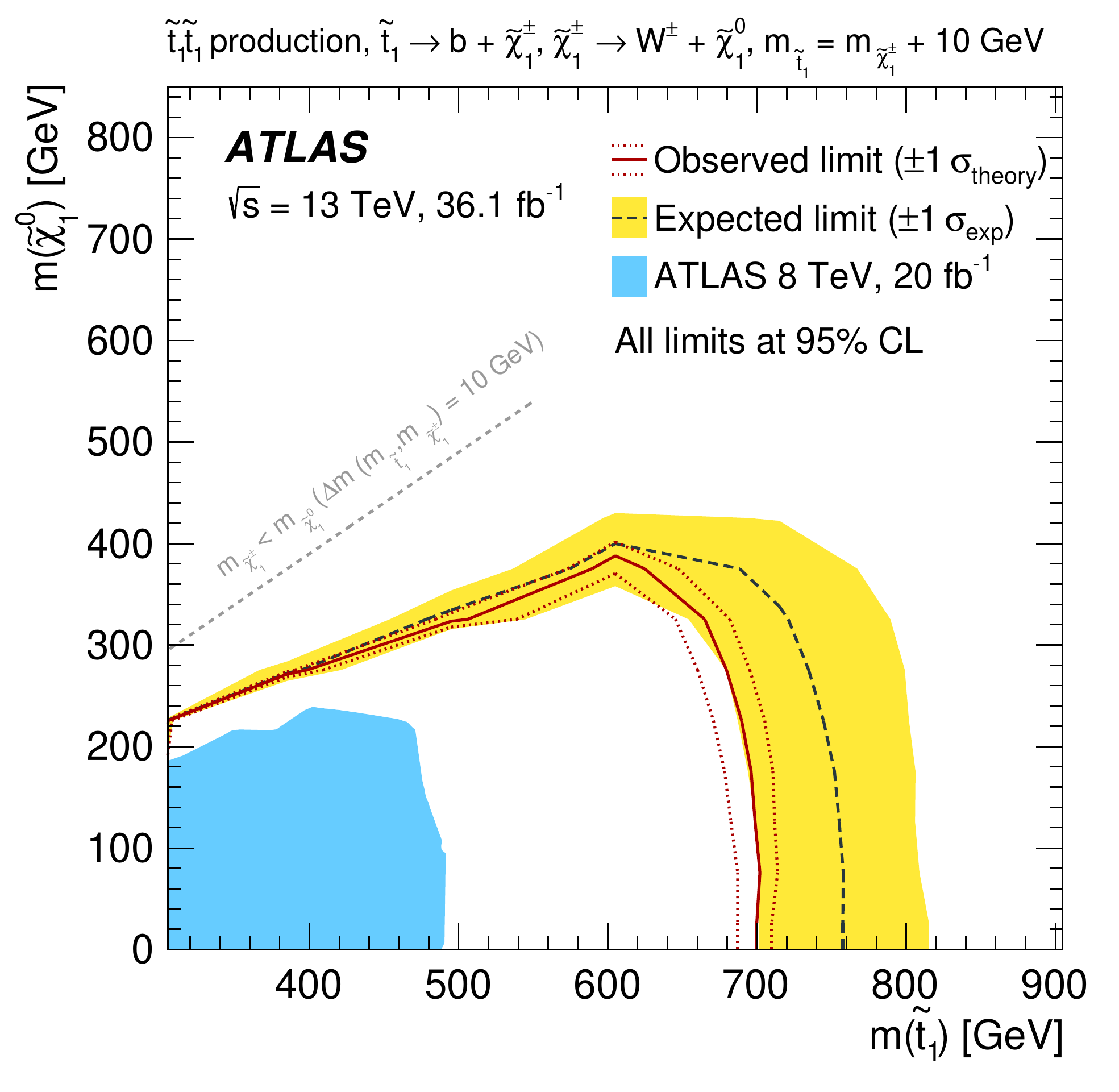}
    \caption{Exclusion contour for a simplified model assuming \stopone pair production, decaying via $\stopone \rightarrow b\chinoonepm$ with 100\% branching ratio. The lightest chargino mass is assumed to be close to the stop mass, $m_{\chinoonepm} = m_{\stopone}-10$~\GeV. The dashed grey line and the shaded yellow band are the expected limit and its $\pm1\sigma$ uncertainty. The thick solid red line is the observed limit for the central value of the signal cross-section. The expected and observed limits do not include the effect of the theoretical uncertainties in the signal cross-section. The dotted lines show the effect on the observed limit when varying the signal cross-section by $\pm1\sigma$ of the theoretical uncertainty. The shaded blue area shows the observed exclusion from the ATLAS $\sqrt{s}=8$~TeV analysis~\cite{Aad:2015pfx}.
    }
    \label{fig:bC1_exclusion}
\end{figure}

Finally, limits are set on a pMSSM model where the wino and bino mass parameters, $M_1$ and $M_2$, are set to $M_2 = 2M_1$ and $m_{\stopone} > m_{\chinoonepm}$. The remaining pMSSM parameters~\cite{Djouadi:1998di,Berger:2008cq} have the following values: $M_3 = 2.2$~\TeV\ (gluino mass parameter), $M_{S} = \sqrt{m_{\stopone} m_{\stoptwo}} = 1.2$~\TeV~(product of top squark masses), $X_{t}/M_{S} = \sqrt{6}$ (mixing parameter between the left- and right-handed states), and $\tan{\beta} = 20$ (ratio of vacuum expectation values of the two Higgs doublets).
The values of $M_3$ and $M_{S}$ have been chosen in order to avoid the current gluino and top squark mass limits, while the value of $X_{t}/M_{S}$ is assumed to obtain a low-mass lightest top squark while maintaining the models consistent with the observed Higgs boson mass of 125~\GeV. Limits are set for both the positive and negative values of $\mu$ (the Higgs mass parameter) as a function of $m_{\stopone}$ and $m_{\ninoone}$, and are shown in Figure~\ref{fig:wino_exclusion}. Top squark masses up to about 700~\GeV\ are excluded for a lightest neutralino of about 280~\GeV. The sensitivity for low values of $m_{\ninoone}$ is limited by the \mttwoL selection acceptance, since $m_{\chinoonepm} - m_{\ninoone}$ is reduced by assuming $M_2 = 2M_1$.

\begin{figure}[!htb]
    \centering
   \includegraphics[width=0.7\textwidth]{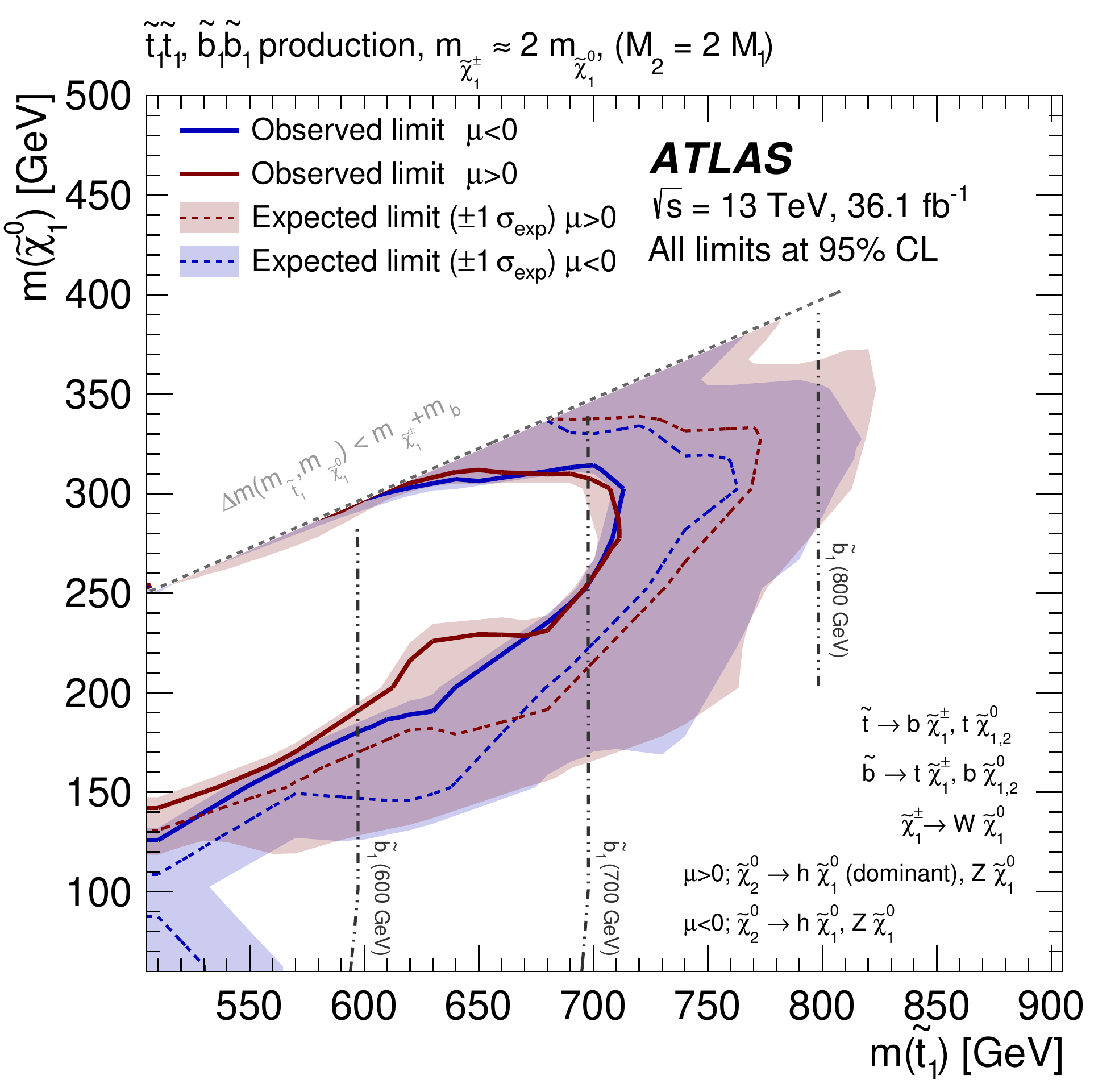}
    \caption{Exclusion contour as a function of $m_{\stopone}$ and $m_{\ninoone}$ in the pMSSM model described in the text. Pair production of \stopone and \sbottomone are considered. Limits are set for both the positive (red in the figure) and negative (blue in the figure) values of $\mu$. The dashed and dotted grey lines indicate constant values of the \sbottomone mass. The signal models included within the shown contours are excluded at 95\% CL. The dashed lines and the shaded band are the expected limit and its $\pm1\sigma$ uncertainty. The thick solid line is the observed limit for the central value of the signal cross-section. The expected and observed limits do not include the effect of the theoretical uncertainties in the signal cross-section.
    }
    \label{fig:wino_exclusion}
\end{figure}

\FloatBarrier
\section{Conclusion}
\label{sec:conclusion}

This article reports a search for direct top squark pair production in final states containing two opposite-charge leptons and large missing transverse momentum, based on a $36.1$~\ifb\ dataset of $\sqrt{s} = 13$~TeV proton--proton collisions recorded by the ATLAS experiment at the LHC in 2015 and 2016. Good agreement was found between the observed events in the data and the expected Standard Model yields.

Model-independent 95\% CL upper limits on the visible cross-section for new phenomena were computed.
The results are also interpreted in terms of simplified models assuming a range of top squark and lightest neutralino masses, with the former decaying into the latter via either a direct two-, three- or four-body decay or via an intermediate chargino state. In the case of top squark decays into $t^{(*)}\ninoone$, top squark masses below 720~\GeV\ are excluded for a massless lightest neutralino. In the three-body decay hypothesis, top squark masses are excluded up to 430~\GeV\ for $m_{\stopone}-m_{\ninoone}$ close to the \Wboson boson mass. In the four-body decay hypothesis, top squark masses are excluded up to 400~\GeV\ for $m_{\stopone}-m_{\ninoone}=40$~\GeV. Both these results extend the coverage of previous searches by about 100~\GeV.
The chargino decay mode, $\stopone \rightarrow b \chinoonepm$, is excluded for top squark masses up to 700~\GeV, assuming that $m_{\stopone} - m_{\chinoonepm} = 10$~\GeV, extending the previous results by almost 200~\GeV. When considering a pMSSM-inspired model including multiple decay chains, top squark masses up to about 700~\GeV\ are excluded for a lightest neutralino of about 280~\GeV. These results extend the region of supersymmetric parameter space excluded by previous LHC searches.

\section*{Acknowledgements}
%

We thank CERN for the very successful operation of the LHC, as well as the
support staff from our institutions without whom ATLAS could not be
operated efficiently.

We acknowledge the support of ANPCyT, Argentina; YerPhI, Armenia; ARC, Australia; BMWFW and FWF, Austria; ANAS, Azerbaijan; SSTC, Belarus; CNPq and FAPESP, Brazil; NSERC, NRC and CFI, Canada; CERN; CONICYT, Chile; CAS, MOST and NSFC, China; COLCIENCIAS, Colombia; MSMT CR, MPO CR and VSC CR, Czech Republic; DNRF and DNSRC, Denmark; IN2P3-CNRS, CEA-DRF/IRFU, France; SRNSF, Georgia; BMBF, HGF, and MPG, Germany; GSRT, Greece; RGC, Hong Kong SAR, China; ISF, I-CORE and Benoziyo Center, Israel; INFN, Italy; MEXT and JSPS, Japan; CNRST, Morocco; NWO, Netherlands; RCN, Norway; MNiSW and NCN, Poland; FCT, Portugal; MNE/IFA, Romania; MES of Russia and NRC KI, Russian Federation; JINR; MESTD, Serbia; MSSR, Slovakia; ARRS and MIZ\v{S}, Slovenia; DST/NRF, South Africa; MINECO, Spain; SRC and Wallenberg Foundation, Sweden; SERI, SNSF and Cantons of Bern and Geneva, Switzerland; MOST, Taiwan; TAEK, Turkey; STFC, United Kingdom; DOE and NSF, United States of America. In addition, individual groups and members have received support from BCKDF, the Canada Council, CANARIE, CRC, Compute Canada, FQRNT, and the Ontario Innovation Trust, Canada; EPLANET, ERC, ERDF, FP7, Horizon 2020 and Marie Sk{\l}odowska-Curie Actions, European Union; Investissements d'Avenir Labex and Idex, ANR, R{\'e}gion Auvergne and Fondation Partager le Savoir, France; DFG and AvH Foundation, Germany; Herakleitos, Thales and Aristeia programmes co-financed by EU-ESF and the Greek NSRF; BSF, GIF and Minerva, Israel; BRF, Norway; CERCA Programme Generalitat de Catalunya, Generalitat Valenciana, Spain; the Royal Society and Leverhulme Trust, United Kingdom.

The crucial computing support from all WLCG partners is acknowledged gratefully, in particular from CERN, the ATLAS Tier-1 facilities at TRIUMF (Canada), NDGF (Denmark, Norway, Sweden), CC-IN2P3 (France), KIT/GridKA (Germany), INFN-CNAF (Italy), NL-T1 (Netherlands), PIC (Spain), ASGC (Taiwan), RAL (UK) and BNL (USA), the Tier-2 facilities worldwide and large non-WLCG resource providers. Major contributors of computing resources are listed in Ref.~\cite{ATL-GEN-PUB-2016-002}.

\clearpage

\printbibliography

\newpage

\begin{flushleft}
{\Large The ATLAS Collaboration}

\bigskip

M.~Aaboud$^\textrm{\scriptsize 137d}$,
G.~Aad$^\textrm{\scriptsize 88}$,
B.~Abbott$^\textrm{\scriptsize 115}$,
O.~Abdinov$^\textrm{\scriptsize 12}$$^{,*}$,
B.~Abeloos$^\textrm{\scriptsize 119}$,
S.H.~Abidi$^\textrm{\scriptsize 161}$,
O.S.~AbouZeid$^\textrm{\scriptsize 139}$,
N.L.~Abraham$^\textrm{\scriptsize 151}$,
H.~Abramowicz$^\textrm{\scriptsize 155}$,
H.~Abreu$^\textrm{\scriptsize 154}$
% ...
\bigskip
\\
% ...
$^{12}$ Institute of Physics, Azerbaijan Academy of Sciences, Baku, Azerbaijan\\
% ...
$^{88}$ CPPM, Aix-Marseille Universit{\'e} and CNRS/IN2P3, Marseille, France\\
% ...
$^{115}$ Homer L. Dodge Department of Physics and Astronomy, University of Oklahoma, Norman OK, United States of America\\
% ...
$^{119}$ LAL, Univ. Paris-Sud, CNRS/IN2P3, Universit{\'e} Paris-Saclay, Orsay, France\\
% ...
$^{137}$ $^{(a)}$ Facult{\'e} des Sciences Ain Chock, R{\'e}seau Universitaire de Physique des Hautes Energies - Universit{\'e} Hassan II, Casablanca; $^{(b)}$ Centre National de l'Energie des Sciences Techniques Nucleaires, Rabat; $^{(c)}$ Facult{\'e} des Sciences Semlalia, Universit{\'e} Cadi Ayyad, LPHEA-Marrakech; $^{(d)}$ Facult{\'e} des Sciences, Universit{\'e} Mohamed Premier and LPTPM, Oujda; $^{(e)}$ Facult{\'e} des sciences, Universit{\'e} Mohammed V, Rabat, Morocco\\
% ...
$^{139}$ Santa Cruz Institute for Particle Physics, University of California Santa Cruz, Santa Cruz CA, United States of America\\
% ...
$^{151}$ Department of Physics and Astronomy, University of Sussex, Brighton, United Kingdom\\
% ...
$^{154}$ Department of Physics, Technion: Israel Institute of Technology, Haifa, Israel\\
$^{155}$ Raymond and Beverly Sackler School of Physics and Astronomy, Tel Aviv University, Tel Aviv, Israel\\
% ...
$^{161}$ Department of Physics, University of Toronto, Toronto ON, Canada\\
% ...
$^{a}$ Also at Department of Physics, King's College London, London, United Kingdom\\
$^{b}$ Also at Institute of Physics, Azerbaijan Academy of Sciences, Baku, Azerbaijan\\
$^{c}$ Also at Novosibirsk State University, Novosibirsk, Russia\\
$^{d}$ Also at TRIUMF, Vancouver BC, Canada\\
$^{e}$ Also at Department of Physics {\&} Astronomy, University of Louisville, Louisville, KY, United States of America\\
% ...
$^{*}$ Deceased
\end{flushleft}

\end{document}